\documentclass{jfm}
\usepackage{graphicx}
\usepackage{epstopdf, epsfig}
\usepackage{amsmath}
\usepackage{xspace}
\usepackage{makecell} %for writing in next line inside a cell in a table
\usepackage{subcaption}

\newcommand{\tc}{\textdegree\xspace}

\title{Contact Angle Hysteresis on Rough Surfaces Part II: Energy Dissipation via Microscale Interface Dynamics}

\author{Pawan Kumar\aff{1}
 \and Dalton J. E. Harvie\aff{1}}
   
\affiliation{\aff{1}Multiphysics Fluid Dynamics Group, Department of Chemical Engineering, University of Melbourne,
Parkville, VIC, 3010, Australia}
\begin{document}

\maketitle

\begin{abstract}
The wetting behaviour of surfaces is important for various applications like super-hydrophobic surfaces, enhanced oil recovery, mining of metal ores and anti-icing surfaces etc. For rough surfaces, which are the rule rather than the exception, designing textured surfaces that have wetting properties tailored to suit these applications generally involves either bio-mimicry or trial and error. Existing wetting theories such as the well-known \cite{wenzel1936resistance} and \cite{cassie1944wettability} models do not predict wetting regimes and importantly, only give a single equilibrium angle rather than a range of stable contact angles (hysteresis) as observed in reality. In this paper, we use a roughness-scale mechanical energy balance (derived in part I) combined with simulations of micro-scale interface dynamics based on open-source software (Surface Evolver \citep{Brakke1992}) to calculate the energy dissipation during the motion of an interface over a chemically homogeneous rough surface. This dissipation is used to predict contact angle hysteresis (CAH) from knowledge of just the surface roughness topography and equilibrium contact angle. We simulate interface dynamics over a surface decorated with a periodic array of round-edge square pillars and show that the energy dissipated varies approximately as $\phi \ln \phi$ with the area fraction ($\phi$), and becomes zero as $\phi \rightarrow 0$. The CAH predicted by our method is in good agreement with the experimental results of \cite{forsberg2010contact}, especially at low area fractions. We also compute CAH for an interface moving at 45\tc to the surface periodicity direction, showing that the experimental measurements are bracketed by the 0\tc and 45\tc advance direction results.
% 243 words
\end{abstract}

% \begin{keywords}
% wetting, contact angle hysteresis, dissipation, Surface Evolver
% \end{keywords}

\section{Introduction}
\label{sec:intro}

The presence of roughness can significantly affect the wettability of a solid \citep{butt2022contact}. The water repelling and self-cleaning property of a lotus leaf \citep{barthlott1997purity}, the ability of rice leaves to direct water droplets in a specific direction \citep{zhu2010mimicking}, water capture from the air by a desert beetle \citep{parker2001water} and the ability of a water strider to walk on water \citep{feng2006design} are some of the examples where this relationship has been used in the nature. In many engineering and industrial applications this relationship plays a very important role as well. The efficiency of industrial floatation cells \citep{chau2009review} and electronic printed devices      
 \citep{meyers2008aqueous}, passive oil-water separation \citep{li2013structured}, self-cleaning surfaces \citep{xu2016biomimetic}, stay dry fabrics \citep{bae2009superhydrophobicity} and novel lab-on-a-chip devices \citep{jia2019effect} are some of the examples where wetting is industrially important. The relationship between roughness and wettability of a surface is, however, not fully understood and we still generally rely on either trial and error or bio-mimicking for designing functional wetting surfaces \citep{yan2011mimicking,sun2005artificial,zhai2006patterned,feng2007superior}.

The spreading of a liquid on a surface in the presence of another immiscible fluid is characterised by the angle that the interface between the two fluids makes with the solid (measured from inside the fluid for which the contact angle is being measured). This angle is called the contact angle and for ideal surfaces (perfectly smooth, chemically homogeneous, rigid and inert to both the fluids), is given by Young's equation {\citep{young1805iii}}
%%%%%Young's Equation
\begin{equation}
	\cos\theta_{\rm{e}} = \left( \frac{\sigma_{\rm{2S}}-\sigma_{\rm{1S}}}{\sigma_{12}} \right).
		\label{eqn:young's_ch6}
\end{equation}
Here, $\theta_{\rm{e}}$ is the Young's angle, and $\sigma_{\rm{1S}}$, $\sigma_{\rm{2S}}$ and $\sigma_{12}$ are, respectively, the fluid-1/solid, fluid-2/solid and fluid-1/fluid-2  interfacial tensions. The contact angle can be measured experimentally and is usually measured at a small distance away from the solid surface. We shall refer to this as the macroscopic contact angle ($\theta_{\rm{m}}$). For ideal surfaces, the macroscopic contact angle is the same as Young's angle, however, this is not true for real surfaces. On a real surface, $\theta_{\rm{m}}$ is measured between the fluid/fluid interface and an average flat surface, which represents an apparent contact angle {\citep{Marmur2006}}. The presence of chemical and/or physical heterogeneities on real surfaces can cause $\theta_{\rm{m}}$ to be significantly different from $\theta_{\rm{e}}$. 

On a rough surface, $\theta_{\rm{m}}$ is observed to depend upon the wetting history of the surface, and has different values depending upon whether a particular state of the system (for example a particular volume of a droplet) is obtained by advancing or receding the three-phase contact line (TPCL, where the fluids 1, 2 and the solid surface meet). When the interface is made to advance slowly\footnote{In the case of a droplet, either by increasing the volume or by tilting the surface on which the drop is deposited. Or by lowering the surface in the pool of liquid in the case of a Wilhelmy plate \citep{law2016surface} type setting.}, the apparent contact angle increases until it reaches the maximum value, at which the TPCL starts advancing. This angle is called the advancing contact angle ($\theta_{\rm{a}}$). Similarly, when the interface is made to recede, the TPCL starts receding only when the apparent contact angle attains a minimum value, known as receding contact angle ($\theta_{\rm{r}}$). The difference between the advancing and receding contact angles is known as contact angle hysteresis ($\Delta \theta_{\rm{cah}}= \theta_{\rm{a}}-\theta_{\rm{r}}$). Contact angle hysteresis (CAH) is  an important parameter for characterising  wetting on  rough surfaces, however, it is still not fully understood.

The earliest relationship between surface roughness and the macroscopic contact angle ($\theta_{\rm{m}}$) was given by {\cite{wenzel1936resistance}}, as 
%%%%%Wenzel Equation
\begin{equation}
	\cos\theta_{\rm{m}} = r \cos\theta_{\rm{e}}.
		\label{eqn:wenzel_ch6}
\end{equation} 
Here, $r$ is the roughness, defined as the ratio of the actual to the apparent surface area of the solid. The Wenzel equation is phenomenological in nature based on the observation that the total surface area of a solid increases due to roughness, and is valid only for the homogeneous (or Wenzel) wetting state (i.e. the droplet is in complete contact with the solid surface and all the surface crevices are filled without any trapping of the surrounding immiscible fluid). There is another possible wetting state, known as the composite (or Cassie or Fakir) wetting state, in which the surrounding fluid is trapped inside the tops of the solid crevices. Here, $\theta_{\rm{m}}$ is governed by Cassie-Baxter equation {\citep{cassie1944wettability}}. Both Cassie-Baxter and Wenzel's equation predict a single equilibrium contact angle value, however, because of $\rm{CAH}$, a rough surface exhibits a number of equilibrium contact angles \citep{huh1977effects,oliver1977resistance,ramiasa2014influence}. As such, these equations cannot predict the hysteresis on a rough surface \citep{good1952thermodynamic, kusumaatmaja2007modeling, pomeau1985contact, gao2007wenzel, patankar2010hysteresis}, and so are not very useful for designing surfaces that require specific wetting behaviour.

Even though we lack a fully predictive wetting model for hysteresis, there are a few fundamental works which can help in explaining the existence of hysteresis on a rough surface. \cite{shuttleworth1948spreading} showed that a droplet may exist in a number of metastable states on a rough surface, which they suggested is responsible for the multiple equilibrium contact angles that characterise CAH. Johnson and Dettre conducted numerical \citep{dettre1964contact} and experimental \citep{johnson1964contact} studies based on the idea of the existence of multiple metastable states \citep{shuttleworth1948spreading}. They hypothesised that the origin of CAH lies in an energy barrier that must be overcome to move between different metastable states and available droplet vibrational energy is responsible for the contact angle hysteresis, however, they did not present any method to quantitatively estimate the droplet vibrational energy or relevant energy barriers. \cite{cox1983spreading} developed a model for hysteresis by treating the interface near the TPCL as a minimal surface (zero mean curvature). The model was developed for sinusoidally undulating surfaces that have a very small slope, which limits its application to real surfaces which usually have more step-like topologies \citep{brandon2003partial,johnson1964contact}. \cite{joanny1984model} presented a model for contact angle hysteresis on a surface containing a dilute number of `strong defects' in the range of a few microns. They suggested that the TPCL pins on a defect when the pinning force due to the defect and the elastic force acting on the distorted TPCL balances out. There are multiple solutions to this force balance and the TPCL can jump from one such position to the other, dissipating energy in the process. This dissipation is related to the contact angle hysteresis as
%%%%%de-Gennes Equation
\begin{equation}
\cos \theta_{\rm{r}} - \cos \theta_{\rm{a}} =n W_{\rm{d}},
\end{equation}
where $n$ is the density of the surface defects and ${W_{\rm{d}}}$ is the energy dissipation during one complete hysteresis cycle around a single defect. Many experimental studies have been done to quantitatively estimate the energy dissipation during hysteresis \citep{priest2009asymmetric, patankar2010hysteresis, ramos2003wetting, ramos2006pinning, reyssat2009contact, delmas2011contact} however, to our knowledge there is no analytical or numerical model available for calculating energy dissipation on generally structured surfaces.  

Surface Evolver (SE) \citep{Brakke1992} is open-source software that has been extensively used for studying contact hysteresis on heterogeneous surfaces \citep{dorrer2008drops, dorrer2007contact, forsberg2010contact, dorrer2007condensation, brandon1997simulated, chen2005anisotropy}. \cite{forsberg2010contact} used SE to calculate advancing contact angles during wetting on a surface with micron-sized pillars of square cross-section. They simulated the liquid-air interface between pillars and a flat surface parallel to the bottom of the pillars. For obtaining the advancing contact angle, the apparent contact angle of the interface was increased until it became unstable, either by depinning from the pillars or by touching the neighbouring pillars. \cite{semprebon2012advancing} used SE to simulate interface profiles pinned on cylindrical pillars and calculated the advancing contact angle of the interface as the maximum possible apparent angle for which an equilibrium interface morphology was possible, given the constraints of equilibrium angle ($\theta_{\rm{e}}$), pillar geometry and pillar separation distance. In all of these works \citep{dorrer2008drops,forsberg2010contact,semprebon2012advancing} the roughness was attributed to pillar structures (square or circular cross-section) with sharp edges, whereas real surfaces comprise of defects with rounded edges. \cite{promraksa2012modeling} used SE to simulate the static equilibrium of a full droplet on a double cosine wave-type surface. The advancing and receding contact angles here were calculated as the maximum and minimum contact angles corresponding to certain metastable states. Other than SE, \cite{mognetti2010modeling} used Lattice Boltzmann simulations to obtain the depinning angles for receding contact lines on a superhydrophobic surface. In all these works, a static interface shape or a static equilibrium shape of a full droplet was calculated rather than the dynamics of the interface over a surface with `strong-defects' \citep{joanny1984model}.

In the accompanying paper (part I) we developed a framework for predicting CAH based upon a mechanical energy balance applied to a control volume moving with the advancing TPCL. Key to this theory is being able to predict interface dynamics around the TPCL from knowledge of the solid topology. In this work, we use a novel numerical method to simulate the microscale dynamics of an interface between two immiscible fluids moving at a very small velocity over a rough solid. The numerical method is integrated into the mechanical energy balance framework presented in part I to develop a fully predictive equation capable of predicting contact angle hysteresis from just the knowledge of surface topology and the equilibrium contact angle. The surface topology considered in this work consists of periodically structured pillars of square cross-section.

\section{Theory}
\label{sec:theory}
\subsection{Dynamics of an interface advancing over a rough solid}
\label{sec:physcial_model}

In mechanical equilibrium, the Young-Laplace equation gives an expression for the pressure difference ($\Delta p$) across an interface between two immiscible fluids as
\begin{equation}
    \Delta p = 2 \kappa \sigma_{12},
    \label{eqn:Young-Laplace_ch6}
\end{equation}
where $\kappa$ is the mean curvature of the interface, $\Delta \rho$ is the density difference between the two fluids and $\sigma_{12}$ is the surface tension of the fluid-fluid interface. 
%If the pressure difference across the interface is constant, equation (\ref{eqn:Young-Laplace_ch6}) results in a surface of constant mean curvature. 
For an interface between two immiscible fluids moving over a rough surface ($h_{\rm{rough}}$) such that the length scale of the surrounding fluid flow ($h_{\rm{surround}}$) is much greater than the surface roughness ($h_{\rm{surround}} \gg h_{\rm{rough}}$), then in a region close to the TPCL, $\Delta p=O(\frac{\sigma_{12}}{h_{\rm{surround}}}) \approx 0$. Putting $\Delta p=0$ in equation (\ref{eqn:Young-Laplace_ch6}) yields $\kappa=0$ which defines a minimal surface. Therefore, the shape of the interface in equilibrium, near the TPCL can be represented by a minimal surface \citep[p. 13]{de2004capillarity}, provided that the length scale of the surface roughness is very small as compared to that of the surrounding fluid flow.

Turning to the dynamics of the interface, the augmented Navier-Stokes equation for the flow of incompressible and immiscible fluids can be written as \citep{popinet2009accurate}
\begin{equation}
    \frac{D}{Dt}\rho\boldsymbol{v}= -\boldsymbol{\nabla}p + \mu \boldsymbol{\nabla}^2 \boldsymbol{v} + \sigma_{12} \kappa \delta_{\rm{S}} \boldsymbol{n}_{\rm{S}} + \rho\boldsymbol{g}.
    \label{eqn:NS_full}
\end{equation}
Here, $\boldsymbol{v}$ is the velocity, $\rho$ is the density, $p$ is the pressure and $\mu$ is the dynamic viscosity of the fluid, $\sigma_{12}$ is the surface tension of the interface, $\delta_{\rm{S}}$ is the surface delta function, $\kappa=-\boldsymbol{\nabla} \cdot \boldsymbol{n}_{\rm{S}}$ is the mean curvature of the interface, $\boldsymbol{n}_{\rm{S}}$ is the unit normal to the interface directed into fluid-1 from fluid-2 and $\boldsymbol{g}$ is the acceleration due to gravity. For the interface moving at very small velocities ($\boldsymbol{v} \approx \boldsymbol{0}$) and in the absence of pressure gradients ($\boldsymbol{\nabla}p = \boldsymbol{0}$), the Navier-Stokes equation simplifies to $\kappa=0$ (upon neglecting the effect of gravity), which is the same solution as given by the Young-Laplace equation. Indeed, we say an interface is in equilibrium when two conditions are met:  Namely that 1) $\kappa=0$ everywhere on the interface, and 2) that the interface is locally intersecting with the rough solid surface at Young's angle. If either of the conditions is not met then the interface will move, as we now explain.

\begin{figure}
     \centering
         \includegraphics[width=0.46\textwidth]{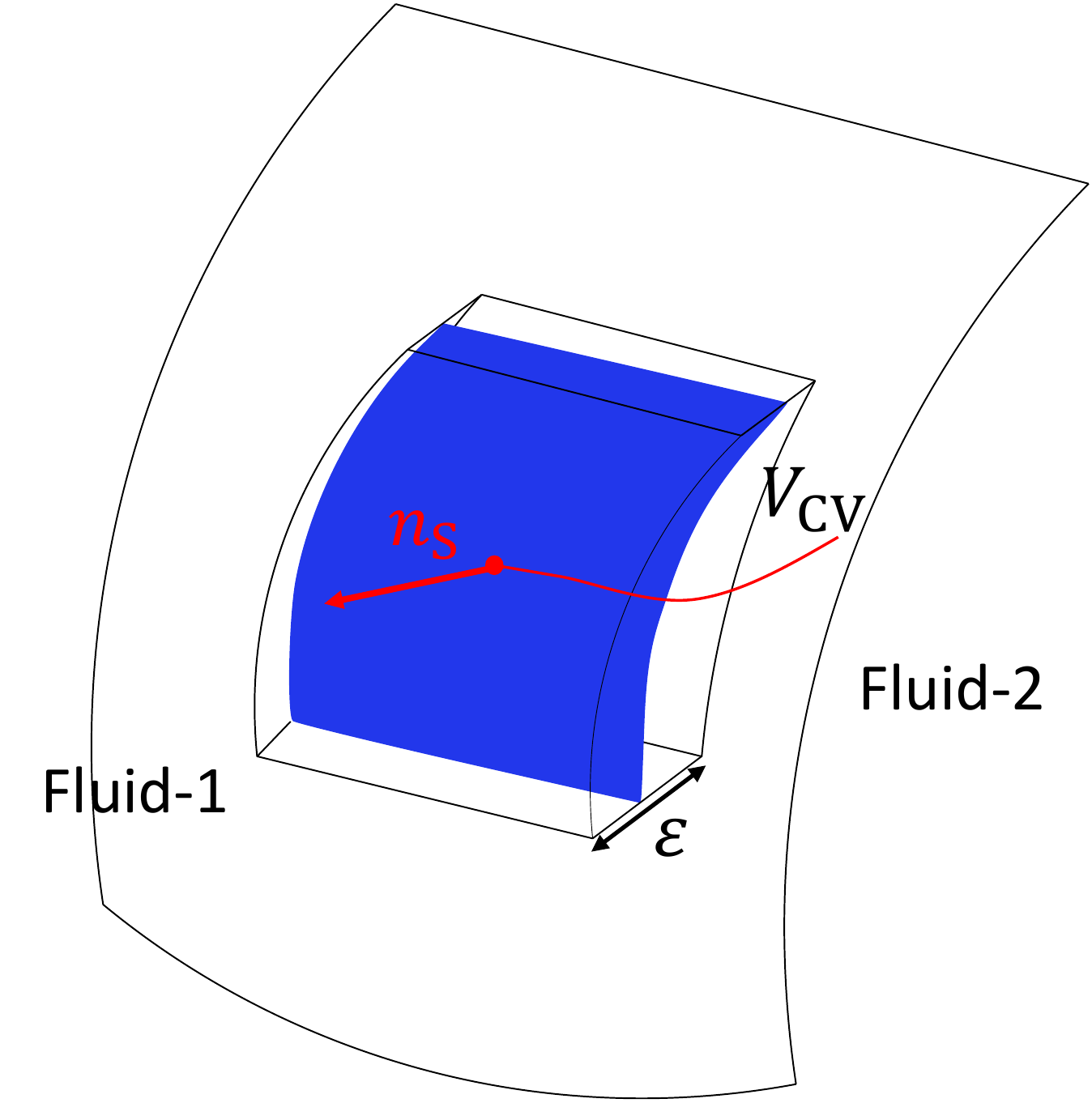}
		\caption{Control volume of small thickness ($\epsilon \gg 1$) chosen around the interface for integrating the Navier-Stokes equation.}
		\label{fig:thin_interface_ch6}
\end{figure}

\begin{figure}
     \centering
         \includegraphics[width=\textwidth]{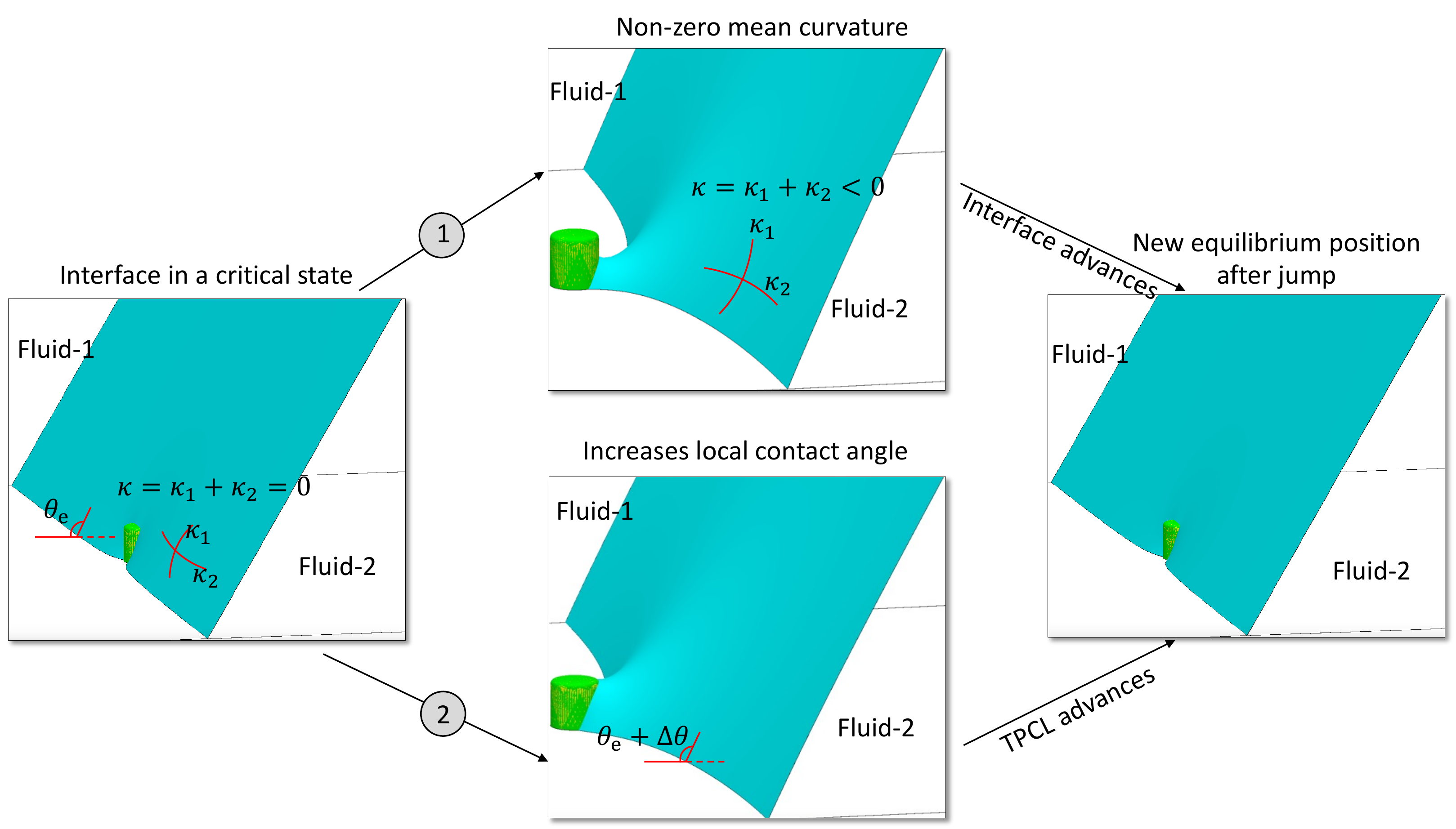}
		\caption{Representation of the phenomena of TPCL jump when the interface is pushed beyond a critical state. Scenario \circleme{1} shows the non-zero mean curvature of the interface and scenario \circleme{2} represents the local contact angle ($\theta_{\rm{local}}$) being greater than Young's angle ($\theta_{\rm{e}}$). The outcome of either of the two scenarios is the jumping of the TPCL towards the bulk of fluid 2, followed by subsequent pinning at the next defect in the direction of the interface motion.}
		\label{fig:reason_for_jump_ch6}
\end{figure}

If the interface is not in equilibrium due to the first condition not being met, then considering the kinematics of the interface near the TPCL we will find that $\kappa<0$ there.  This is the first scenario shown in figure \ref{fig:reason_for_jump_ch6}, which illustrates how interface jumps occur. Violation of the first equilibrium condition occurs when more liquid is added to a drop in equilibrium or when a Wilhelmy plate is pushed into a pool of liquid, for example. To understand the interface behaviour under these circumstances, we refer back to the Navier-Stokes equation (\ref{eqn:NS_full}) and take the dot product of this equation with the unit normal to the interface ($\boldsymbol{n}_{\rm{S}}$), i.e.
\begin{equation}
    \left(\frac{D}{Dt}\rho\boldsymbol{v}\right) \cdot \boldsymbol{n}_{\rm{S}} = -\boldsymbol{\nabla}p \cdot \boldsymbol{n}_{\rm{S}} + \mu \boldsymbol{\nabla}^2 \boldsymbol{v} \cdot \boldsymbol{n}_{\rm{S}} + \sigma_{12} \kappa \delta_{\rm{S}} \boldsymbol{n}_{\rm{S}} \cdot \boldsymbol{n}_{\rm{S}} + \rho\boldsymbol{g} \cdot \boldsymbol{n}_{\rm{S}}.
    \label{eqn:NS_dot1}
\end{equation}
Integrating equation (\ref{eqn:NS_dot1}) over a small control volume ($V_{\rm{CV}}$) around the interface as shown in figure \ref{fig:thin_interface_ch6}, and assuming for simplicity that $\boldsymbol{n}_{\rm{S}}$ does not change significantly with time, gives
\begin{equation}
\begin{split}
        \int_{{V_{\rm{CV}}}}\left(\frac{D}{Dt}\rho\boldsymbol{v}\right) \cdot \boldsymbol{n}_{\rm{S}} dV  &= -\int_{{V_{\rm{CV}}}}\boldsymbol{\nabla}p \cdot \boldsymbol{n}_{\rm{S}} dV + \int_{{V_{\rm{CV}}}}\mu \boldsymbol{\nabla}^2 \boldsymbol{v} \cdot \boldsymbol{n}_{\rm{S}} dV\\
        &+ \int_{{V_{\rm{CV}}}}\sigma_{12} \kappa \delta_{\rm{S}} dV + \int_{{V_{\rm{CV}}}}\rho\boldsymbol{g} \cdot \boldsymbol{n}_{\rm{S}} dV.
    \label{eqn:NS_dot2}
    \end{split}
\end{equation}
The left-hand side of the equation (\ref{eqn:NS_dot2}) can be simplified using the Reynolds transport theorem, noting that the control volume moves at the local velocity ($\boldsymbol{v}$), and assuming the fluid is incompressible,
\begin{equation}
    \int_{{V_{\rm{CV}}}}\frac{D}{Dt}\rho\boldsymbol{v} \cdot \boldsymbol{n}_{\rm{S}} dV = \frac{d}{dt} \int_{{V_{\rm{CV}}}} \rho (\boldsymbol{v} \cdot \boldsymbol{n}_{\rm{S}})dV.
    \label{eqn:NS_dot3}
\end{equation}
Simplifying the pressure gradient term in equation (\ref{eqn:NS_dot2}) using the Gauss-Ostrogradskii Divergence theorem to convert the volume integral to a surface integral over the control volume surface ($S_{\rm{CV}}$), gives
\begin{equation}
    \int_{{V_{\rm{CV}}}}\boldsymbol{\nabla}p \cdot \boldsymbol{n}_{\rm{S}} dV =  \int_{{S_{\rm{CV}}}} p(\boldsymbol{n}_{\rm{CV}} \cdot \boldsymbol{n}_{\rm{S}}) dS
    = \Delta p A,
    \label{eqn:NS_dot4}
\end{equation}
where $\Delta p$ is the pressure difference across the interface and $A$ is the area of the interface. In equation (\ref{eqn:NS_dot4}) we have neglected pressure contributions originating from the small (thickness $\epsilon$) sides of the control volume.
Using equations (\ref{eqn:NS_dot3}), (\ref{eqn:NS_dot4}) and (\ref{eqn:NS_dot1}) hence gives,
\begin{equation}
    \frac{d}{dt} \int_{{V_{\rm{CV}}}} \rho (\boldsymbol{v} \cdot \boldsymbol{n}_{\rm{S}})dV = -\Delta p A + \int_{{V_{\rm{CV}}}}\mu \boldsymbol{\nabla}^2 \boldsymbol{v} \cdot \boldsymbol{n}_{\rm{S}} dV + \int_{{V_{\rm{CV}}}}\sigma_{12} \kappa \delta_{\rm{S}} dV + \int_{{V_{\rm{CV}}}}\rho\boldsymbol{g} \cdot \boldsymbol{n}_{\rm{S}} dV.
    \label{eqn:NS_dot5}
\end{equation}
We next assume for simplicity that the curvature of the interface in the small $V_{\rm{CV}}$ is uniform, and denote the interfacial velocity in the direction of $\boldsymbol{n}_{\rm{S}}$ as $v_{\rm{S}}=\boldsymbol{v} \cdot \boldsymbol{n}_{\rm{S}}$ and the volume of the control volume as depicted in figure \ref{fig:thin_interface_ch6} as $\delta V$. Also, since we are interested in the interface dynamics near the TPCL, the system under consideration is small (less than the capillary length $r_{\rm{cv,grav}}$ from part I) and the effect of gravity ($\boldsymbol{g}$) can be neglected. This gives
\begin{equation}
    \frac{d}{dt}(\rho v_{\rm{S}})\delta V = -\Delta p A + \int_{{V_{\rm{CV}}}}\mu \boldsymbol{\nabla}^2 \boldsymbol{v} \cdot \boldsymbol{n}_{\rm{S}} dV + \sigma_{12} \kappa A.
    \label{eqn:NS_dot6}
\end{equation}
This equation shows that the local acceleration of the interface is governed by the pressure difference across the interface, the curvature of the interface, and viscous dissipation occurring local to the interface. Noting further based on the system geometry that the Laplace pressure difference across the interface is given by $\Delta p = \mathcal{O} (\sigma_{12}/ h_{\rm{surround}}) \approx 0$, equation (\ref{eqn:NS_dot6}) can be further simplified to
\begin{equation}
    \frac{d}{dt}(\rho v_{\rm{S}})\delta V  =  \int_{V_{\rm{CV}}} \left[ \mu \boldsymbol{\nabla}^2 \boldsymbol{v}\right] \cdot \boldsymbol{n}_{\rm{S}} dV + \sigma_{12} \kappa A.
    \label{eqn:NS_interface_velocity2}
\end{equation}
This shows that for a negative $\kappa$, the acceleration of the interface in the direction of $\boldsymbol{n}_{\rm{S}}$ is negative, i.e. the interface accelerates in a direction opposite to the surface normal pointing from fluid-2 to fluid-1. Likewise, for a positive $\kappa$, the interface accelerates in the direction of $\boldsymbol{n}_{\rm{S}}$. The former scenario represents an advancing interface while the latter refers to the receding motion of the interface. Hence, equation (\ref{eqn:NS_interface_velocity2}) shows that if an interface is in equilibrium ($\boldsymbol{v}=\boldsymbol{0}$) but is then perturbed to be just out of equilibrium (so that $\kappa \neq 0$), the interface will accelerate in the direction of $\kappa \boldsymbol{n}_{\rm{S}}$, and further the interface will keep moving until a new equilibrium topology is found (at which $\kappa=0$ again).  More specifically, if the TPCL is pinned while the macroscopic interface continues to advance, eventually $\kappa < 0$ near the TPCL and locally the interface will accelerate towards fluid 2 again to restore equilibrium conditions.

Under the second scenario the interface is not in equilibrium because the local contact angle is not everywhere equal to the equilibrium angle, as depicted by the second case in figure \ref{fig:reason_for_jump_ch6}. Specifically, we assume that after advancing the interface we have $\theta_{\rm{local}}>\theta_{\rm{e}}$, where $\theta_{\rm{local}}$ is the local contact angle. This is another possibility when more liquid is added to a drop which is in equilibrium or when a Wilhelmy plate is pushed into a pool of liquid. Since $\theta_{\rm{local}}$ is not the same as the equilibrium angle ($\theta_{\rm{e}}$), a driving capillary force equalling $\sigma_{12}(\cos\theta_{\rm{e}}-\cos\theta_{\rm{local}})$ \citep{ramiasa2014influence} accelerates the TPCL in the direction of interface advancement.
 
In summary, two conditions must be satisfied for an interface to be in equilibrium \citep{bartell1953effect}, being 1) zero mean curvature ($\kappa=0$) everywhere on the interface and 2) the interface intersecting the solid surface locally at Young's angle ($\theta_{\rm{local}}=\theta_{\rm{e}}$). The above analysis shows however that if either condition is not met then the interface will move at capillary driven speeds in the direction of the macroscopic interface motion until a new equilibrium position is found.  We refer to this rapid motion of the interface as a \textit{jump}, and the overall behaviour of the TPCL as \textit{stick-slip}.  We denote the state of an interface in equilibrium but where an infinitesimal movement of the macroscopic interface will result in an acceleration of the interface as the first critical state (that is, before the jump), and the first equilibrium interface position found after this while the macroscopic interface continues to move as the second critical state (that is, after the jump). Therefore, an interface moving at infinitesimal small velocity under the action of an infinitesimal pressure gradient can be modeled as a series of minimal surfaces with occasional contact line jumps occurring between our defined first and second critical states.

\subsection{Simulating interfacial dynamics}
With a physical model for interface advance established, we now consider how these dynamics can be simulated.
\subsubsection{Simulation domain}\label{sec:domain_bc}
\begin{figure}
     \centering
         \includegraphics[width=\textwidth]{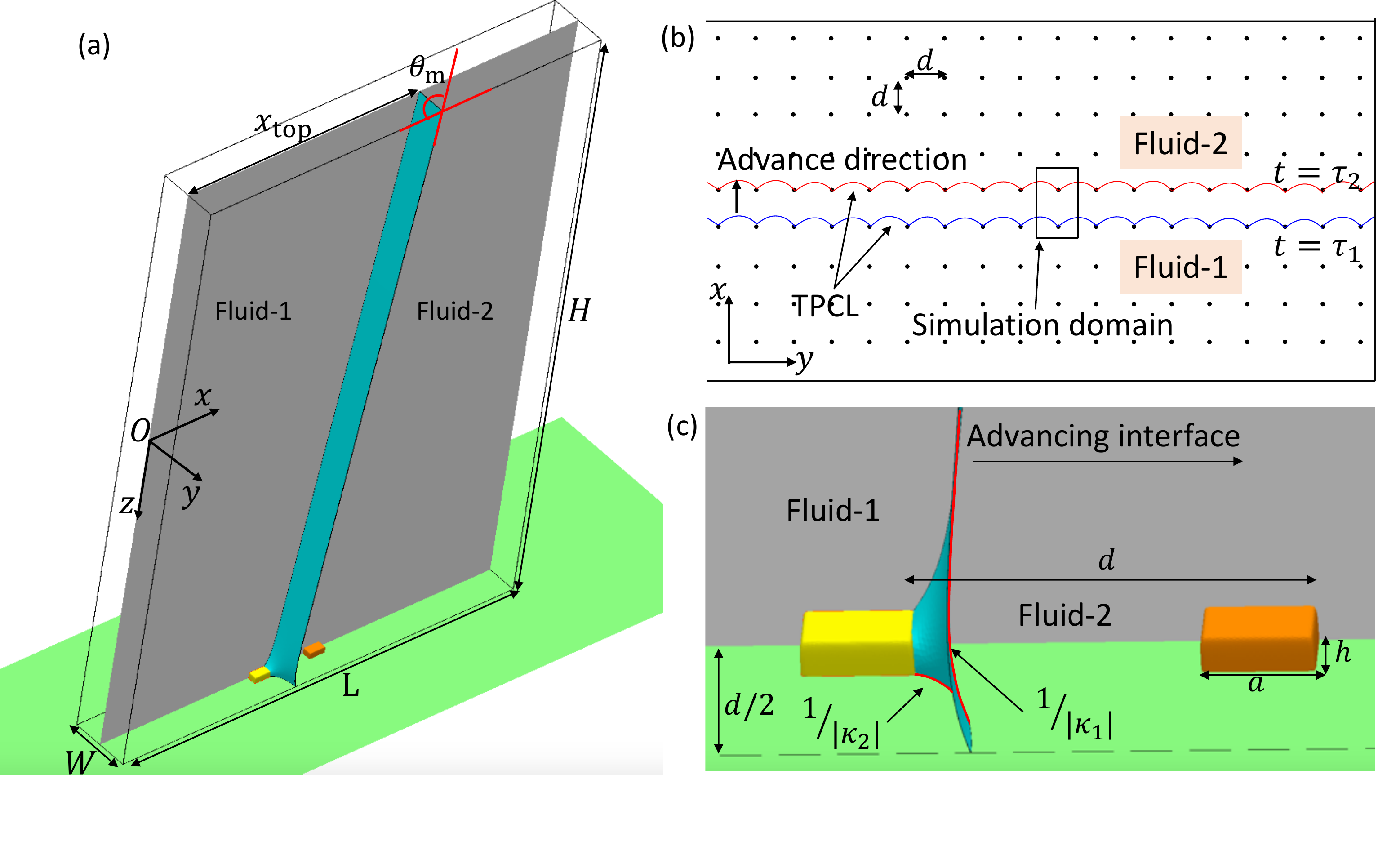}
		\caption{(a) Three-dimensional view of the simulation domain. Fluid-1 is advancing towards the right while fluid-2 is receding. The macroscopic contact angle ($\theta_{\rm{m}}$) is measured at the top of the domain. The symmetry plane passing through the pillar centers and parallel to the $x-z$ plane is shown in grey. (b) Advancing TPCL on a surface with roughness in the form of micro-scale pillars (side $a$ and height $h$) arranged in a square array with inter-pillar spacing $d$. The blue line shows the TPCL pinned at the pillars at an instant of time ($t=\tau_1$) and the red line shows the TPCL  pinned  on the next row of pillars at an instant of time ($t=\tau_2$) depicting the TPCL jump. (c) Zoomed view of the interface near the TPCL capturing the principal curvatures $\kappa_1$ and $\kappa_2$ in the $x-z$ and $x-y$ planes respectively.}
		\label{fig:physical_model_ch6}
\end{figure}

To simulate the interfacial dynamics we choose a simulation domain in the form of a rectangular channel, as shown in figure \ref{fig:physical_model_ch6}(a). The bottom surface of the simulation domain is decorated with pillars of square cross-section (side $a$ and height $h$) arranged in a square array with a spacing $d$ (see figure \ref{fig:physical_model_ch6}(b)). The height, width and length of the simulation domain are $H$, $W$ and $L$ respectively. All lengths are non-dimensionalized with respect to the pillar side, i.e.
\begin{equation}
\begin{split}
&W = \frac{W^{'}}{a^{'}}, \quad H = \frac{H^{'}}{a^{'}}, \quad L = \frac{L^{'}}{a^{'}}, \\
&d = \frac{d^{'}}{a^{'}}, \quad h=\frac{h^{'}}{a^{'}}, \quad \text{and} \quad a = \frac{a^{'}}{a^{'}}=1.\\
\end{split}
\label{eqn:channel_dimensions_ch6}
\end{equation}
Here, $W^{'}$, $H^{'}$ and $L^{'}$ are the dimensional channel width, height and length respectively and $a^{'}$, $h^{'}$, $d^{'}$ are the dimensional pillar side, height and inter-pillar distance respectively. 
Since the surface is periodic and the interface is moving in the direction of surface periodicity, the equilibrium interface morphology will be periodic in nature with the period the same as the inter-pillar distance ($d$). Why these interface morphologies are symmetric is discussed later. For now, because of this periodic nature, the width of the simulation domain ($W$) is set equal to the distance between pillar centers ($d$) and the length of the simulation domain is chosen to be long enough so as to capture at least one TPCL jumping event after the simulation start.
%contain the motion of the interface in the duration $\Delta t=\tau_2 - \tau_1$. 
The height of the simulation domain is chosen such that all the deformations in the interface are contained within it (see sensitivity analysis in Appendix \ref{sec:channel_height:appendix}). A typical simulation domain is chosen such that the pillar is located at the center of the domain, i.e. at $y=0$ and the domain walls are at $y=-W/2$ and $y=W/2$ respectively.

\subsubsection{Energy minimization}
As discussed in \S\ref{sec:physcial_model}, physically the interface moves in one direction, progressing through a series of equilibrium stages. These equilibrium interface morphologies are obtained by minimising the total interfacial energy within the simulation domain which comprises the interfacial energies of the fluid-1/fluid-2, fluid-1/solid and fluid-2/solid interfaces respectively. The total interfacial energy ($E^{'}$) within the simulation domain can be written as
\begin{equation}
    E^{'} = \sum_{i<j} \sigma_{ij}A_{ij} = \sigma_{12}A_{12} + \sigma_{\rm{1S}}A_{\rm{1S}} + \sigma_{\rm{2S}}A_{\rm{2S}}.
    \label{eqn:system_energy_1}
\end{equation}
Here, $\sigma_{ij}$ and $A_{ij}$ represent the interfacial tension and area of the $ij{\rm{th}}$ interface i.e., $12$, $\rm{1S}$ and $\rm{2S}$ representing fluid-1/fluid-2, fluid-1/solid and fluid-2/solid interfaces. If the total area of the bottom surface is $A_{\rm{t}}$, then $A_{\rm{1S}}+A_{\rm{2S}}=A_{\rm{t}}$, i.e. $A_{\rm{2S}}=A_{\rm{t}}-A_{\rm{1S}}$, and therefore equation (\ref{eqn:system_energy_1}) can also be written as
\begin{equation}
    E^{'} = \sigma_{12}A_{12} + (\sigma_{\rm{1S}}-\sigma_{\rm{2S}})A_{\rm{1S}} + \sigma_{\rm{2S}}A_{\rm{t}}.
        \label{eqn:system_energy_2}
\end{equation}
Defining $E=E^{'}-\sigma_{\rm{2S}}A_{\rm{t}}$, where $\sigma_{\rm{2S}}A_{\rm{t}}$ is a constant for a given fluid-2/solid combination and simulation domain geometry, the total relative energy ($E$) within the domain can be written using equation (\ref{eqn:system_energy_2}) and Young's equation (\ref{eqn:young's_ch6}) as
\begin{equation}
    E = \sigma_{12}(A_{\rm{12}} - A_{\rm{1S}}\cos\theta_{\rm{e}}).
    \label{eqn:domain_energy}
\end{equation}
For the purpose of brevity in the rest of the paper we refer to the total relative energy (equation (\ref{eqn:domain_energy})) as total energy.

The minimal surface morphology of interface is obtained by minimising the total energy (equation (\ref{eqn:domain_energy})) within the simulation domain (as shown in figure \ref{fig:physical_model_ch6}(a)). The energy minimization is carried out in Surface Evolver (SE) \citep{Brakke1992}, which is an open-source software often used for studying surface wetting problems \citep{irannezhad2023fluid, pour2019equilibrium, semprebon2012advancing, forsberg2010contact,dorrer2008drops,dorrer2007contact,dorrer2007condensation,chen2005anisotropy,brandon1997simulated}. SE uses the method of gradient descent for arriving at the state of minimum energy. A surface is represented by a set of triangular facets and the minimum energy configuration is obtained by moving  vertices  in the direction of a negative gradient of energy while maintaining the applied  constraints of constant volume, fixed pressure, fixed edges, vertices or any other user-defined constraints.  An initial geometry is loaded into the software, which then evolves towards a minimum energy shape with subsequent refinement and iterations. 
%SE offers a unique capability of modifying the constraints or making  changes in the original geometry, and system parameters during the evolution process, which provides an interactive approach towards the study of surfaces.
  \begin{figure}
      \centering
      \includegraphics[width=0.85\textwidth]{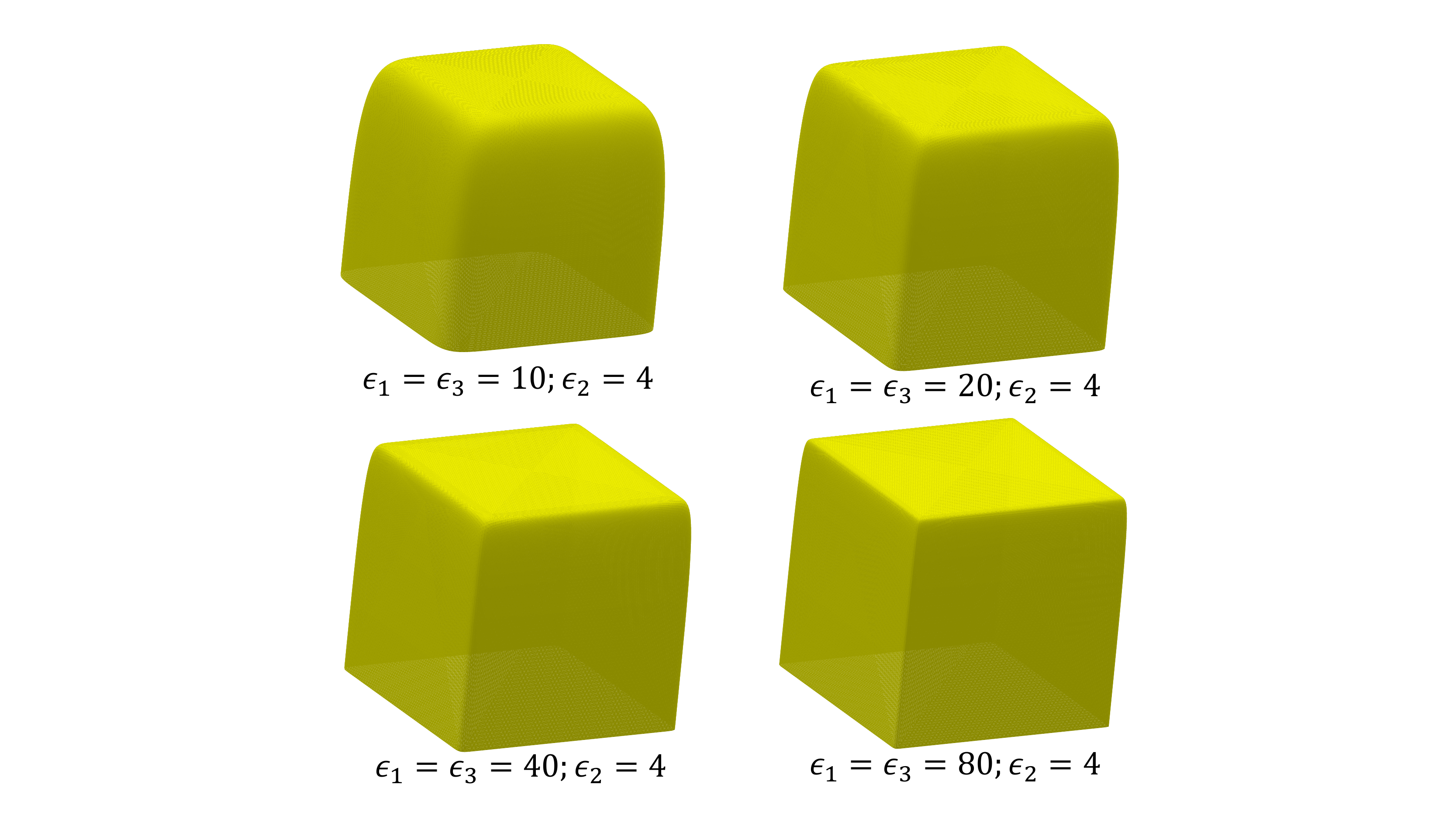}
      \caption{Use of superquadrics for modeling pillars. The shape of the cross-section is fixed by the coefficient $\epsilon_2$, which is chosen as 4 for square pillars. The sharpness of the edges can be controlled by the coefficients $\epsilon_1$ and $\epsilon_3$. The higher the value of $\epsilon_1$, $\epsilon_3$, the sharper the pillar edge. In this work we use $\epsilon_1$=$\epsilon_3$=40 and $\epsilon_2$=4.}
      \label{fig:defect_shape_ch6}
  \end{figure}
The SE algorithm minimises the total energy of the simulation domain as given in equation (\ref{eqn:domain_energy}), by adding the energy of individual facets of the interface mesh via
\begin{equation}
E =  \left(\sigma_{12} \sum_{ j=1}^{j={N_1}} \frac{1}{2}|\boldsymbol{e_0} \times \boldsymbol{e_1}|_j \right)_{12} - \cos\theta_{\rm{e}}  \left(\sigma_{12} \sum_{ k=1}^{{k=N_2}} \frac{1}{2}|\boldsymbol{e_0} \times \boldsymbol{e_1}|_k \right)_{1\rm{S}}.
\label{eqn:evolver_energy_ch6}
\end{equation}
Here, $\boldsymbol{e_0}, \boldsymbol{e_1}$ represents the facet edge vectors and ${N}_1, {N}_2$ represents the total number of facets on the fluid-1/fluid-2 and fluid-1/solid interface respectively. The Surface Evolver documentation \citep{brakke1994surface} gives more details regarding the implementation of equation (\ref{eqn:evolver_energy_ch6}). 
\subsubsection{Pillar geometry}
For modeling the pillars we use three-dimensional shapes known as superquadrics \citep{Barr1981}: In our application, these are rectangular prisms with rounded edges. In general, superquadrics can be represented by the following equation
%%%%%Equation: Superquad equation
\begin{equation}
\left(\frac{x-x_1}{s_1}\right)^{\epsilon_1}+\left(\frac{y-y_1}{s_2}\right)^{\epsilon_2}+\left(\frac{z-z_1}{s_3}\right)^{\epsilon_3}=1.
\label{eqn:superquad_ch6}
\end{equation}
where $\epsilon_1,\epsilon_2$ and $\epsilon_3$ control the shape of the superquadric  cross-section, and $s_1, s_2, s_3$ control the aspect ratio of the pillars respectively. The values of $x_1, y_1, z_1$ place the pillar inside the domain. 
The main advantage of using superquadrics for modeling pillar structures is that they create smooth surfaces with controlled sharpness of the edges (see figure \ref{fig:defect_shape_ch6}). As well as being numerically more convenient (avoiding infinite curvatures associated with sharp corners), a pillar with slightly rounded edges better represents the actual profile of fabricated surfaces. For the current simulations we use $\epsilon_1=40, \epsilon_2=4, \epsilon_3=40$, as shown in figure \ref{fig:defect_shape_ch6}.
\subsubsection{Simulating interface advancement}
\label{sec:int_dyn_sim}
As already discussed in \S\ref{sec:physcial_model}, an advancing interface moving at an infinitesimal velocity under an infinitesimal pressure gradient can be represented by a series of minimal surfaces, which are obtained by minimising the total interfacial energy within the simulation domain (equation (\ref{eqn:domain_energy})) at each simulated time point. Instead of the full interface, we simulate a portion of a larger interface which is moving continuously under the action of macroscopic flow at large distances away from the TPCL. From the point of view of our simulations, this implies that the interface is moving at a constant rate at the top of the domain (assuming the domain is sufficiently high such that the separation of length scales inherent in the physical system is ensured) whereas the TPCL moves in its characteristic \textit{stick-slip} motion. Therefore, the advancing motion of the interface within the simulation domain can be attained by gradually advancing the interface top. In terms of boundary conditions, this means that the interface top is constrained to a specific value ($x_{\rm{top}}$) at a certain time point, which is then gradually incremented in steps. The other boundary conditions are the orthogonality between the simulation domain walls and the interface, i.e.
\begin{equation}
    \boldsymbol{n}_{\rm{S}} \cdot \boldsymbol{j} = 0,
    \label{eqn:bc_wall}
\end{equation}
along the domain walls, representing periodicity in the direction perpendicular to the interface advancement direction, where $\boldsymbol{n}_{\rm{S}}$ is the unit normal vector to the fluid-fluid interface (directed into fluid-1 from fluid-2) and $\boldsymbol{j}$ is the unit vector in the direction of $y$ axis (see figure \ref{fig:physical_model_ch6}(a)). Along the base of the computational domain, the boundary condition of Young's angle is applied, that is, equation (\ref{eqn:young's_ch6})
\begin{equation}
\boldsymbol{n}_{\rm{S}} \cdot \boldsymbol{n}_{\rm{solid}} = -\cos\theta_{\rm{e}},
\label{eqn:surface_normals_ch6}
\end{equation}
where $\boldsymbol{n}_{\rm{solid}}$ is the unit normal vector to the solid surface at the TPCL (directed out of the surface).

\begin{figure}
     \centering
         \includegraphics[width=\textwidth]{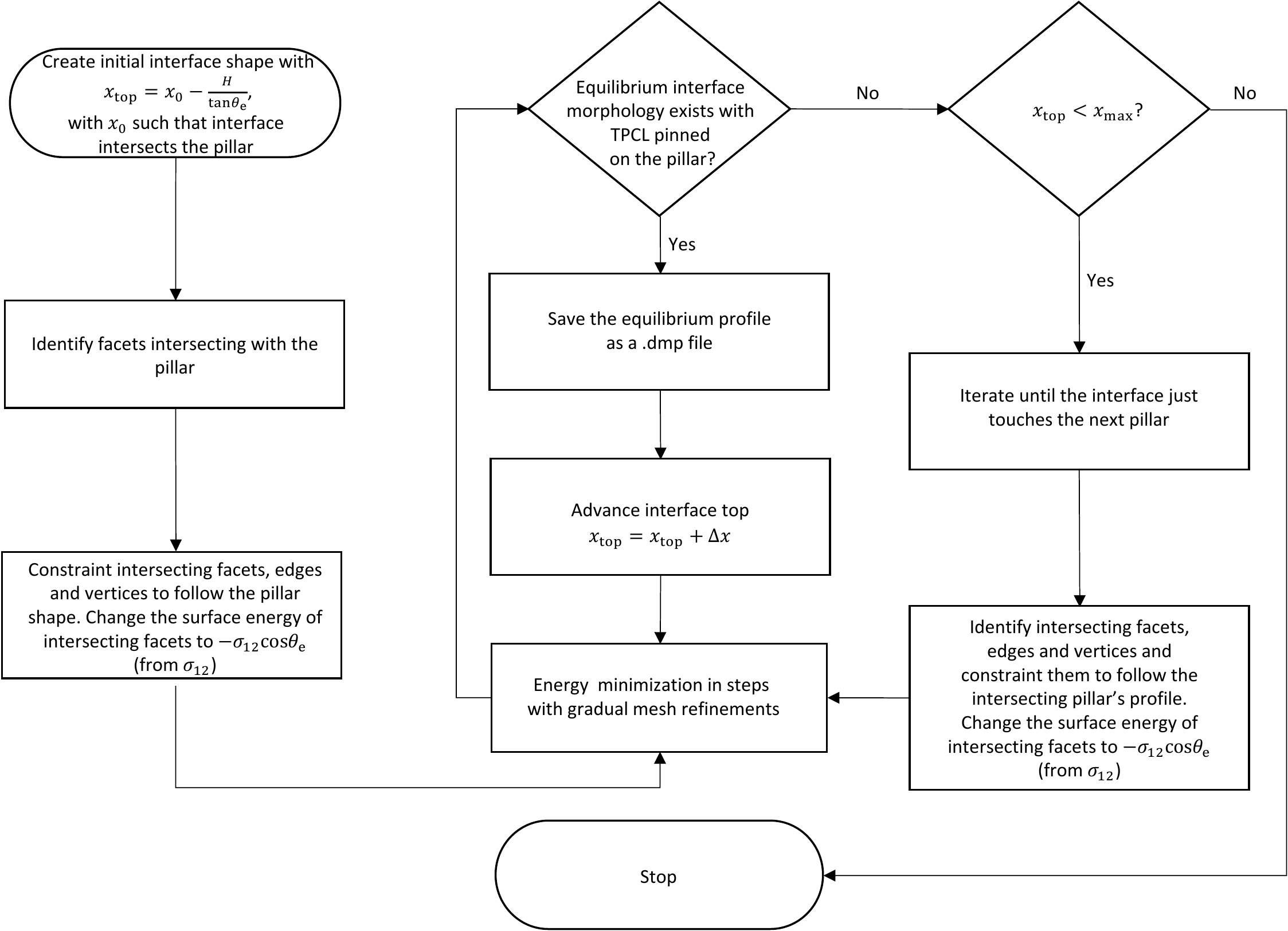}
		\caption{Flowchart of the algorithm used for simulating the advancement of an interface by the incremental advance method.}
		\label{fig:algo_xtop_main}
\end{figure}

Figure \ref{fig:algo_xtop_main} schematically illustrates our interface advance algorithm. A typical simulation starts with a flat interface constrained according to the above-mentioned boundary conditions and positioned close to a pillar such that the interface is just touching it. We use a starting interface position ($x_{\rm{top}}$) such that the macroscopic contact angle ($\theta_{\rm{m}}$) is equal to the Young's angle  ($\theta_{\rm{e}}$), and therefore, we choose $x_{\rm{top}}=x_0 - H/\tan \theta_{\rm{e}}$, where $x_0$ represents the TPCL position. Now, the intersection between the interface mesh and the pillar is identified. The intersecting facets of the interface and the associated edges and vertices are constrained to follow the profile of the pillar (equation (\ref{eqn:superquad_ch6})) and the surface energy of the intersecting facets (of the interface) are changed to $-\sigma_{12}\cos\theta_{\rm{e}}$ (from $\sigma_{12}$) representing their transition from a fluid-1/fluid-2 to fluid-1/solid interface. Now the energy minimization is carried out in a number of steps with gradual mesh refinements and an equilibrium interface morphology is obtained. This is the initial equilibrium shape with $x_{\rm{top}}$ chosen such that $\theta_{\rm{m}}=\theta_{\rm{e}}$. In order to advance the interface, $x_{\rm{top}}$ is increased by a small amount ($\Delta x$), which is the step size of the interface advancement. With $x_{\rm{top}}$ changed to $x_{\rm{top}}+\Delta x$ at the top and keeping the other boundary conditions unchanged, energy minimization is carried out once again to obtain the minimal surface morphology of the interface at this new position. This process is repeated until the interface reaches the first critical state, where it is not possible to advance the interface and find a minimal surface shape such that the TPCL is still pinned on the first pillar. The position of the interface top and the macroscopic contact angle when the interface is in this critical state are represented as $x_{\rm{s}}$ and $\theta_{\rm{s}}$ respectively. The accuracy with which $x_{\rm{s}}$ and $\theta_{\rm{s}}$ can be predicted depends upon the step size ($\Delta x$), with better accuracy achieved using smaller step sizes (see Appendix \ref{sec:step-size:appendix}).

Up until this point, we have discussed the advancing motion of an interface starting from an equilibrium position with $\theta_{\rm{m}}=\theta_{\rm{e}}$ to the first critical state. Now, we discuss how the interface advance is simulated after reaching this state. Any further increment in the $x_{\rm{top}}$ now results in the depinning of the TPCL from the pillar\footnote{In this paper, we focus on the interface motion such that the TPCL depins from a pillar before it gets pinned again at the next suitable location and the critical interface morphologies are such that the interface is not touching the next pillar in the direction of the motion.} and following our physical model of interface advance the TPCL now moves forward until becoming pinned again at the next pillar in the direction of advancement. To find an equilibrium interface morphology, we start with finding the intersection between the interface and the next pillar in the direction of advancement followed by constraining the intersecting facets (and edges and vertices of the interface) to the shape of the new pillar. The surface energy of the intersecting facets of the interface is once again changed to $-\sigma_{12} \cos \theta_{\rm{e}}$. With the interface touching the new pillar and the above-mentioned changes to the intersecting facets of the interface, energy minimization is carried out to obtain the equilibrium interface morphology. We call this equilibrium interface morphology after the TPCL has just depinned from a pillar as the `second-critical state' and refer to the corresponding interface position and the macroscopic contact angle as $x_{\rm{s}^{'}}$ and $\theta_{\rm{s}^{'}}$ respectively. The morphological transition between the two critical states captures a typical TPCL jumping event. To further advance the interface, subsequent $x_{\rm{top}}$ increments and energy minimization are carried out to obtain the next equilibrium, first and second critical interface morphologies and this process is repeated until we reach the desired maximum displacement of the interface ($x_{\rm{max}}$). We call this method of simulating interface dynamics by gradual increments in $x_{\rm{top}}$ the `incremental advance method'. In our simulations, we were able to find an equilibrium interface morphology on a neighbouring pillar in the direction of interface motion every time the TPCL executes a jumping event as long as the step size is small enough (see Appendix \ref{sec:step-size:appendix}). This is expected since the surface is structured and all the pillars have the same geometry \citep{huh1977effects}. Equilibrium morphologies depicting a typical interface advancement are shown in Appendix \ref{sec:morphologies_appendix}.

Now that we have described the method of simulating a typical interface advancement, we discuss the role of the step size ($\Delta x$) on the accuracy with which the advancement of a real interface can be captured by our method. A typical interface advancement on a rough surface is characterised by the \textit{stick-slip} motion executed by the TPCL. This has been discussed in detail in part I and is briefly discussed here. The TPCL during a jumping event moves with the capillary velocity ($\boldsymbol{v}_{\rm{cap}}$) and the jumping event is completed in a very short duration of time ($\tau_{\rm{cap}}$), such that
\begin{equation}
    v_{\rm{cap}} =  \text{min} \left ( \frac{\sigma_{12}}{\mu}, \sqrt{\frac{\sigma_{12}}{\rho d}} \right ),
\label{eq:vcap}
\end{equation}
and
\begin{equation}
    \tau_{\rm{cap}} = \frac{d}{v_{\rm{cap}}},
\end{equation}
where $\rho$ and $\mu$ are chosen as the maximum of the fluid's densities and viscosities (see part I). In a typical scenario of a water droplet spreading in air on a surface where the roughness is of the order of a few microns, $v_{\rm{cap}}$ is approximately 3 m/s and $\tau_{\rm{cap}} \approx$ 3 $\mu$s. Therefore, during a TPCL jumping event, the surrounding interface doesn't significantly move macroscopically while the TPCL moves from one set of pillars to the next. In the context of our simulations this means that during the TPCL jump $x_{\rm{top}}$ should not change, i.e. both of the critical states should be calculated using the same $x_{\rm{top}}$ (i.e. $x_{\rm{s}^{'}} \to x_{\rm{s}}$). In order to ensure this, we ideally require an infinitesimally small step size to be used in our simulations, (i.e. $\Delta x \to 0$), however this would result in a correspondingly large number of equilibrium interface morphologies between the initial and the first critical state and hence an impractically large computational time. However, as we discuss in \S\ref{sec:dissipation}, for calculating the dissipation in energy during the interface advancement, only the two critical states of the interface during the advancing motion need be computed. Therefore, we propose another method which is more efficient in finding these two states compared to the method of incremental advance as described above. We call this method the `critical-state method', and it is described in Appendix  \ref{sec:second-critical:appendix}. In this method, the interface top is constrained by a certain macroscopic angle ($\theta_{\rm{m}}$) and is advanced by gradually incrementing $\theta_{\rm{m}}$, i.e. $\theta_{\rm{m}} + \Delta \theta$, where $\Delta \theta$ is the step size. For the case of an interface advancing on a structured surface in the direction of surface periodicity, the critical-state method is consistent with the incremental advance method (see Appendix \ref{sec:algo_comparison:appendix}). Most of the results presented in this paper are simulated using this critical-state method.

\begin{figure}
     \centering
         \includegraphics[width=0.90\textwidth]{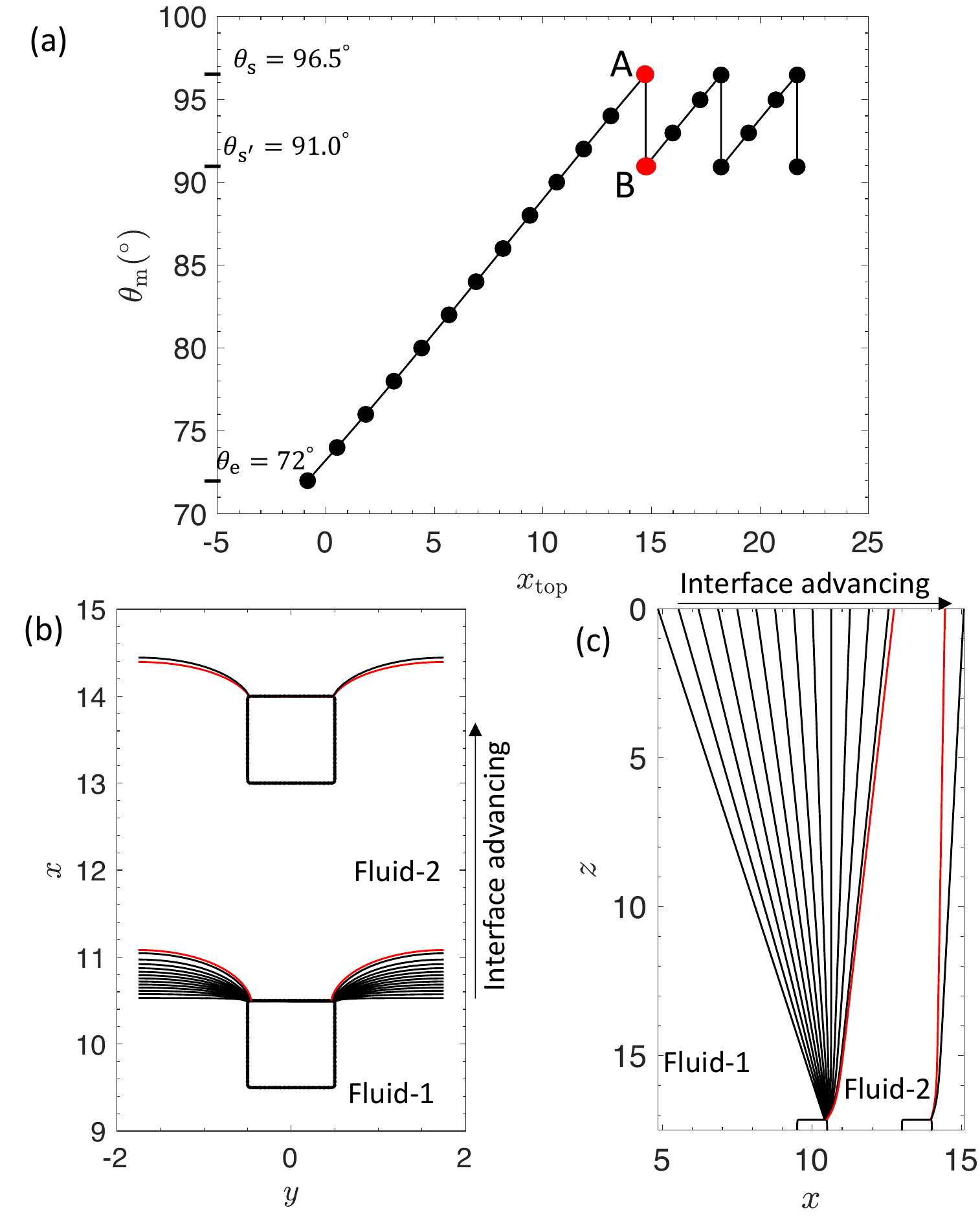}
		\caption{(a) Variation in macroscopic contact angle ($\theta_{\rm{m}}$) measured at the simulation domain's top with the position of the interface top ($x_{\rm{top}}$) during advancement of the interface (for $\theta_{\rm{e}}=72$\textdegree, $\phi=0.08$, $h/a=0.35$). Points `A' and `B' represent the first and second critical states respectively. (b) TPCL lying on the domain base corresponding to the equilibrium states shown in (a). (c) Projection on the symmetry plane of the equilibrium interface morphologies corresponding to the states shown in (a). The TPCL and the interface projection on the symmetry plane corresponding to the first and second critical states are shown in red in (b) and (c) respectively.}
		\label{fig:theory_thm_ch6}
\end{figure}

In figure \ref{fig:theory_thm_ch6} we show example results for the typical advancement of an interface between two immiscible fluids (fluid-1 and fluid-2) over a rough surface similar to the one as shown in figure \ref{fig:physical_model_ch6}(b). We simulate the motion of the interface in the direction of surface periodicity. Suppose at a particular instant of time ($t=\tau_{1}$), the TPCL is shown by the solid blue lines in figure \ref{fig:physical_model_ch6}(b). The interface is then made to advance continuously, such that at the time ($t=\tau_2$), the TPCL is represented by the solid red line in figure \ref{fig:physical_model_ch6}(b), after having executed a jump. Figure \ref{fig:theory_thm_ch6}(a) shows the variation in macroscopic contact angle ($\theta_{\rm{m}}$) with the interface position ($x_{\rm{top}}$) as it advances within the simulation domain.
The portion of the TPCL lying on the domain base and the projection of the interface on the symmetry plane (see figure \ref{fig:physical_model_ch6}(a)) for the interface morphologies during the advancement of the interface is shown in figures \ref{fig:theory_thm_ch6}(b) and (c) respectively. For every value of $x_{\rm{top}}$ we have a unique contact line shape. This is because as we change $x_{\rm{top}}$, the curvature in $x-z$ plane, i.e. $\kappa_1$ (figure \ref{fig:physical_model_ch6}(c)) changes by a small amount. This change in $\kappa_1$ is balanced by the same change in $\kappa_2$ (curvature in $x-y$ plane, see figure \ref{fig:physical_model_ch6}(c)), but in the opposite direction so that the mean curvature remains zero. Here, both the curvatures, i.e., $\kappa_1$ and $\kappa_2$ are measured in the same neighbourhood, therefore, everywhere on the interface, $\kappa_2$ adjusts so as to balance the changes in $\kappa_1$ when $x_{\rm{top}}$ changes. However, when the interface reaches a critical state (for example, state `A' in figure \ref{fig:theory_thm_ch6}(a)), it is not possible to find a minimal surface shape under the constraint of the local angle being equal to Young's angle everywhere on the TPCL, and the interface accelerates in the direction of $x_{\rm{top}}$ increment. When the interface is at first critical state, the macroscopic angle attains its maximum value ($\theta_{\rm{s}}$). Upon a small advancement of the interface from this first critical state, the interface starts to accelerate but pins again at the next closest pillar in the direction of advancement.

\begin{figure}
     \centering
         \includegraphics[width=0.40\textwidth]{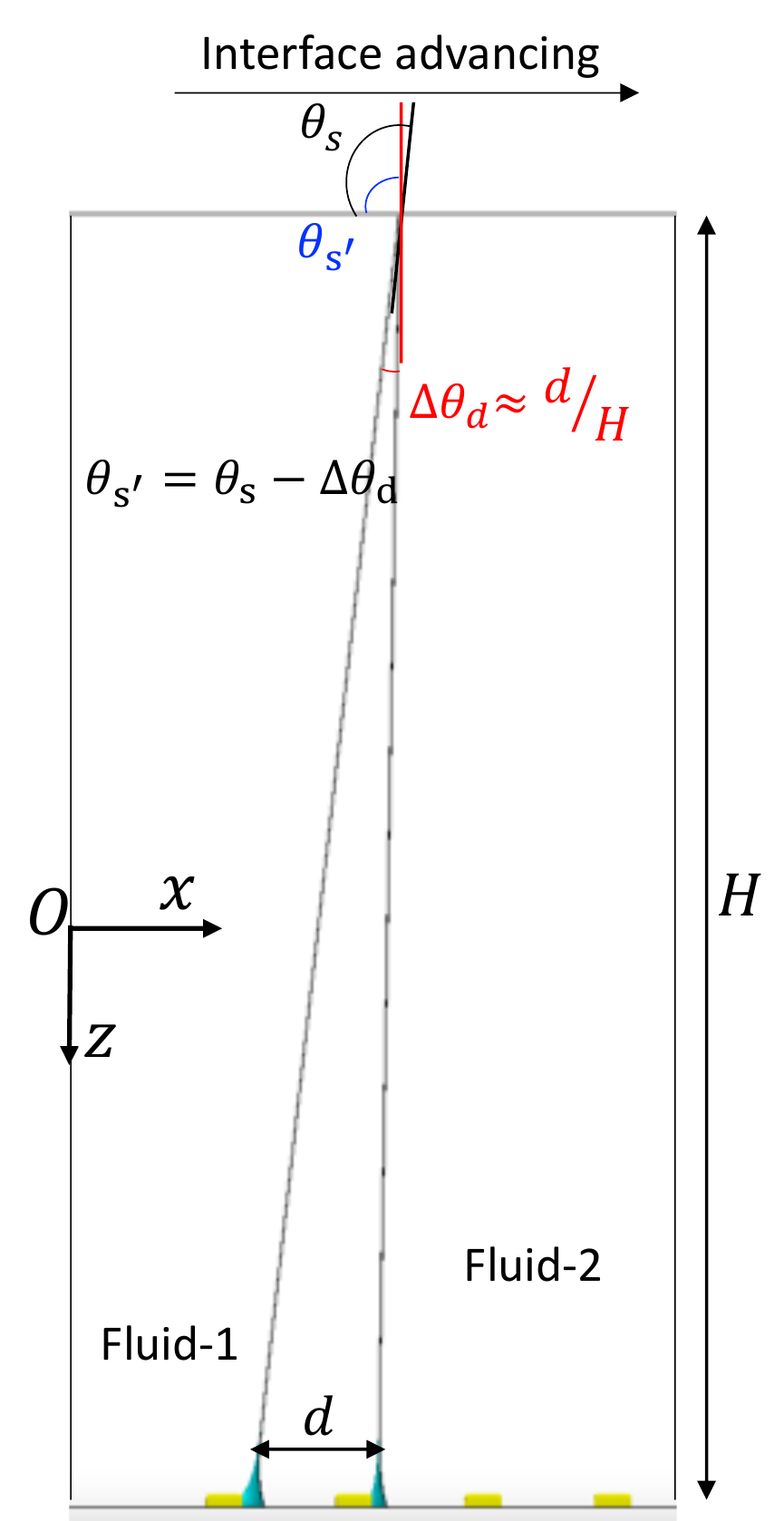}
		\caption{Equilibrium interface morphologies capturing a TPCL jump. The macroscopic contact angle decreases from $\theta_{\rm{s}}$ to $\theta_{\rm{s}^{'}}$ during the morphological transition from the critical to the second-critical state. The difference between the two angles ($\Delta \theta_{\rm{d}}=\theta_{\rm{s}}-\theta_{\rm{s}^{'}}$) decreases as the domain height is increased (see \S\ref{sec:meb_vs_SE}).}
		\label{fig:effect_of_jump_ch6}
\end{figure}

Indeed, on a structured surface with pillars of the same geometry and chemical properties, an interface moving in the direction of surface periodicity is able to find an equilibrium morphology on the next closest pillar every time the TPCL executes a jump, provided that the macroscopic flow velocities are very small ($\boldsymbol{v}_{\rm{CV}} \approx \boldsymbol{0}$) and $\Delta \theta$ is small. This is illustrated in figure \ref{fig:effect_of_jump_ch6}, where we show the equilibrium interface morphologies during the two critical states. The interface motion during the TPCL jump behaves as if it is pivoted at the top of the simulation domain and this motion results in a decrease in the interface's curvature in the $x-z$ plane. The interface remains pinned on this new defect until the curvature in the $x-z$ plane is increased to the value at which the interface arrives at the first critical state again and the TPCL executes a jump again. This is repeated again and again as the interface advances over the surface. The TPCL moves from one row of pillars to the next row, pinning every time it encounters a fresh row. This motion captures in the well-known \textit{stick-slip} behaviour of the TPCL.

\begin{figure}
     \centering
         \includegraphics[width=\textwidth]{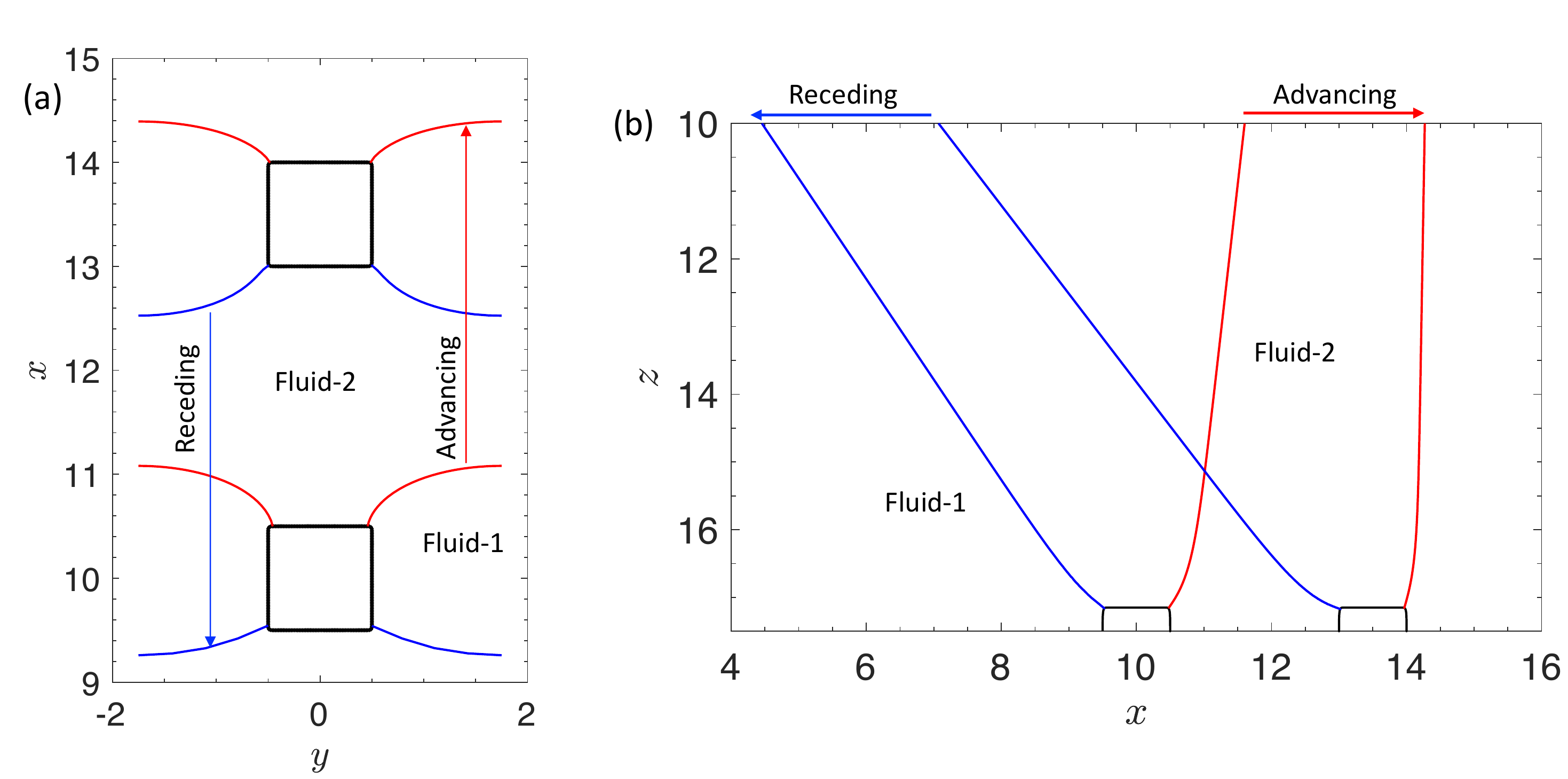}
		\caption{(a) TPCL on the domain base capturing first and second critical state interface morphologies during advancing (red) and receding (blue) motion. (b) Projection of the equilibrium interface morphologies on the symmetry plane during advancing (red) and receding (blue) motion capturing first and second critical states. The interface is traversing on a surface with $\theta_{\rm{e}}=72$\tc and $h/a=0.35$.}
		\label{fig:adv_rec_combined_ch6}
\end{figure}

Another point regarding interface movement is that the morphology of an equilibrium interface depends upon its history. Specifically, interface behaviour during advancing and receding motion is different (hysteresis). Here we show that the numerical model also predicts different interface morphologies depending on its history. We start with the interface in equilibrium and instead of advancing it, we recede the interface gradually (by gradual decrements in $\theta_{\rm{m}}$). In figure \ref{fig:adv_rec_combined_ch6}(a) we plot the portion of the TPCL which is on the domain base capturing the first and second critical states during the advancing and receding motion of the interface. We observe that these two critical states are completely different during the advancing and receding motion. In figure \ref{fig:adv_rec_combined_ch6}(b) we show the projection of interface morphologies on the symmetry plane capturing the first and second critical states, which also shows the difference in the interface morphologies during advancing and receding motion. Therefore, the interface morphologies and the macroscopic contact angle in critical states not only depend upon the surface roughness and Young's angle but also upon the history of the interface. This is consistent with our physical interpretation, discussed in \S\ref{sec:physcial_model} regarding how an interface advances.

\subsection{Calculating energy dissipation from interface dynamics}
\label{sec:dissipation}
Now that we have established an efficient numerical method for predicting interfacial dynamics, we want to use it to find the dissipation in energy ($D$) as the interface advances within the simulation domain. Referring back to part I, this dissipation in energy is directly related to CAH, and is given by
\begin{equation}
    D = -\sum_{k=1}^N \sum_{i<j} \sigma_{ij} \frac{\widehat{\Delta A_{ij}}_k}{A_{\rm{CV}}},
    \label{eqn:diss_paper1}
\end{equation}
where $\widehat{{\Delta A_{ij}}_k}$ is the change in the area of the $ij$ interface within the simulation domain when the TPCL is executing its $k$th jumping event, $\sigma_{ij}$ is the interfacial tension of the $ij$ interface and $A_{\rm{CV}}$ is the area traversed by the TPCL projected on the $x-y$ plane during which it executes $N$ number of jumping events. 

\begin{figure}
     \centering
         \includegraphics[width=\textwidth]{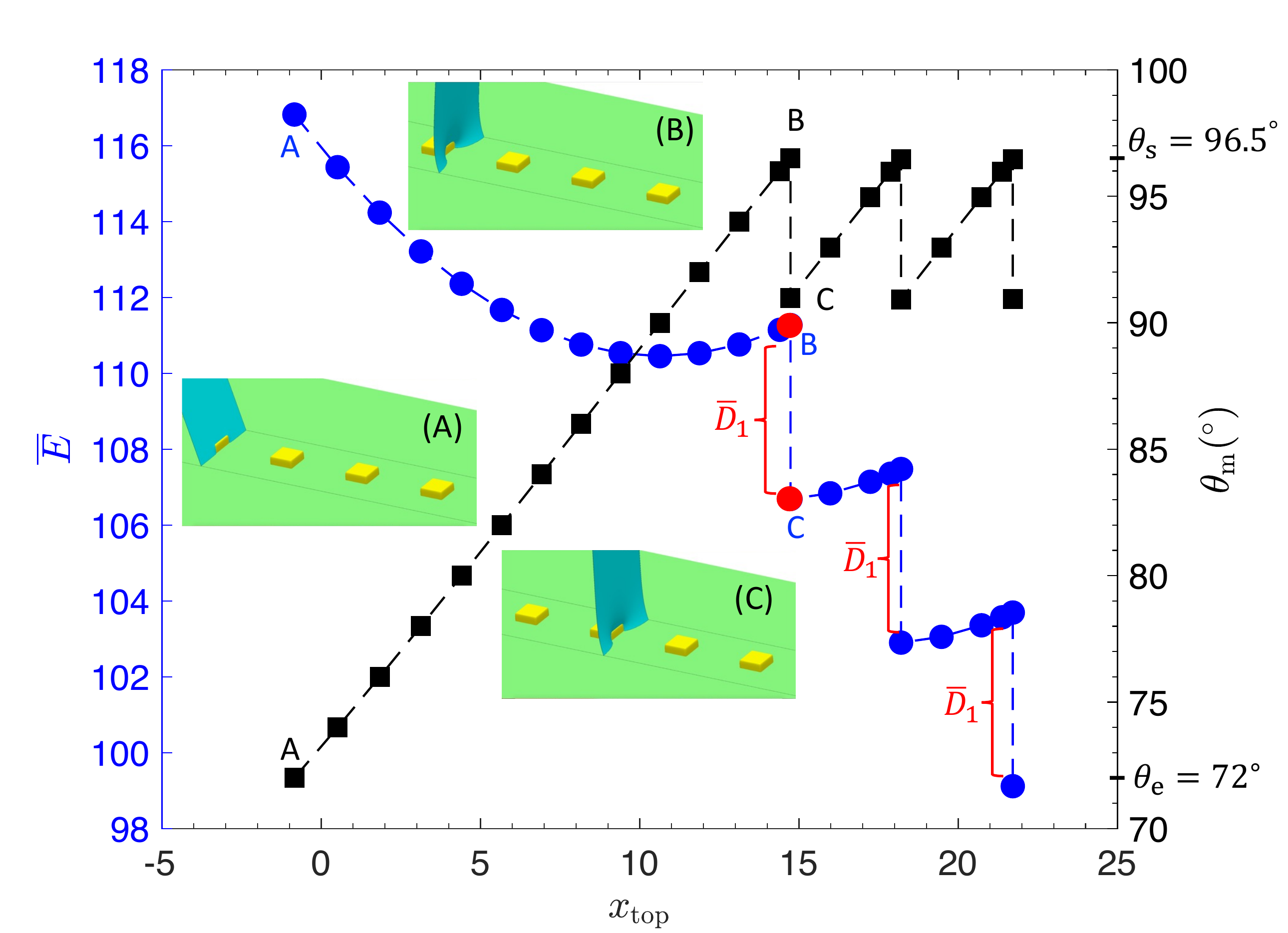}
		\caption{Variation of total non-dimensionalized energy within the simulation domain ($\overline{E}$, shown by blue circles) with interface position ($x_{\rm{top}}$). The advancing motion of the interface has been captured on a surface with $\theta_{\rm{e}}=72$\textdegree, $\phi=0.08$ and pillar aspect ratio of 0.35. As the interface advances, the TPCL moves in a \textit{stick-slip} fashion executing jumps and dissipating energy. Here we have shown three such dissipation events. We observe that during the motion of an interface in the surface periodicity direction of a structured surface, the magnitude of non-dimensional energy dissipation ($\overline{D}_1$, per pillar) is the same during each TPCL jumping event. Insets (A)-(C) show different equilibrium morphologies of the interface during advancement. Inset (A) shows the equilibrium interface morphology when $\theta_{\rm{m}}=\theta_{\rm{e}}$. This represents a typical starting point of an interface advancement simulation. Inset (B) shows the equilibrium interface morphology during a critical state (the interface is advancing from the left to the right side). Inset (C) shows the equilibrium interface morphology during a second-critical state. Insets (B) and (C) represent the first TPCL jumping event. Variation in macroscopic contact angle ($\theta_{\rm{m}}$, in degrees) corresponding to the equilibrium states shown in the total non-dimensional energy ($\overline{E}$) plot is shown by black squares. When the interface reaches a critical state (for example state `B'), the macroscopic contact angle attains the maximum value ($\theta_{\rm{s}}$).}
		\label{fig:dissipation_def_ch6}
\end{figure}

Before going into the details of this energy dissipation calculation, we first examine some aspects of the nature of total interfacial energy variation as an interface advances within the simulation domain. Figure \ref{fig:dissipation_def_ch6} shows the variation in total non-dimensional energy ($\overline{E}$) within the simulation domain as a function of $x_{\rm{top}}$. The total energy is non-dimensionalized by the cross-sectional area of the pillars and the surface tension of the fluid-fluid interface i.e.
\begin{equation}
    \overline{E}=\frac{E}{\sigma_{12}a^2}.
    \label{eqn:non_dim_energy_def}
\end{equation}
As the interface advances within the domain,  we observe a decreasing trend in the total non-dimensional energy. This is because of the particular nature of the surface under consideration that has $\theta_{\rm{e}}<90$\textdegree.
However, there is a sudden drop in the total non-dimensional energy at certain points, such as 'B' and 'C'. We also observe a drop in the macroscopic contact angle at the same times. These points represent first and second critical states respectively (see figure \ref{fig:theory_thm_ch6}). Also in figure \ref{fig:dissipation_def_ch6}, we observe that the drop in total non-dimensional energy (dissipation) is the same during each TPCL jumping event. This is because the dynamics of each jump are the same as the interface is moving over a structured surface and in the direction of the surface periodicity. 

\begin{figure}
     \centering
         \includegraphics[width=\textwidth]{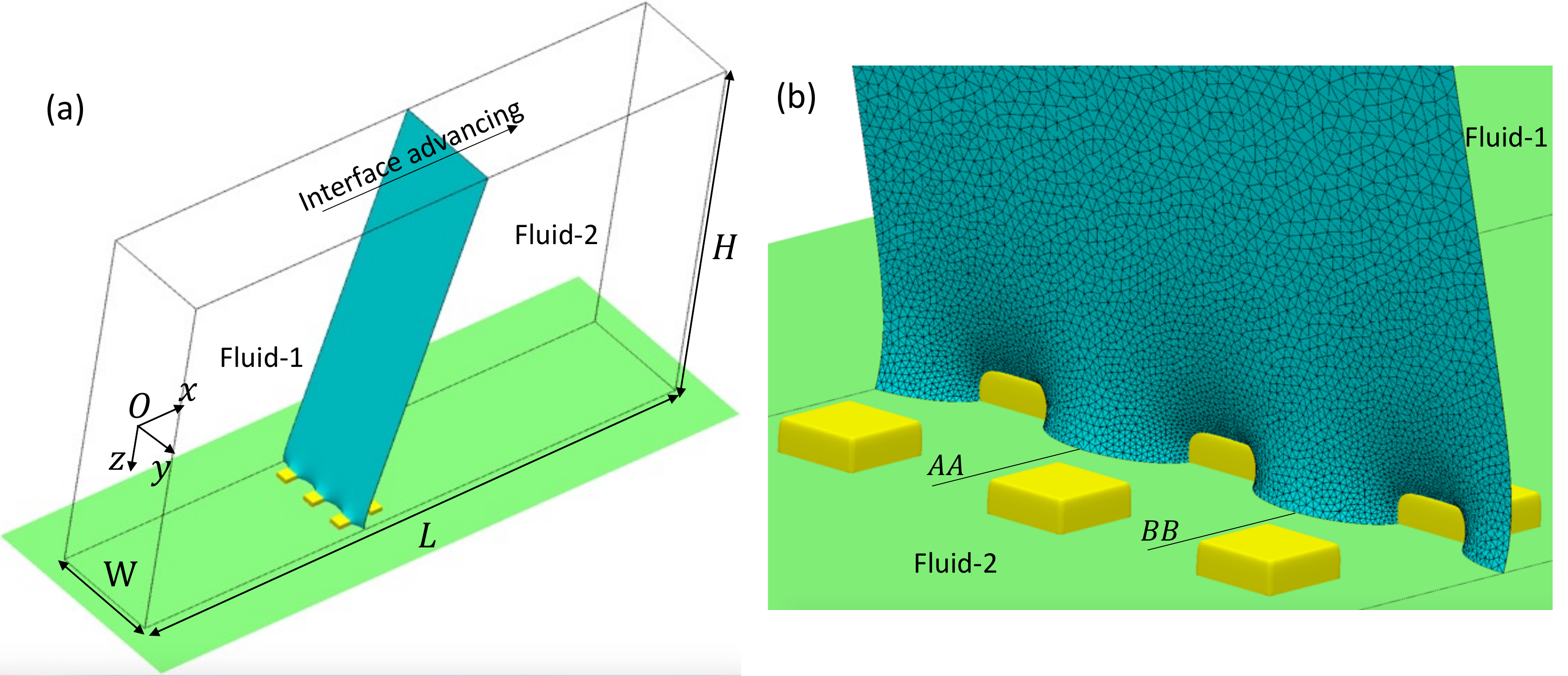}
		\caption{(a) Three-dimensional view of the simulation domain with three pillars in a row ($\phi=0.13, h/a=0.35, \theta_{\rm{e}}=72$\textdegree). The interface is pinned on the first row and represents the stable morphology just before depinning from the pillars (critical state). (b) Zoomed-up view of the pinned interface. Equilibrium interface morphology is symmetric around each pillar. The zoomed is flipped about the $z$ axis to show the fluid-1 tongue protruding between the pillars.}
		\label{fig:interface_symmetry_ch6}
\end{figure}

\begin{figure}
 	\centering
 		\includegraphics[width=\textwidth]{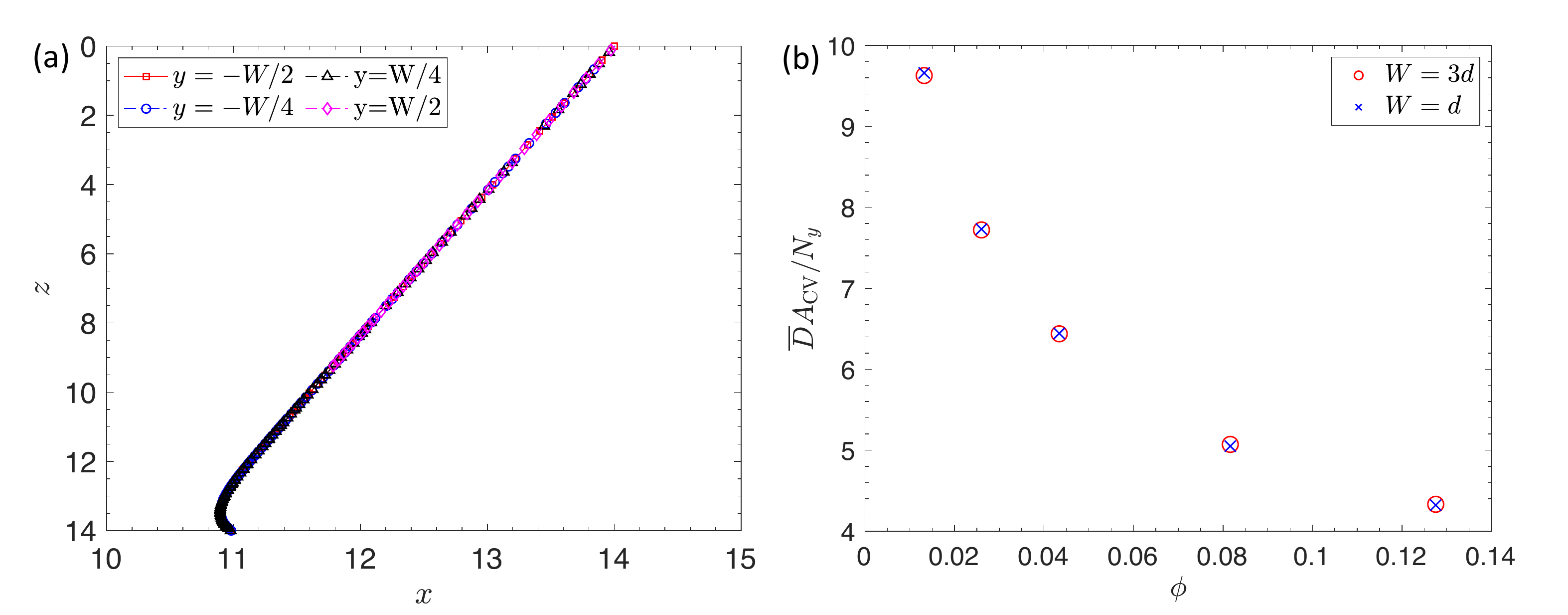}
 		\caption{(a) Equilibrium interface morphologies projected on a plane parallel to $x-z$ plane and passing through domain's front and back walls and through section `AA' and `BB' as shown in figure \ref{fig:interface_symmetry_ch6}(b). Similar profiles are obtained at the four locations. (b) Total non-dimensional energy dissipation ($\overline{D}A_{\rm{CV}}$) divided by the number of pillars along the row ($N_y$) ($h/a=0.35, \theta_{\rm{e}}=72$\textdegree). The energy dissipation per pillar is independent of the number of pillars used in the simulation.}
 		\label{fig:six_pillar_projection_ch6}
 \end{figure}

We also examine the role of the simulation domain's width in calculating interface dynamics. Simulations can be performed in a computational domain of any width which is an integer multiple of the inter-pillar distance. In figure \ref{fig:dissipation_def_ch6} we have shown the interface dynamics calculated using a simulation domain which has a width equal to the distance between neighbouring pillars ($W=d$), while in figure \ref{fig:interface_symmetry_ch6} we show the equilibrium morphology of an interface pinned on a row of three pillars. The shapes are the same. Further, figure \ref{fig:six_pillar_projection_ch6} shows the interface profile ($\phi=0.13$, $\theta_{\rm{e}}=72$\tc and $h/a=0.35$) projected on a plane parallel to the $x-z$ plane and passing through $y=-W/2$, $y=-W/4$ (i.e., passing through section BB in figure \ref{fig:interface_symmetry_ch6}(b)), $y=W/4$ (i.e., passing through section AA in figure \ref{fig:interface_symmetry_ch6}(b)) and $y=W/2$ from the three pillar width simulation. All four profiles are identical, confirming the symmetry in interface morphology around each pillar. 
 
Another aspect of this periodicity in interface morphologies is that the dissipation in energy from an interface pinned on a single pillar is the same as the energy dissipation from an interface pinned on a row of $N_y$ pillars divided by the number of pillars ($N_y$). In order to verify this, we calculate the non-dimensional energy dissipation ($\overline{D}$, as the difference in total non-dimensional interfacial energy within the domain between the first and second critical states) when an interface pinned on a row of three pillars along the simulation domain's width executes a jump. The dissipation is calculated for a single TPCL jumping event from one row to the next (3 pillars in each row). We plot the total dissipation divided by the number of pillars with the pillar area fraction ($\phi$) in figure \ref{fig:six_pillar_projection_ch6}(b) for both the single pillar and three pillar width simualtions. We observe that for a structured surface (and when the interface is moving in the direction of surface periodicity) the dissipation per pillar is independent of the number of pillars used in the simulation.
%Therefore, the total energy dissipated during the interface movement can be calculated from the energy dissipation due to a single TPCL jumping event between two single pillars in consecutive rows (i.e. dissipation per pillar per dissipation event).

Hence, the energy dissipated when the TPCL pinned on a single pillar executes a jump to the next pillar can be used to calculate the total dissipation in energy during the travel of an interface that executes $N$ jumps over an array of pillars that is $N_y$ pillars wide.  Specially for this situation equation (\ref{eqn:diss_paper1}) becomes
\begin{equation}
    D = -N_y N \sum_{i<j} \sigma_{ij} \frac{\widehat{\Delta A_{ij}}_1}{A_{\rm{CV}}},
    \label{eqn:dissipation_intro}
\end{equation}
where $\widehat{\Delta {A_{ij}}_1}$ is the change in interfacial areas during a single TPCL jumping event from one pillar to the next. Equation (\ref{eqn:dissipation_intro}) can, therefore be written as
\begin{equation}
\begin{split}
 D &= -N_y N \left(\frac{\sigma_{12}\Delta A_{12}+\sigma_{\rm{1S}}\Delta A_{\rm{1S}} + \sigma_{\rm{2S}}\Delta A_{\rm{2S}}}{A_{\rm{CV}}}\right),\\
 &= -N_y N \left(\frac{\sigma_{12}\Delta A_{12}-(\sigma_{\rm{2S}} - \sigma_{\rm{1S}}) \Delta A_{\rm{1S}} }{A_{\rm{CV}}}\right).
 \label{eqn:dissipation_intro2}
 \end{split}
\end{equation}
%
%The total change in the energies of the simulation domain due to the changes in all three interface areas (12, $\rm{1S}$ and $\rm{2S}$) during a single TPCL jumping event (i.e, $\Delta E$) can be calculated using equation (\ref{eqn:domain_energy}) when applied to a simulation domain with a single pillar along the domain's width and along the direction of the interface advancement. Using equations (\ref{eqn:diss_paper1}) and (\ref{eqn:domain_energy})
%
In equation (\ref{eqn:dissipation_intro2}) we have used the relationship $\Delta A_{\rm{2S}}=-\Delta A_{\rm{1S}}$. Using equations (\ref{eqn:domain_energy}), (\ref{eqn:young's_ch6}) and (\ref{eqn:dissipation_intro2}), we can write
\begin{equation}
    \begin{split}
     D &= -N_y N \frac{\Delta E}{A_{\rm{CV}}} \\
       &= \sigma_{12} \phi \overline{D}_1,
    \end{split}
    \label{eqn:dissipation_ch6_1}
\end{equation}
where $\phi = N_y N A_{\rm{p}}/A_{\rm{CV}}$ is the pillar area fraction, $A_{\rm{p}}$ is the cross-sectional area of the pillar tops ($a^2$ in the present case) and $\overline{D}_1$ is the non-dimensional dissipation per pillar per dissipation event, defined by
\begin{equation}
    \overline{D}_1 = \frac{-\Delta E}{\sigma_{12}A_{\rm{p}}}
    \label{eqn:diss_per_pillar}
\end{equation}
From equations (\ref{eqn:dissipation_ch6_1}) and (\ref{eqn:diss_per_pillar}) we can relate $D$, which is the total dissipation in energy per unit area traversed by the TPCL ($A_{\rm{CV}}$) during interface advancement over a structured surface in the direction of surface periodicity, to the non-dimensional energy dissipation per pillar per dissipation event as calculated in single pillar column simulations. The total dissipation ($\overline{D}$) can also be non-dimensionalized by the interfacial tension $\sigma_{12}$, i.e.
\begin{equation}
 \overline{D}=\frac{D}{\sigma_{12}}=\phi \overline{D}_1. 
 \label{eqn:dissipation_ch6_3}
\end{equation}
In the rest of the paper, we refer to the dissipation per pillar per dissipation event as dissipation per pillar for brevity.

We finish our description of the numerical method with two final comments regarding efficiency. As well as the periodicities discussed above, the fluid-fluid interface also possesses a reflection symmetry with respect to a plane passing through the centre of the pillars and parallel to the $x-z$ plane. Mathematically this is expressed as $\boldsymbol{n}_{\rm{S}} \cdot \boldsymbol{j}=0$ along this plane (figure \ref{fig:physical_model_ch6}(a), symmetry plane is shown in grey colour). This symmetry allows us to reduce the computational expense by simulating only half of the simulation domain width (compared to a single pillar simulation) without affecting the outcome of the simulations. Therefore, we use a domain which is only $d/2$ wide and $30d$ high for the majority of our simulations ((see Appendix \ref{sec:channel_height:appendix})). The dissipation for the full domain (i.e., $H=30d, W=d$) is then obtained as twice the dissipation calculated using this half domain (i.e., $H=30d, W=d/2$). Also, to increase computational efficiency and accuracy the simulations are performed using a non-uniform mesh which is dynamically adjusted based on distance to the rear face of each pillar, according to equation (\ref{eqn:mesh_refinement_scheme_ch6}). Appendix \ref{sec:mesh_resolution:appendix} gives more details of this dynamic mesh generation technique.

\section{Results and discussions}
\label{sec:results}

In this section, we present our results for the advancing and receding motion of an interface on a surface with a structured array of square pillars. We discuss the effect of pillar area fraction ($\phi$) and aspect ratio ($h/a$) on the macroscopic contact angle and dissipation in energy. Based on the numerical results, we develop a relationship between the pillar area fraction and the energy dissipation for both the advancing and receding interface. Finally, using the mechanical energy balance, we present a predictive equation for the contact angle hysteresis on such surfaces.

\subsection{Advancing interface}
\subsubsection{Interface behaviour}
\begin{figure}
     \centering
     \includegraphics[width=\textwidth]{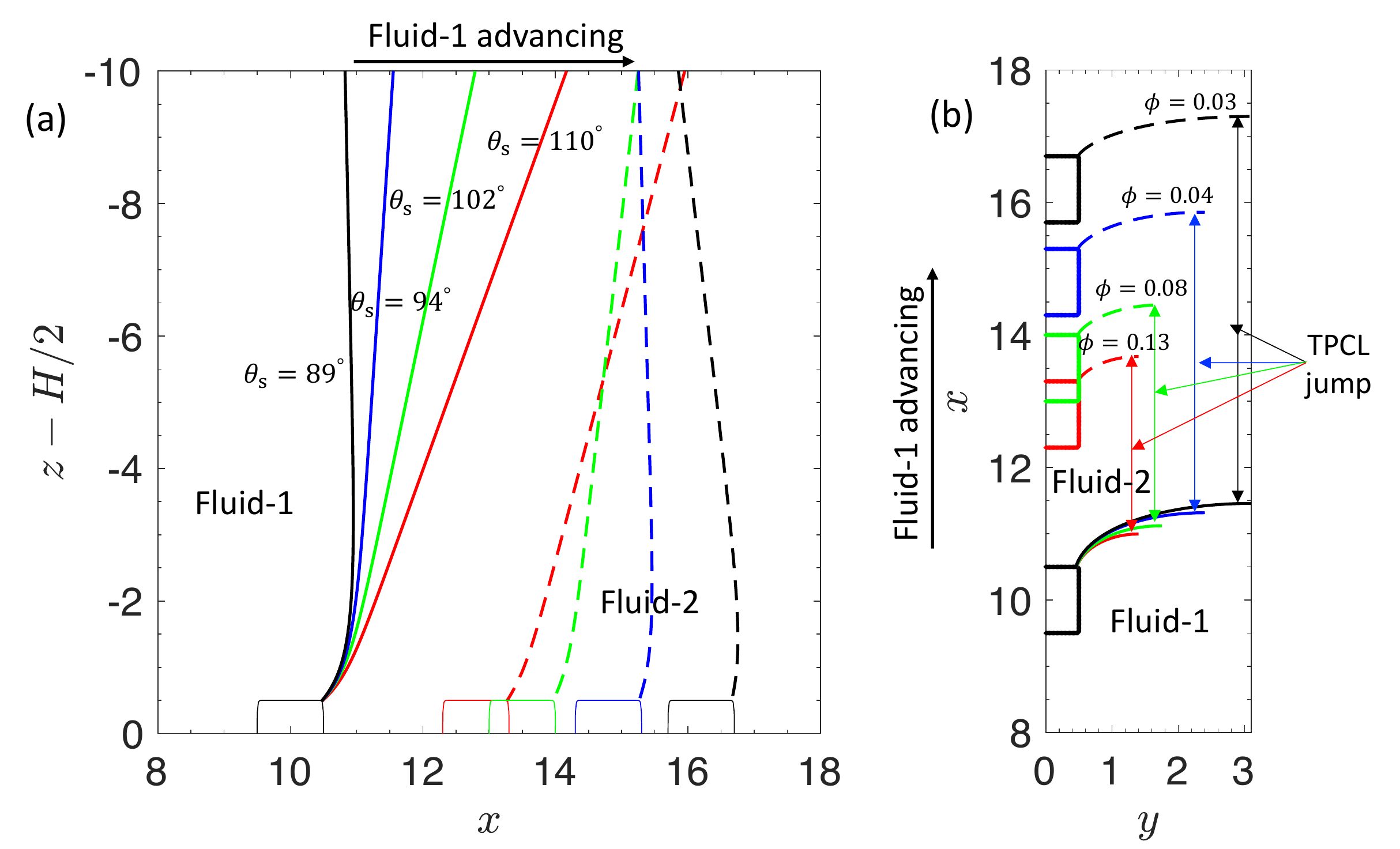}
     \caption{(a) Equilibrium interface profiles projected on the symmetry plane passing through the center of the pillar and parallel to $x-z$ plane. The results are calculated for $\theta_{\rm{e}}=72$\tc and pillar aspect ratio ($h/a$) of 0.5. The critical contact angle ($\theta_{\rm{s}}$) exhibited by the pinned interface varies with area fraction ($\phi$) ($110$\textdegree, $102$\textdegree, $94$\textdegree, $89$\tc for pillar area fractions $\phi=0.13$ (red), $\phi=0.08$ (green), $\phi=0.04$ (blue), $\phi=0.03$ (black) respectively). (b) Shows the TPCL on the domain bottom corresponding to the equilibrium interface morphologies as depicted in (a). The TPCL curvature increases as the pillar area fraction increases.}
     \label{fig:proj_width_ch6}
 \end{figure}

Figure \ref{fig:proj_width_ch6}(a) shows the equilibrium interface profile projected on the symmetry plane (passing through pillar center and parallel to the $x-z$ plane) for a pinned interface with $\theta_{\rm{e}}=72$\textdegree, $h/a=0.50$, $\phi=0.03,0.04,0.08$ and $0.13$. The solid lines represent the interface profile at the first critical state and the dashed lines represent the equilibrium interface profiles after the TPCL jump i.e. the second-critical state. We observe that $\theta_{\rm{s}}$ increases with the increase in pillar area fraction and from figure \ref{fig:proj_width_ch6}(b), where we have plotted the TPCL lying on the domain base, we can see that the curvature in the $x-y$ plane adjacent to the pillar increases with the increase in pillar area fraction ($\phi$).

\begin{figure}
    \centering
    \includegraphics[width=\textwidth]{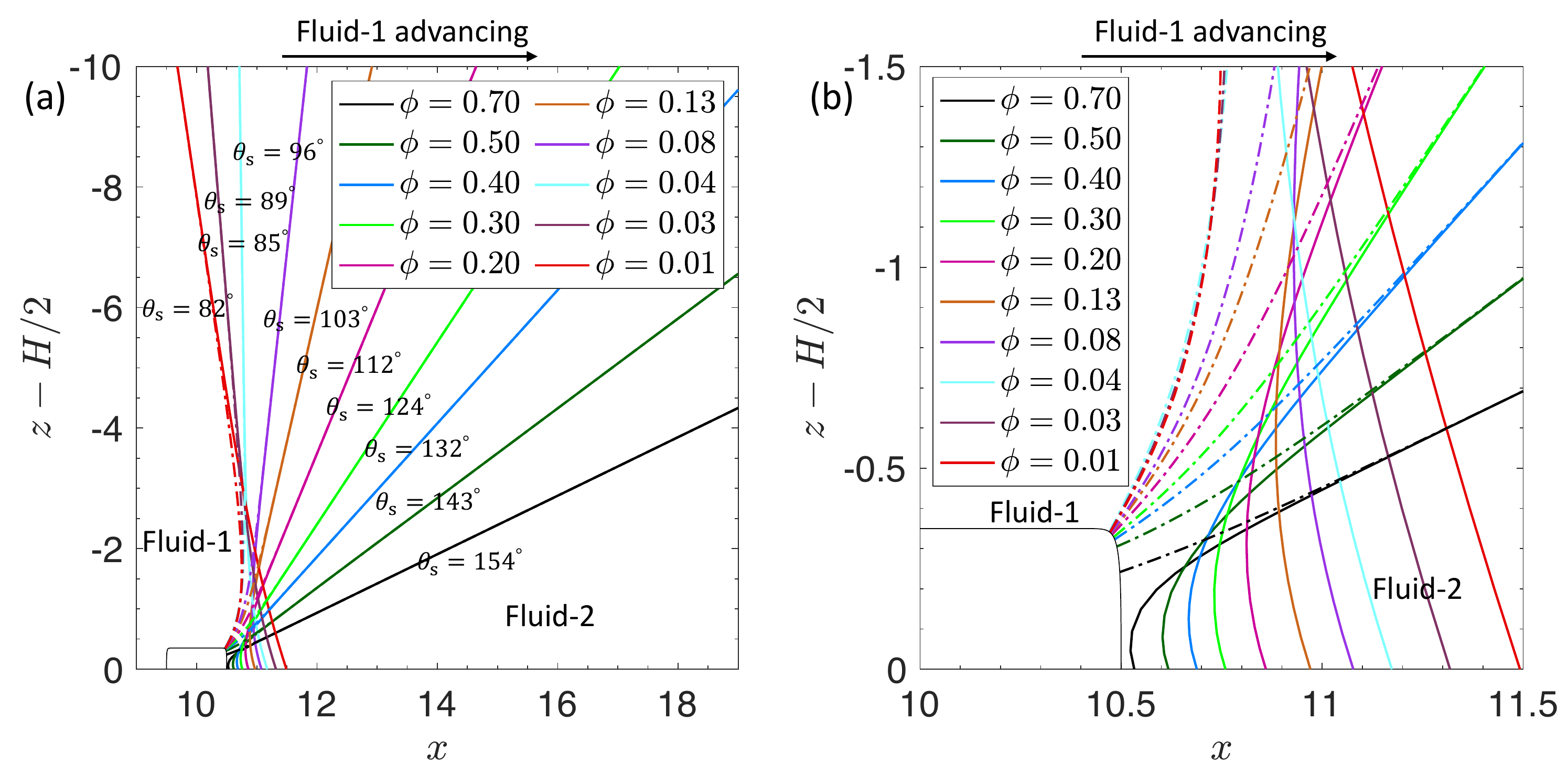}
    \caption{Equilibrium interface morphologies just before depinning (first critical state) as projected on the symmetry plane (shown by dashed-dotted lines) and the domain wall ($y=W/2$, shown by solid lines) for an advancing interface on a surface with $\theta_{\rm{e}}=72$\textdegree, $h/a=0.35$ and area fractions $\phi=0.01, 0.03, 0.04, 0.08, 0.13, 0.20, 0.30, 0.40, 0.50, 0.70$. (b) Zoomed view of interface morphologies near the TPCL.}
    \label{fig:front_back_ch6}
\end{figure}

Figure \ref{fig:front_back_ch6}(a) shows the interface morphology during a first critical state projected on the simulation domain wall ($y=W/2$) as well as the symmetry plane for $\theta_{\rm{e}}=72$\textdegree, $h/a=0.35$ and for a large variety of area fractions between $\phi=0.01$ to $\phi=0.70$. The maximum macroscopic contact angle ($\theta_{\rm{s}}$) for different pillar area fractions ($\phi$) is also shown. As previously noted, there is a direct relationship between $\theta_{\rm{s}}$ and $\phi$, i.e. $\theta_{\rm{s}}$ increases with $\phi$. Other information that can be inferred from figure  \ref{fig:front_back_ch6}(b), is the nature of the projection on the domain wall. Since the interface in equilibrium represents a minimal surface, at every point the mean curvature is zero. Therefore, a high curvature of the interface in the $x-y$ plane is balanced by a similar magnitude of curvature, but in the opposite direction, in the $x-z$ plane, with both measured in the same neighbourhood. As noted earlier, the curvature in the $x-y$ plane next to the pillar increases with $\phi$, therefore, the curvature of the interface in the $x-z$ plane as well increases with $\phi$, which is the reason we observe large $\theta_{\rm{s}}$ values at higher area fractions. A high curvature in the $x-y$ plane also means that the TPCL intersects the domain wall at a point closer to the pillar on which the interface is pinned. Therefore, in figure \ref{fig:front_back_ch6}(b) the interface projection on the domain front wall for the pillar area fraction ($\phi=0.70$) intersects the domain bottom very close to the pillar on which the interface is pinned (this can also be observed in the TPCL projections for the first critical interface morphologies shown as solid lines in figure \ref{fig:proj_width_ch6}(b)). 

\begin{figure} 
     \centering
     \includegraphics[width=\textwidth]{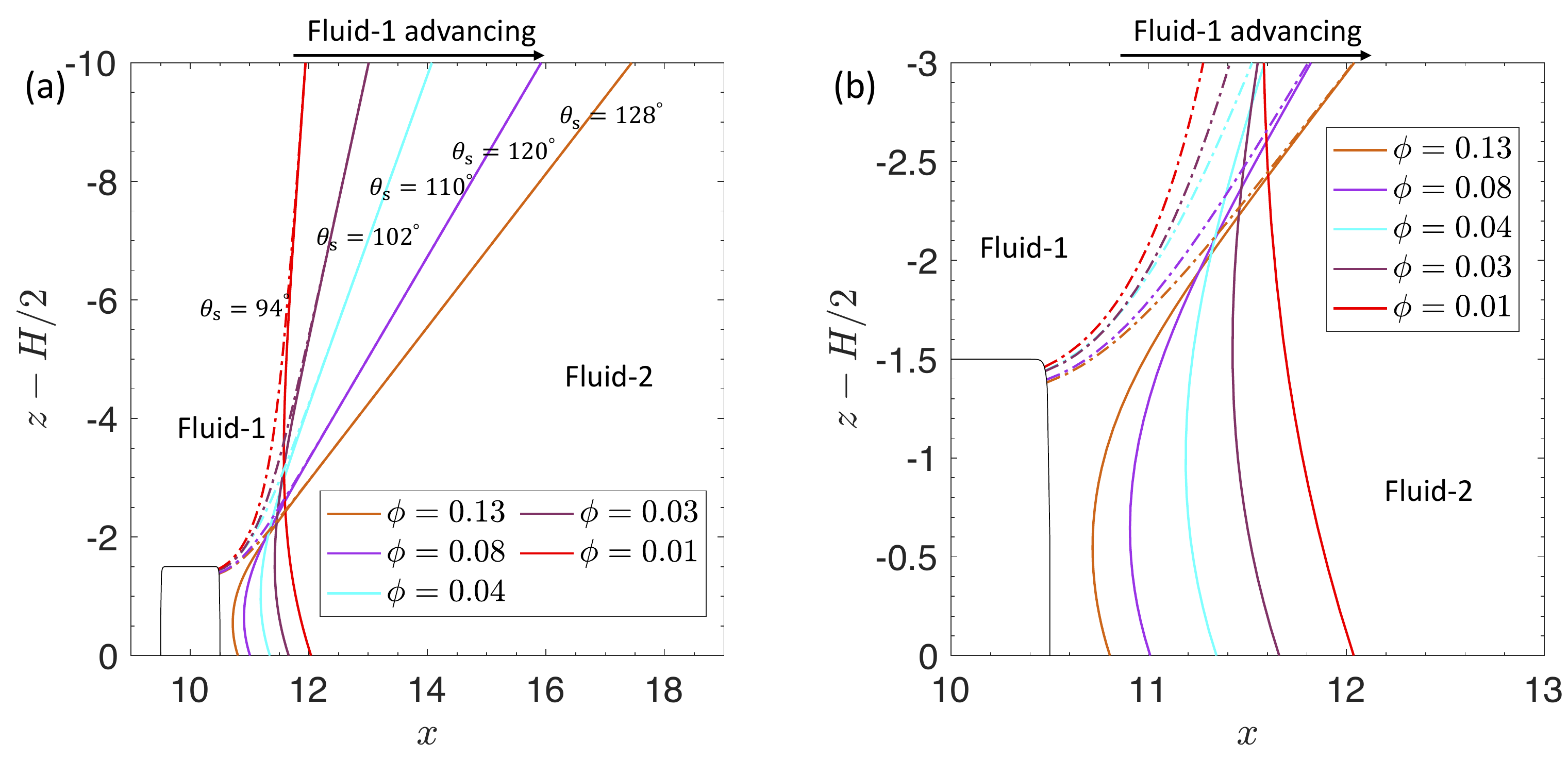}
     \caption{Equilibrium interface morphologies just before depinning (first critical state) as projected on the symmetry plane (shown by dashed-dotted lines) and the domain wall ($y=W/2$, shown by solid lines) for an advancing interface on a surface with $\theta_{\rm{e}}=72$\textdegree, $h/a=1.5$ and area fractions $\phi=0.01, 0.03, 0.04, 0.08, 0.13$. (b) Zoomed view of interface morphologies near the TPCL.}
     \label{fig:front_back_1d5_ch6}
 \end{figure}

\begin{figure}
    \centering
    \includegraphics[width=\textwidth]{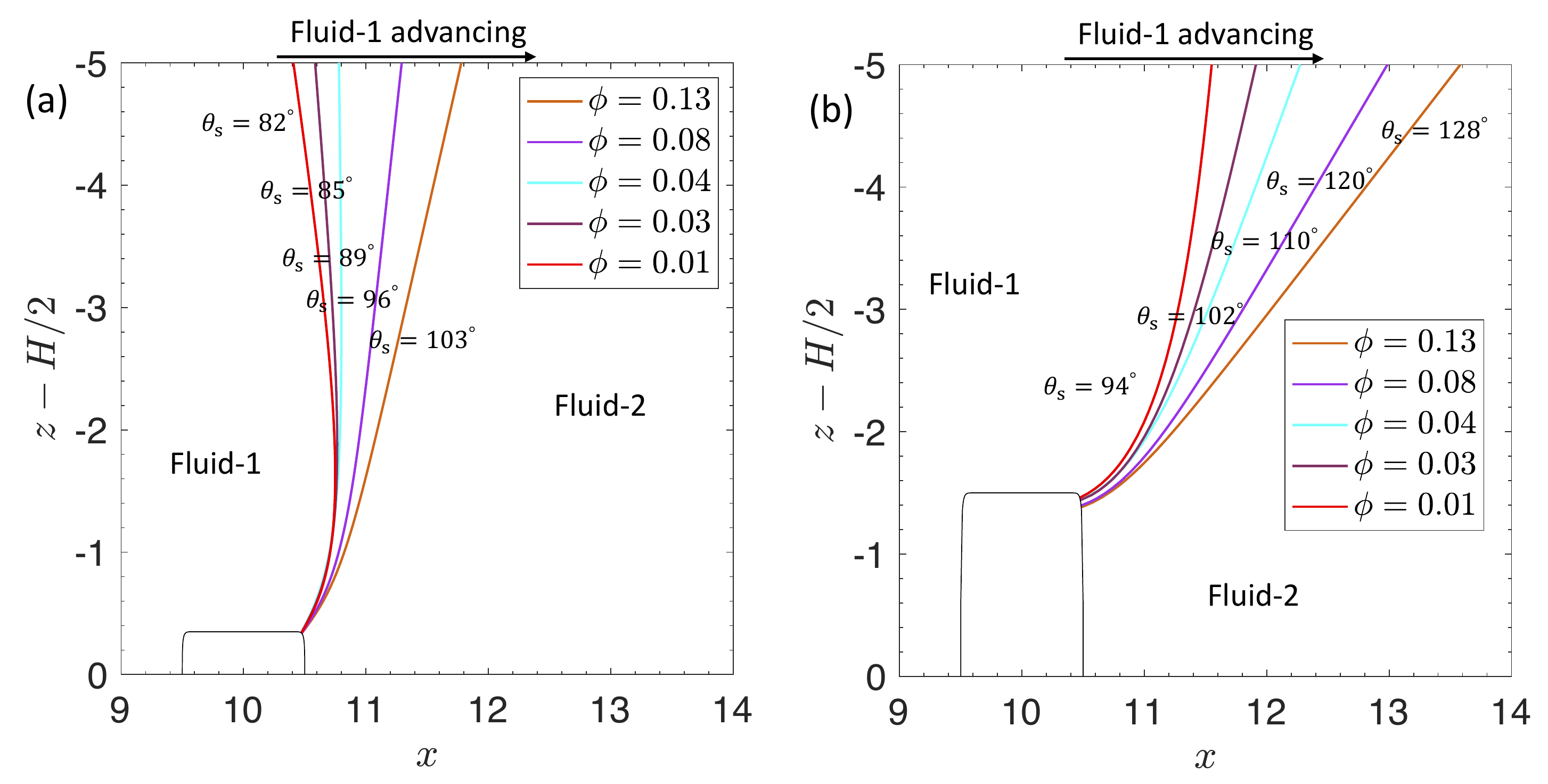}
    \caption{First critical equilibrium interface morphologies as projected on the symmetry plane for pillar aspect ratio ($h/a$) 0.35 (a) and 1.5 (b) for an advancing interface on a surface with $\theta_{\rm{e}}=72$\tc and area fraction $\phi=0.01, 0.03, 0.04, 0.08, 0.13$. The interface curvature as well as the maximum macroscopic contact angle ($\theta_{\rm{s}}$) increases with the pillar aspect ratio ($h/a$).}
    \label{fig:proj_asp_ratio_ch6}
\end{figure}

Apart from the pillar area fraction, we observe that the pillar aspect ratio also affects the interface curvature and the maximum macroscopic contact angle ($\theta_{\rm{s}}$). In figure \ref{fig:front_back_1d5_ch6} we show the projection of interface morphologies in the first critical state for a pillar aspect ratio ($h/a$) of 1.5 ($\theta_{\rm{e}}=72$\tc and $\phi=0.01,0.03,0.04,0.08$ and $0.13$). We observe that qualitatively, the interface in equilibrium behaves in a similar manner on both the pillar aspect ratios ($h/a=0.35, 1.5)$, however quantitatively the principal curvatures in both the $x-z$ and $x-y$ planes are comparatively higher for higher aspect ratio pillars.
% Although, qualitatively the interface in equilibrium behaves in a similar manner on both the pillar aspect ratios ($h/a=0.35, 1.5)$, quantitatively the pinning strength is much higher in case of the higher aspect ratio. Therefore, the maximum macroscopic contact angles ($\theta_{\rm{s}}$) are higher for higher aspect ratio even at the same pillar area fraction. 
This is reflected in a higher magnitude of the maximum macroscopic contact angle ($\theta_{\rm{s}}$) for higher aspect ratios which is shown in figure \ref{fig:proj_asp_ratio_ch6}, where we plot the first critical interface morphologies projected on the symmetry plane for pillar aspect ratios 0.35 and 1.5 and area fractions 0.01, 0.03, 0.04, 0.08 and 0.13 respectively. 

\begin{figure} 
    \centering
    \includegraphics[width=0.80\textwidth]{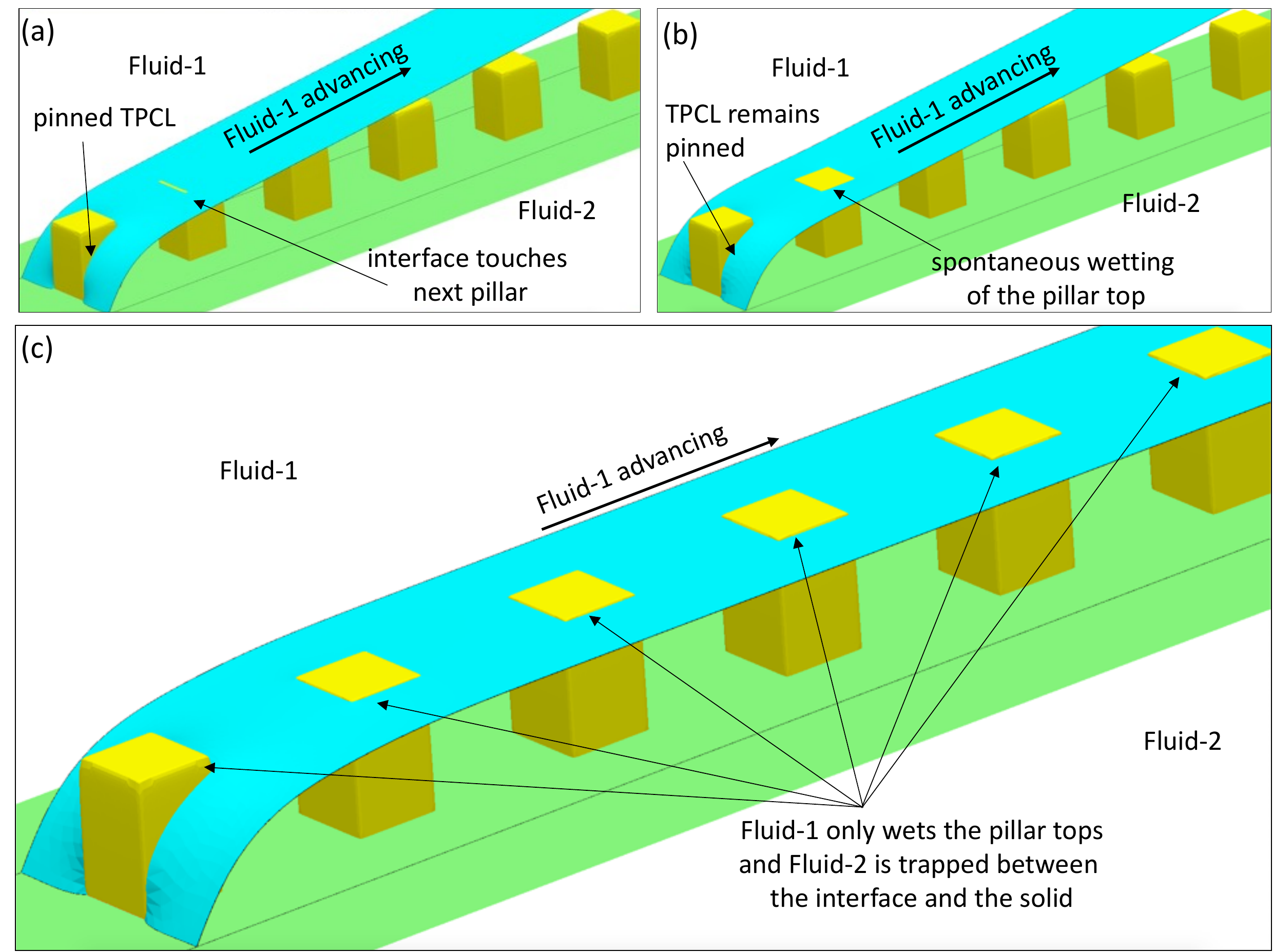}
    \caption{Equilibrium interface morphologies capturing the advancing motion when the interface while pinned on a pillar touches the next pillar in the direction of advancement. (a) The interface is touching the next pillar and the fluid-1 spontaneously spreads on the top of the pillar as shown in (b). However, the TPCL remains pinned on the first pillar while the fluid-1 gradually wets the pillar tops. During this process, a thin film of the surrounding fluid (fluid-2) gets trapped in between the interface and the solid surface as shown in (c). The interface morphologies shown here are simulated with $\theta_{\rm{e}}=121$\textdegree, $\phi=0.13$ and $h/a=1.5$.}
    \label{fig:wetting_transition}
\end{figure}

An interesting equilibrium interface morphology arises when the pillar area fraction is increased beyond a certain critical point, at which we observe that the interface makes contact with the next pillar before depinning from the previous pillar (shown in figure \ref{fig:wetting_transition}(a)). The area fraction at which this morphological transition occurs depends upon the pillar geometry and Young's angle. Even though such interface morphologies are not the focus of this paper, we show one case as an example of the capability of the present numerical method to deal with such morphologies. In figure \ref{fig:wetting_transition} we show the advancing motion of an interface over a surface with $\theta_{\rm{e}}=121$\tc and $h/a=1.5$. 
The interface, before depinning from the first pillar makes contact with the next pillar in the direction of interface advancement and a portion of the fluid-1/fluid-2 interface is converted to the fluid-1/solid interface. The equilibrium interface morphology is shown in figure \ref{fig:wetting_transition}(b) where we observe that fluid-1 only wets the pillar top. This process is repeated again, resulting in an equilibrium interface morphology with fluid-1 wetting the pillar tops while fluid-2 is trapped between the solid surface and the interface (figure \ref{fig:wetting_transition}(c)). It is also possible to have other equilibrium interface morphologies, such as the fluid-1 wetting the pillar tops as well as some portion of the pillar sides for a different set of  $\theta_{\rm{e}}$ and $h/a$ values. Essentially this is modeling the transition from a Wenzel to Cassie wetting state (for certain Young's angles). The study of such cases is left to future work.

\subsubsection{Energy dissipation during advancing motion of the interface}
\label{sec:dissipation_advancing}
\begin{figure}
    \centering
    \includegraphics[width=0.65\textwidth]{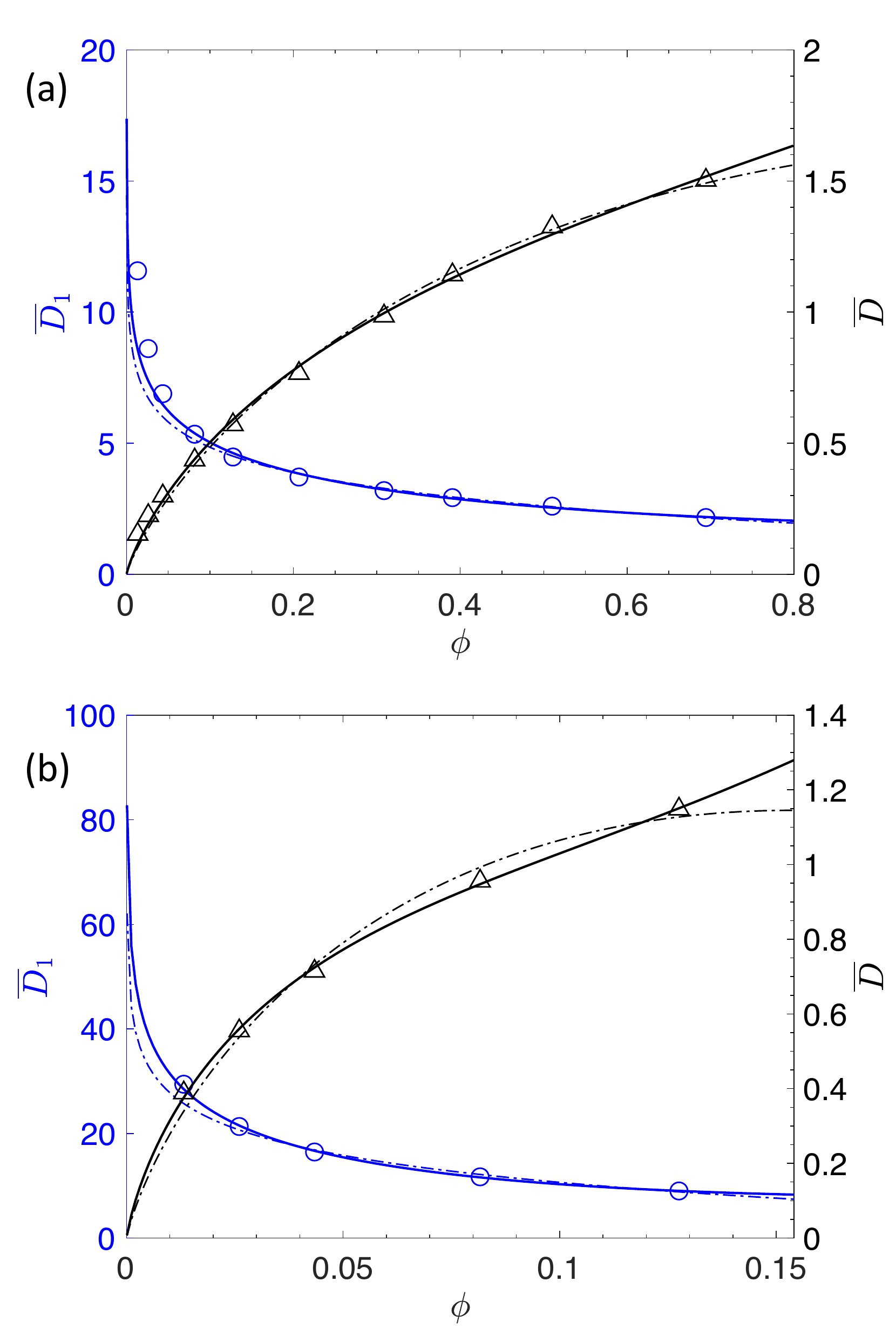}
    \caption{Variation in the total non-dimensional energy dissipation ($\overline{D}$, shown by triangles) and non-dimensional energy dissipation per pillar ($\overline{D}_1$, shown by circles) with the pillar area fraction ($\phi$) for an interface advancing in the surface periodicity direction is shown. Results have been simulated for Young's angle $72$\tc and pillar aspect ratio of (a) 0.35 and (b) 1.5 respectively. Fitting equations of dilute (equation (\ref{eqn:total_diss_dilute})) and non-dilute (equation (\ref{eqn:total_diss_non_dilute})) form for $\overline{D}$ are shown by dashed-dotted and solid black lines respectively. Fitting equations for $\overline{D}_1$ are shown by dashed-dotted (dilute form, equation (\ref{eqn:diss_dilute})) and solid (non-dilute, equation (\ref{eqn:diss_non_dilute})) blue lines respectively.}
    \label{fig:adv_diss_ch6}
\end{figure}

In this section, we develop a relationship between the energy dissipation and pillar area fraction for an interface advancing in the direction of surface periodicity. Figure \ref{fig:adv_diss_ch6} shows the non-dimensional energy dissipation per pillar ($\overline{D}_1$) and total non-dimensional energy dissipation ($\overline{D}$) plotted against pillar area fraction ($\phi$) (pillar aspect ratio 0.35 and 1.5, and $\theta_{\rm{e}}=72$\textdegree). 
The non-dimensional dissipation per pillar is obtained from the interface dynamics simulations for different area fractions and pillar aspect ratios, while the total non-dimensional dissipation is obtained using, $\overline{D}=\phi \overline{D}_1$ from equation (\ref{eqn:dissipation_ch6_3}). 

For fitting $\overline{D}_1$ to an appropriate correlation, we start with an equation form proposed by \cite{joanny1984model} for the dissipation in energy due to a single strong defect on a dilute surface, i.e.
\begin{equation}
 \overline{D}_1 = A \ln \phi + C.
 \label{eqn:diss_dilute}
\end{equation}
Here $A$ and $C$ are constants that depend upon the pillar aspect ratio ($h/a$) and Young's angle ($\theta_{\rm{e}}$) of the surface. The form of the relationship between $\overline{D}_1$ and $\phi$ (equation (\ref{eqn:diss_dilute})) arises from the relationship between $\phi$ and $d$ (inter pillar distance) on a structured surface \citep{joanny1984model}. When a liquid surface is perturbed by the presence of a pillar, the perturbation is carried to infinity, which can be captured by a function of logarithmic nature. Under these conditions, the geometry of the pillar is of secondary importance to the distance between pillars, provided that the pillar can actually pin the interface and the distance between pillars is relatively large (i.e., a dilute surface). The interface is perturbed when it touches a pillar, however, the perturbation is arrested by the surrounding pillars which impose a cutoff length on the deformation caused in the interface. Measured with respect to the undistorted TPCL profile, if $x$ is the position of the distorted TPCL in the direction of interface advance at a certain location ($y$) along the domain width, then it is related to the inter-pillar distance ($d$) and the location of the pillar center (i.e., $y_1$, see equation (\ref{eqn:superquad_ch6})) as \citep{nadkarni1992investigation}
\begin{equation}
    x = \lambda \ln \left( \frac{d}{|y-y_1|}\right),
\end{equation}
where $\lambda$ is a constant depending upon the pillar geometry, equilibrium angle ($\theta_{\rm{e}}$) and fluid/fluid interfacial tension ($\sigma_{12}$). Also, since the pillar area fraction is a function of the inter-pillar distance (that is, $\phi=1/d^2$, noting that $d$ is non-dimensionalised by $a$),
%Now, the perturbation in the interface can be represented by a logarithmic function of the distance between the pillars ($d$). 
the TPCL profile of a pinned interface can be represented by the logarithmic function of the pillar area fraction ($\phi$). This is shown in figure \ref{fig:proj_width_ch6}(b), where we have plotted the TPCL for different area fractions. As the energy dissipation depends upon the change in fluid-fluid and fluid-solid interfacial areas, which varies logarithmically with $\phi$, $\overline{D}_1$ displays a logarithmic dependence on $\phi$. Hence, under the dilute surface assumption, the total non-dimensional dissipation ($\overline{D}$) can be obtained using equations (\ref{eqn:diss_dilute}) and (\ref{eqn:dissipation_ch6_3}) as
\begin{equation}
    \overline{D} = A\phi\ln\phi + C\phi
    \label{eqn:total_diss_dilute}
\end{equation}

At higher $\phi$, however, the relative distance between pillars is less and the shape of the pillars becomes more important in determining the form of the dissipation and we expect $\overline{D}$ to deviate from the dilute form. Therefore we incorporate a term $B\phi$ into the expression, rationalised as the first Taylor series correction to the dilute form as an alternative correlation form for $\overline{D}_1$,
\begin{equation}
 \overline{D}_1 = A \ln \phi + B \phi + C.
 \label{eqn:diss_non_dilute}
\end{equation}
Here, $B$ is a constant depending upon the pillar geometry and Young's angle. Based on equation (\ref{eqn:diss_non_dilute}), a `non-dilute' form of the total non-dimensional dissipation can similarly be written as 
\begin{equation}
 \overline{D} = A\phi\ln\phi + B\phi^2 + C\phi
 \label{eqn:total_diss_non_dilute}
\end{equation}

We calculate the coefficients $A$, $B$ and $C$ by fitting equations (\ref{eqn:total_diss_dilute}) and (\ref{eqn:total_diss_non_dilute}) to the variation in total non-dimensional dissipation ($\overline{D}$) with pillar area fractions generated from our numerical results. Here, $\overline{D}$ is calculated from the interface dynamics simulations over a single pillar as $\overline{D}=\phi \overline{D}_1$. The values of different fitting coefficients and the $R^2$ values of the fit (for $\overline{D}$) are given in table \ref{tab:adv_diss_par_ch6}. Overall the dilute and non-dilute correlation forms fit the results accurately, with the non-dilute form producing slightly higher $R^2$ values.
\begin{table}
  \begin{center}
\def~{\hphantom{0}}
  \begin{tabular}{lccc}
      \makecell{Equation}  & \quad $h/a=0.35$   & \quad \quad  $h/a=1.5$  \\[3pt]
      \makecell{$\overline{D}=A\phi\ln\phi + C\phi$\\(dilute surface)}  & \makecell{$A=-1.39$\\$C=1.64$\\$R^2=0.996$} & \quad \quad \makecell{$A=-7.44$\\$C=-6.49$\\$R^2=0.987$} \\
      \makecell{$\overline{D}=A\phi\ln\phi + B \phi^2 + C\phi$\\(non-dilute surface)}  & \makecell{$A=-1.80$\\$B=1.09$\\$C=0.77$\\$R^2=0.998$} & \quad \quad \makecell{$A=-11.26$\\$B=53.16$\\$C=-20.95$\\$R^2=0.999$}
  \end{tabular}
  \caption{Fitting parameters for the total non-dimensional energy dissipation ($\overline{D}$) dependence on the pillar area fraction ($\phi$) for an advancing interface ($\theta_{\rm{e}}=72$\textdegree), as suggested in equations (\ref{eqn:total_diss_dilute}) and (\ref{eqn:total_diss_non_dilute}).}
  \label{tab:adv_diss_par_ch6}
  \end{center}
\end{table}

An observation from the peculiar nature of the $\overline{D}_1$ variation in equations (\ref{eqn:diss_dilute}) and (\ref{eqn:diss_non_dilute}), is that despite the presence of $\ln \phi$, which makes the dissipation per pillar non-vanishing even on very dilute surfaces ($\phi \to 0$), the behaviour of total dissipation ($\overline{D}$) is different, becoming zero as the pillar area fraction approaches zero. In fact, as shown by the simulation results $\overline{D}_1$ becomes very large as the pillar area fraction is reduced, as can be seen in figure \ref{fig:adv_diss_ch6}. The total non-dimensional energy dissipation ($\overline{D}$) becomes zero on very dilute surfaces due to the limit of $\phi \ln \phi$, i.e.
\begin{equation}
\begin{split}
    \lim_{\phi \to 0} \overline{D} &=  \lim_{\phi \to 0} (A\phi \ln \phi + B \phi^2 + C \phi),\\
    &=\lim_{\phi \to 0} (A\phi \ln \phi),\\
    &=\lim_{\phi \to 0} \frac{A \ln \phi}{1/ \phi} \stackrel{\text{H}}{=} (A/ \phi)(-\phi^2),\\
    &=0.
    \label{eqn:diss_limit1}
    \end{split}
\end{equation}
Based on the nature of $\overline{D}$ and $\overline{D}_1$ with respect to $\phi$, it may appear that the behaviour of an interface on a surface with a dilute number of strong defects is very different from the surface having a single strong defect. According to equation (\ref{eqn:diss_non_dilute}), if an interface moving with a small macroscopic flow velocity ($\boldsymbol{v}_{\rm{CV}}$) interacts with a single strong defect then it should dissipate an infinite amount of energy when released from the defect. However, in reality, physical surfaces have finite sizes and (usually) more than a single defect, so have finite cutoff lengths. For example for a small droplet, the cutoff length is the droplet size. If a droplet of a few millimeters in diameter spreads on a surface with a single strong defect \citep{joanny1984model} of a few micrometers in size, the area fraction would be of the order of 10$^{-6}$. The non-dimensional dissipation per pillar, according to the equation (\ref{eqn:diss_non_dilute}) and values as given in table \ref{tab:adv_diss_par_ch6} is then of the order of 10$^{-5}$, which is almost zero, as predicted by equation (\ref{eqn:total_diss_non_dilute}). Also, from our simulations we observe that the maximum macroscopic contact angle ($\theta_{\rm{s}}$) decreases as the distance between pillars increases, therefore we expect $\theta_{\rm{s}}$ to approach $\theta_{\rm{e}}$ as $\phi \to 0$.

\subsubsection{Advancing contact angle}
In this section, we develop a predictive equation for the advancing contact angle of an interface in the Wenzel wetting state on a rough surface. Writing the mechanical energy balance equation from part I
\begin{equation}
\sum_{i<j} \sigma_{ij}\frac{\overrightarrow A_{ij}-\overleftarrow A_{ij}}{\sigma_{12} A_{\rm{CV}}} -\cos\theta_{\rm{meb}}-\overline{D}  = 0.
\label{eqn:meb_equation_ch6}
\end{equation}
Here, $\overrightarrow A_{ij}$ and $\overleftarrow A_{jj}$ are $ij$ interface areas entering and leaving the control volume respectively, as it moves over the surface, $A_{\rm{CV}}$ is the projection of the area swept by the control volume on a flat surface parallel to the average solid surface and $\theta_{\rm{meb}}$ is the macroscopic angle of the interface. For the present case - that is, Wenzel wetting state over a surface with square cross-sectioned pillars arranged in a square array and the interface advancing in surface periodicity direction - the various area terms can be expanded as
\begin{equation}
	\begin{split}
		\overrightarrow A_{2S} &= A_{\rm{CV}} + nA_{\rm{CV}}(4ah), \quad \overrightarrow A_{1S} = 0, \quad \overrightarrow A_{12} = 0,  \\
		\overleftarrow A_{2S} &= 0, \quad \overleftarrow A_{1S} = A_{\rm{CV}} + nA_{\rm{CV}}(4ah), \quad \overleftarrow A_{12} = 0. 
	\end{split}
	\label{eqn:areas_ch6}
\end{equation} 
Here, $n$ is the number of pillars per unit area. Substituting the values from equation (\ref{eqn:areas_ch6}) into equation (\ref{eqn:meb_equation_ch6}) yields
\begin{equation}
		\cos \theta_{\rm{meb,a}} = (1+4 \phi (h/a)) \cos \theta_{\rm{e}} - \overline{D}.
		\label{eqn:meb_equation2_ch6}
\end{equation}
Here, $\theta_{\rm{meb,a}}$ is the advancing contact angle of the interface calculated using the mechanical energy balance equation (\ref{eqn:meb_equation_ch6}). From equations (\ref{eqn:meb_equation2_ch6}), (\ref{eqn:total_diss_dilute}) and (\ref{eqn:total_diss_non_dilute}), the expression for advancing contact angle can hence be written as
\begin{equation}
	\cos \theta_{\rm{meb,a}} = r \cos \theta_{\rm{e}} - \overline{D},
	\label{eqn:meb_equation3_ch6}
\end{equation}
where $r=(1+4\phi h/a$) is the Wenzel roughness ratio. Equation (\ref{eqn:meb_equation3_ch6}) only depends upon the pillar aspect ratio ($h/a$), contact angle on the flat surface ($\theta_{\rm{e}}$) and the surface area fraction of the pillars ($\phi$). The constants $A$, $B$ and $C$ can be predicted from the energy minimization (example table \ref{tab:adv_diss_par_ch6}), which makes equation (\ref{eqn:meb_equation3_ch6}) fully predictive in nature.

\subsection{Receding interface}

In this section, we present the results for the receding motion of an interface in the direction of surface periodicity and understand its dynamics, and the dissipation in energy during the TPCL jumps. Referring back to figure \ref{fig:physical_model_ch6}, the advancing motion of fluid-1 is accompanied by the receding motion of fluid-2 by the same amount. Therefore, we simulate the receding motion of an interface with Young's angle $\theta_{\rm{e}}$ by simulating the advancing motion of the surrounding fluid with Young's angle $180$\textdegree$-\theta_{\rm{e}}$. Here we present the simulation results for the receding motion of fluid-1 ($\theta_{\rm{e}}=72$\textdegree) by simulating the advancing of fluid-2 with $\theta_{\rm{e}}=108$\textdegree. However, in the following discussion, we present the results as if fluid-1 is receding with $\theta_{\rm{e}}=72$\textdegree.
%we refer to $\theta_{\rm{e}}$ as measured from the inside of fluid-1, that is, the receding motion of fluid-1 is presented as a receding interface with $\theta_{\rm{e}}=72$\textdegree. 

\subsubsection{Interface behaviour}
\begin{figure}
	\centering
		\includegraphics[width=0.60\textwidth]{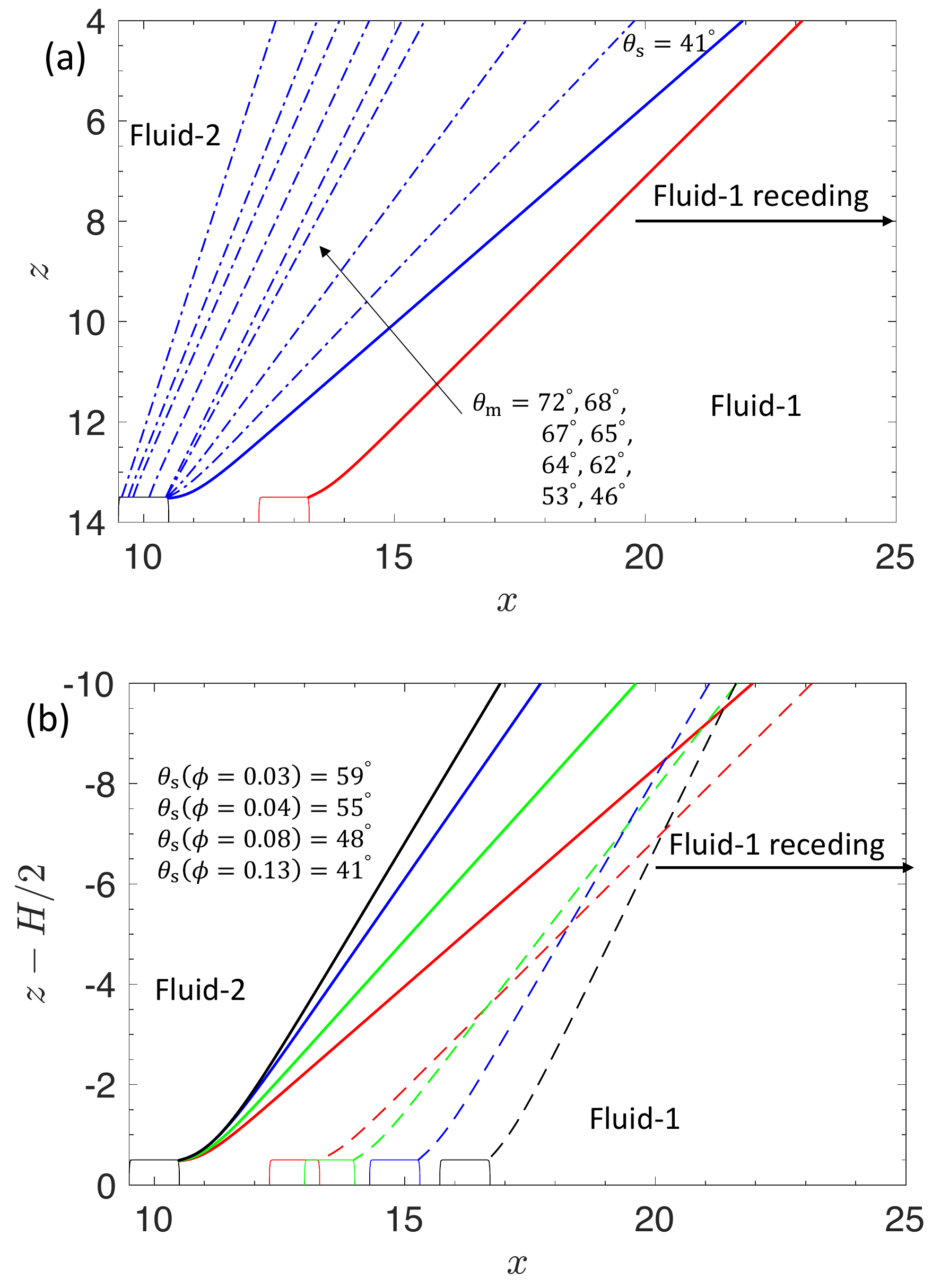}
			\caption{Equilibrium morphologies during the receding motion of an interface projected on the symmetry plane. The results are calculated for $\theta_{\rm{e}}=72$\tc and pillar aspect ratio ($h/a$) of 0.5 (all the angles are measured from fluid-1 side). (a) The interface can find an equilibrium morphology while pinned at the pillar front and slides on the pillar top when it recedes, finally reaching the first critical state at the rear face of the pillar. Equilibrium morphologies before the interface reach the critical state are shown by dashed-dotted lines. The corresponding macroscopic contact angle values for the equilibrium interface profiles are also shown ($\phi=0.13$). At $\theta_{\rm{s}}=41$\textdegree, the interface reaches the first critical state. The projection of the equilibrium interface morphologies in the first and second critical states are shown by solid blue and red lines respectively. (b) Equilibrium interface morphologies projected on the symmetry plane capturing the interfacial morphologies in the first (solid lines) and second critical states (dashed lines), for area fractions, $\phi=0.13$ (red), $\phi=0.08$ (green), $\phi=0.04$ (blue), and $\phi=0.03$ (black) respectively.}
			\label{fig:xz_projection_receding_ch6}
\end{figure}

Figure \ref{fig:xz_projection_receding_ch6}(a) shows the equilibrium morphologies of a pinned interface projected on the symmetry plane as fluid-1 recedes ($\theta_{\rm{e}}=72$\tc and $h/a=0.50$). Some of the equilibrium interface profiles before depinning are plotted as dashed-dotted lines. The blue solid line represents the interface's first critical state, after which any further advancement will result in the TPCL jump. The equilibrium interface profile after depinning (i.e. the second critical state) is shown by the solid red line. In figure \ref{fig:xz_projection_receding_ch6}(b) we show the equilibrium interface morphologies projected the symmetry plane for different area fractions ($\phi=0.13, 0.08, 0.04$, $0.03 $ and $\theta_{\rm{e}}=72$\textdegree, $h/a=0.50$). The first and second critical states are shown by the solid and dashed lines respectively. 
%%%%%Plot: Interface projection on wall receding interface

\begin{figure}
\centering
	\includegraphics[width=\textwidth]{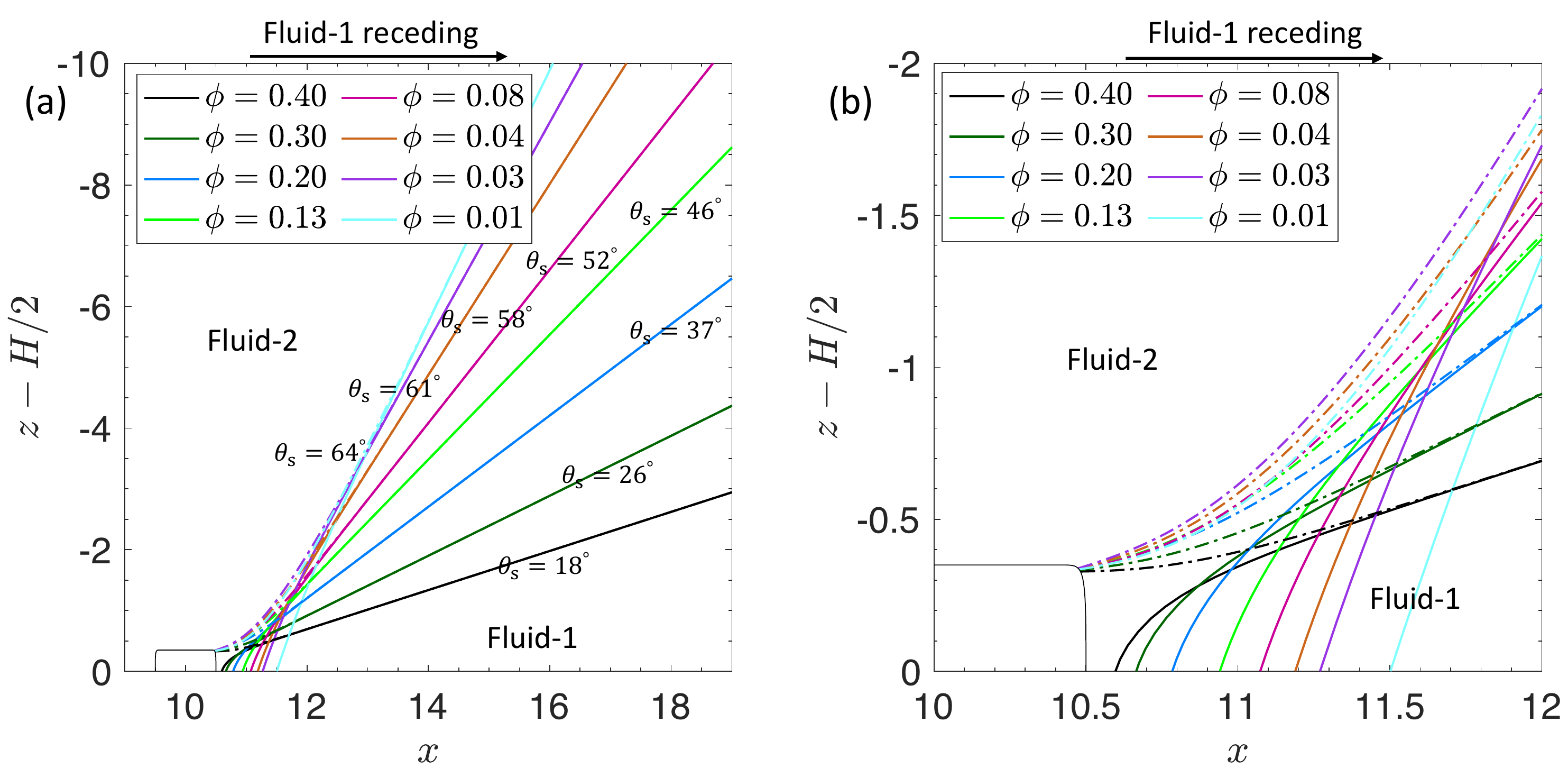}
	\caption{Equilibrium interface morphologies just before depinning (first critical state) as projected on the symmetry plane (shown by dashed-dotted lines) and the domain wall ($y=W/2$, shown by solid lines) for a receding interface on a surface with $\theta_{\rm{e}}=72$\textdegree, $h/a=0.35$ and area fractions $\phi=0.01, 0.03, 0.04, 0.08, 0.13, 0.20, 0.30, 0.40$ respectively. (b) Zoomed view of the interface morphology near the TPCL.}
	\label{fig:front_back_rec_ch6}
\end{figure}

A comparison between the equilibrium interface morphologies during the advancing (figure \ref{fig:theory_thm_ch6}(c)) and receding (figure \ref{fig:xz_projection_receding_ch6}(a)) motion of an interface with $\theta_{\rm{e}}=72$\tc reveals that during the receding motion, the interface can exist in an equilibrium state such that a portion of the TPCL lies on the pillar tops. The TPCL slides on the pillar top before it finally gets pinned at the rear face of the pillar where the interface exists in first critical state before the TPCL jumps. However, the first critical interface morphologies during advancing as well as the receding motion exist at the rear face of the interface. Figure \ref{fig:front_back_rec_ch6} shows the first critical state morphologies of the interface for a range of area fractions, projected on the symmetry plane as well as the domain wall ($y=W/2)$ for the receding motion of fluid-1 on a surface with $\theta_{\rm{e}}=72$\tc and pillar aspect ratio ($h/a$) of 0.35. Similar to the advancing interface case (figure \ref{fig:front_back_ch6}), in the case of a receding interface as well, we observe the interface curvature to increase with the pillar area fraction.

\subsubsection{Energy dissipation during receding motion of the interface}
\begin{figure}
    \centering
    \includegraphics[width=0.65\textwidth]{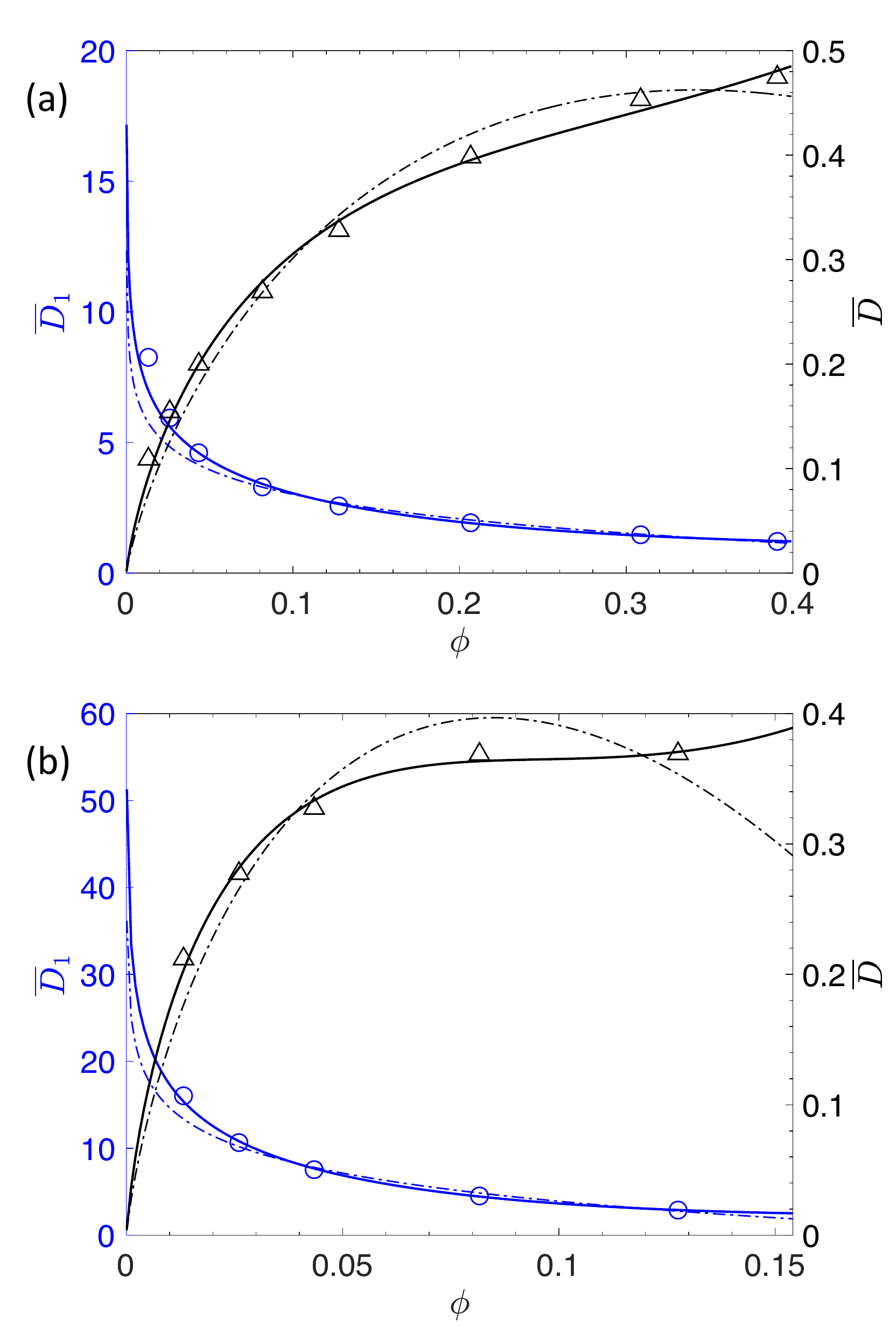}
    \caption{Variation in the total non-dimensional energy dissipation ($\overline{D}$, shown by triangles) and non-dimensional energy dissipation per pillar ($\overline{D}_1$, shown by circles) with the pillar area fraction ($\phi$) for an interface receding in the direction of surface periodicity. Results have been simulated for Young's angle $72$\tc and pillar aspect ratio of (a) 0.35 and (b) 1.5 respectively. Fitting equations of dilute (equation (\ref{eqn:total_diss_dilute})) and non-dilute (equation (\ref{eqn:total_diss_non_dilute})) form for $\overline{D}$ are shown by dashed-dotted and solid black lines respectively. Fitting equations for $\overline{D}_1$ are shown by dashed-dotted (dilute form, equation (\ref{eqn:diss_dilute})) and solid (non-dilute, equation (\ref{eqn:diss_non_dilute})) blue lines respectively.}
    \label{fig:rec_diss_ch6}
\end{figure}

Figure \ref{fig:rec_diss_ch6} shows the variation in total non-dimensional energy dissipation ($\overline{D}$) and non-dimensional energy dissipation per pillar ($\overline{D}_1$) with the pillar area fraction ($\phi$) for a receding interface on a surface with pillars of aspect ratio ($h/a$) 0.35 and 1.5 and Young's angle ($\theta_{\rm{e}}=72$\textdegree). 
The interface is moving in the direction of surface periodicity. Qualitatively, the variation in $\overline{D}$ and $\overline{D}_1$ with $\phi$ for a receding interface is similar to the variation in both the dissipation ($\overline{D}_1$ and $\overline{D}$) for an advancing interface (figure \ref{fig:adv_diss_ch6}), which is expected as the receding motion of an interface can also be identified as the advancing motion of the surrounding fluid. Therefore, we use the fitting equations of the forms as given in equations (\ref{eqn:diss_dilute}), (\ref{eqn:total_diss_dilute}) for $\overline{D}_1$ and $\overline{D}$ under the dilute surface assumption and (\ref{eqn:diss_non_dilute}), (\ref{eqn:total_diss_non_dilute}) for $\overline{D}_1$ and $\overline{D}$ under non-dilute surface assumption respectively. In table \ref{tab:rec_diss_par_ch6} we show the fitting coefficients and $R^2$ value for the total non-dimensional dissipation according to equations (\ref{eqn:total_diss_dilute}) and (\ref{eqn:total_diss_non_dilute}) for a receding interface with $\theta_{\rm{e}}=72$\tc and pillar aspect ratios ($h/a$) 0.35 and 1.5 respectively. As per the advancing cases, the two forms for $\overline{D}$ and $\overline{D}_1$, are able to accurately fit the numerically generated results, with the non-dilute form again producing slightly higher $R^2$ values.
\begin{table}
  \begin{center}
\def~{\hphantom{0}}
  \begin{tabular}{lccc}
      \makecell{Equation}  & \quad $h/a=0.35$   & \quad \quad  $h/a=1.5$  \\[3pt]
      \makecell{$\overline{D}=A\phi\ln\phi + C\phi$\\(dilute surface)}  & \makecell{$A=-1.35$\\$C=-0.10$\\$R^2=0.975$} & \quad \quad \makecell{$A=-4.67$\\$C=-6.85$\\$R^2=0.859$} \\
      \makecell{$\overline{D}=A\phi\ln\phi + B \phi^2 + C\phi$\\(non-dilute surface)}  & \makecell{$A=-2.09$\\$B=3.54$\\$C=-2.12$\\$R^2=0.995$} & \quad \quad \makecell{$A=-7.46$\\$B=38.78$\\$C=-17.40$\\$R^2=0.991$}
  \end{tabular}
  \caption{Fitting parameters for the total non-dimensional energy dissipation ($\overline{D}$) dependence on the pillar area fraction ($\phi$) for a receding interface ($\theta_{\rm{e}}=72$\textdegree), as suggested in equations (\ref{eqn:total_diss_dilute}) and (\ref{eqn:total_diss_non_dilute}).}
  \label{tab:rec_diss_par_ch6}
  \end{center}
\end{table}

\begin{figure}
    \centering
    \includegraphics[width=0.65\textwidth]{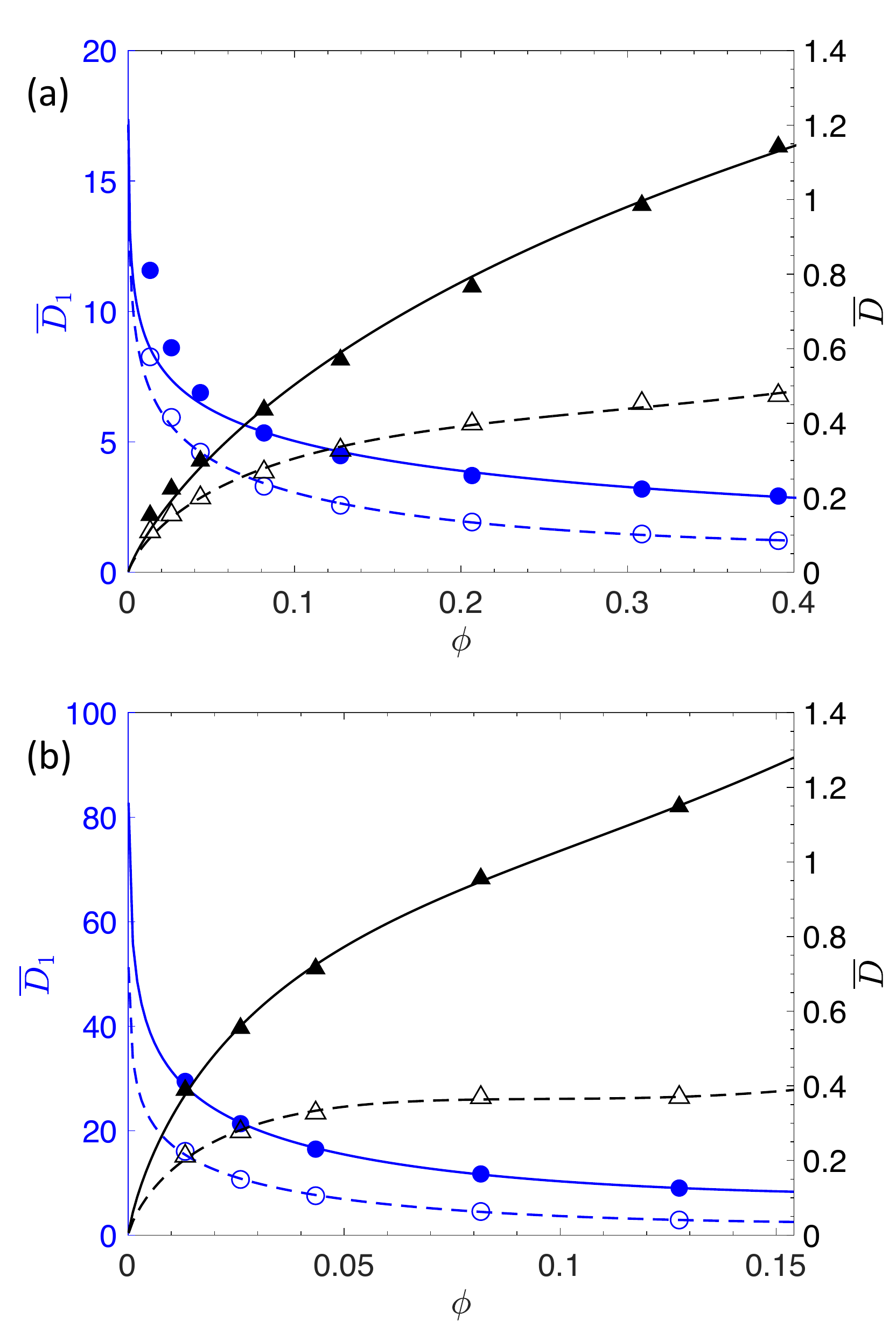}
    \caption{Variation in the non-dimensional energy dissipation per pillar ($\overline{D}_1$, shown by circles) and total non-dimensional energy dissipation ($\overline{D}$, shown by triangles) with the pillar area fraction ($\phi$) for an interface advancing (shown by filled symbols) and receding (shown by empty symbols) in the direction of surface periodicity on a surface with $\theta_{\rm{e}}=72$\tc and pillar aspect ratios ($h/a$) 0.35 (a) and 1.5 (b) respectively. The black lines are for equation (\ref{eqn:total_diss_non_dilute}) and blue lines for equation (\ref{eqn:diss_non_dilute}) with solid and dashed representing the advancing and receding motion of the interface.}
    \label{fig:diss_compare_ch6}
\end{figure}

From figures \ref{fig:adv_diss_ch6} and \ref{fig:rec_diss_ch6}, we observe that the non-dimensional dissipation per pillar ($\overline{D}_1$) decreases while the total non-dimensional dissipation ($\overline{D}$) increases with the pillar area fraction ($\phi$) for both the advancing and receding interfaces. Also, both of $\overline{D}_1$ and $\overline{D}$ increase with the pillar aspect ratio ($h/a$) for advancing as well as receding interfaces. However, the magnitude of non-dimensional dissipation per pillar and total non-dimensional dissipation is different for the advancing and receding motion, even at the same pillar area fraction. This is shown in figure \ref{fig:diss_compare_ch6} where we plot the variation in $\overline{D}_1$ and $\overline{D}$ with $\phi$ for advancing and receding motion of the interface on a surface with $\theta_{\rm{e}}=72$\tc and pillar aspect ratios 0.35 and 1.5 respectively. We observe that the non-dimensional dissipation per pillar and total non-dimensional dissipation is higher for the advancing interface as compared to its receding motion on the same surface. The energy dissipation is, therefore, asymmetric in nature which may result in different rates at which the advancing contact angle increases or the receding contact angle decreases with an increase in the pillar area fraction.

\subsubsection{Receding contact angle}
Since the receding motion of fluid-1 is equivalent to the advancing motion of fluid-2, the equation for the receding contact angle ($\theta_{\rm{meb,r}}$) can be obtained from the equation for the advancing contact angle (equation (\ref{eqn:meb_equation3_ch6})) by replacing $\theta_{\rm{meb,a}}$ with (180\textdegree$- \theta_{\rm{meb,r}}$) and $\theta_{\rm{e}}$ with (180\textdegree$-\theta_{\rm{e}}$) in equation (\ref{eqn:meb_equation3_ch6}), i.e.
\begin{equation}
	\begin{split}
	\cos \theta_{\rm{meb,r}}&= r \cos \theta_{\rm{e}} + \overline{D},
	\label{eqn:meb_rec_equation1_ch6}
\end{split}
\end{equation}
where $r=(1+4\phi h/a$) is the Wenzel roughness ratio. The total non-dimensional energy dissipation ($\overline{D}$) is as per the form presented in equations (\ref{eqn:total_diss_dilute}) and (\ref{eqn:total_diss_non_dilute}) with the coefficients as given in table \ref{tab:rec_diss_par_ch6}.

\subsection{Relationship between $\theta_{\rm{s}}$, $\theta_{\rm{s}^{'}}$ and $\theta_{\rm{meb}}$}
\label{sec:meb_vs_SE}
\begin{figure}
	\centering
		\includegraphics[width=0.65\textwidth]{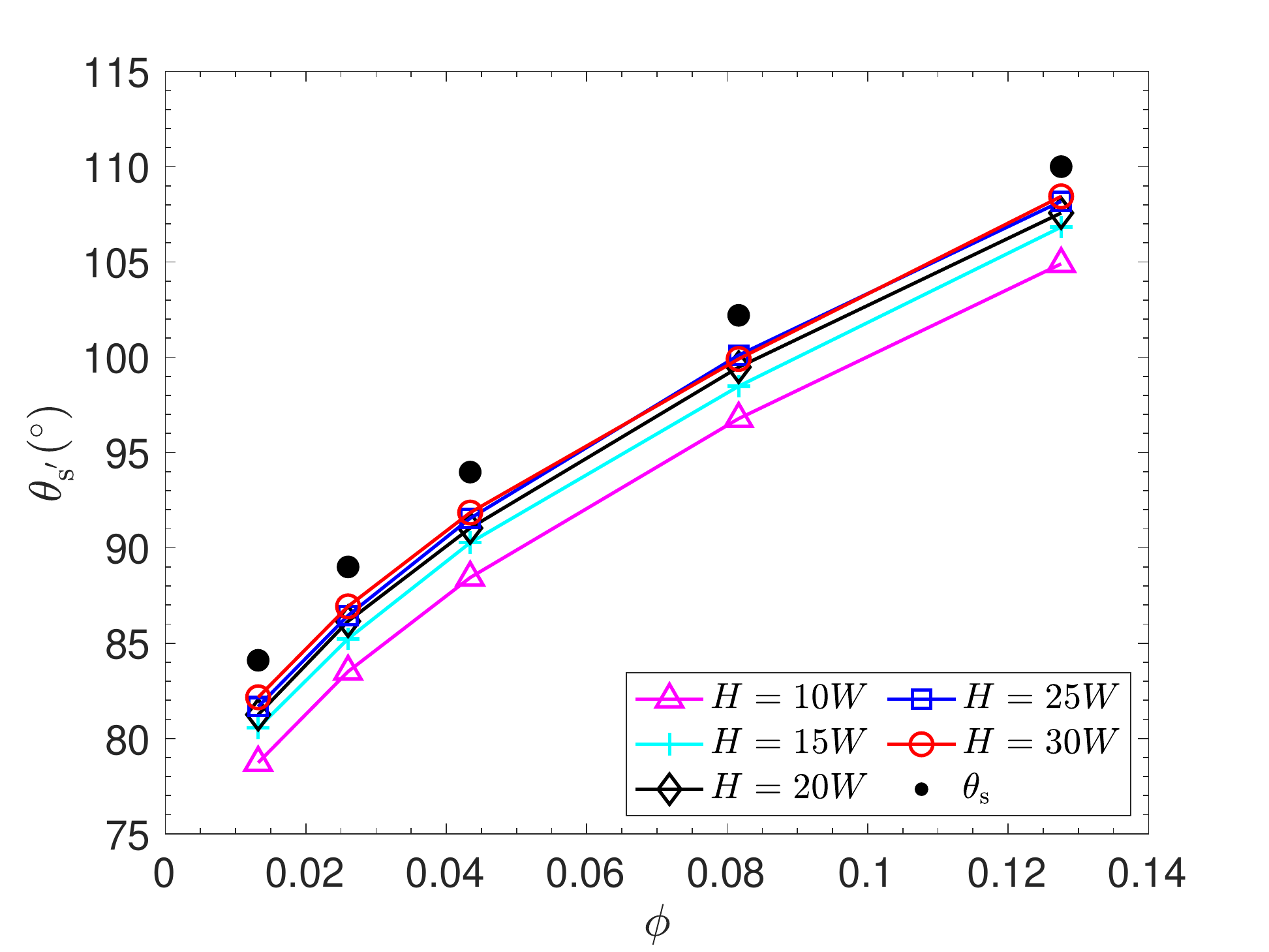}
		\caption{Variation in the macroscopic contact angle corresponding to the second critical state ($\theta_{\rm{s^{'}}}$) with the pillar area fraction ($\phi$) for $\theta_{\rm{e}}=72$\tc and $h/a=0.5$. The (first) critical contact angles ($\theta_{\rm{s}}$) are also shown as black circles.}
		\label{fig:th_s_height_ch6}
\end{figure}

Based on the results of our interface dynamics simulations, we propose three main parameters for characterizing a typical interface advancement on a structured surface in the direction of surface periodicity: the (first) critical angle ($\theta_{\rm{s}}$), the macroscopic contact angle after the TPCL depinning i.e. the second critical angle ($\theta_{\rm{s}^{'}}$) and the dissipation in energy ($\overline{D}_1$ or $\overline{D}$). From figure \ref{fig:th_s_height_ch6} we observed that the critical angle ($\theta_{\rm{s}}$) is independent of domain height provided that the domain is high enough so as to ensure an approximately planar interface near the top. However, $\theta_{\rm{s}^{'}}$ depends upon the domain height, which is shown in figure \ref{fig:th_s_height_ch6} where we plot the variation in $\theta_{\rm{s}^{'}}$ with $\phi$ for different domain heights ($H$) expressed as a function of the domain width ($W$).
We observe that $\theta_{\rm{s}^{'}}$ increases as the domain height is increased and approaches the (first) critical angle ($\theta_{\rm{s}}$, shown as black circles) as the domain height becomes large. Therefore, we expect that as the domain height approaches infinity, the interface should advance with a constant macroscopic angle ($\theta_{\rm{s}}$). Also, since the mechanical energy balance framework discussed in part I relates the dissipation in energy during interface advancement to a single macroscopic contact angle, we expect the advancing contact angle based on the balance of mechanical energy (equation (\ref{eqn:meb_equation3_ch6})) to approach $\theta_{\rm{s}}$ under the assumption of separation of length scales i.e. large domain heights.

\begin{figure}
	\centering
		\includegraphics[width=0.65\textwidth]{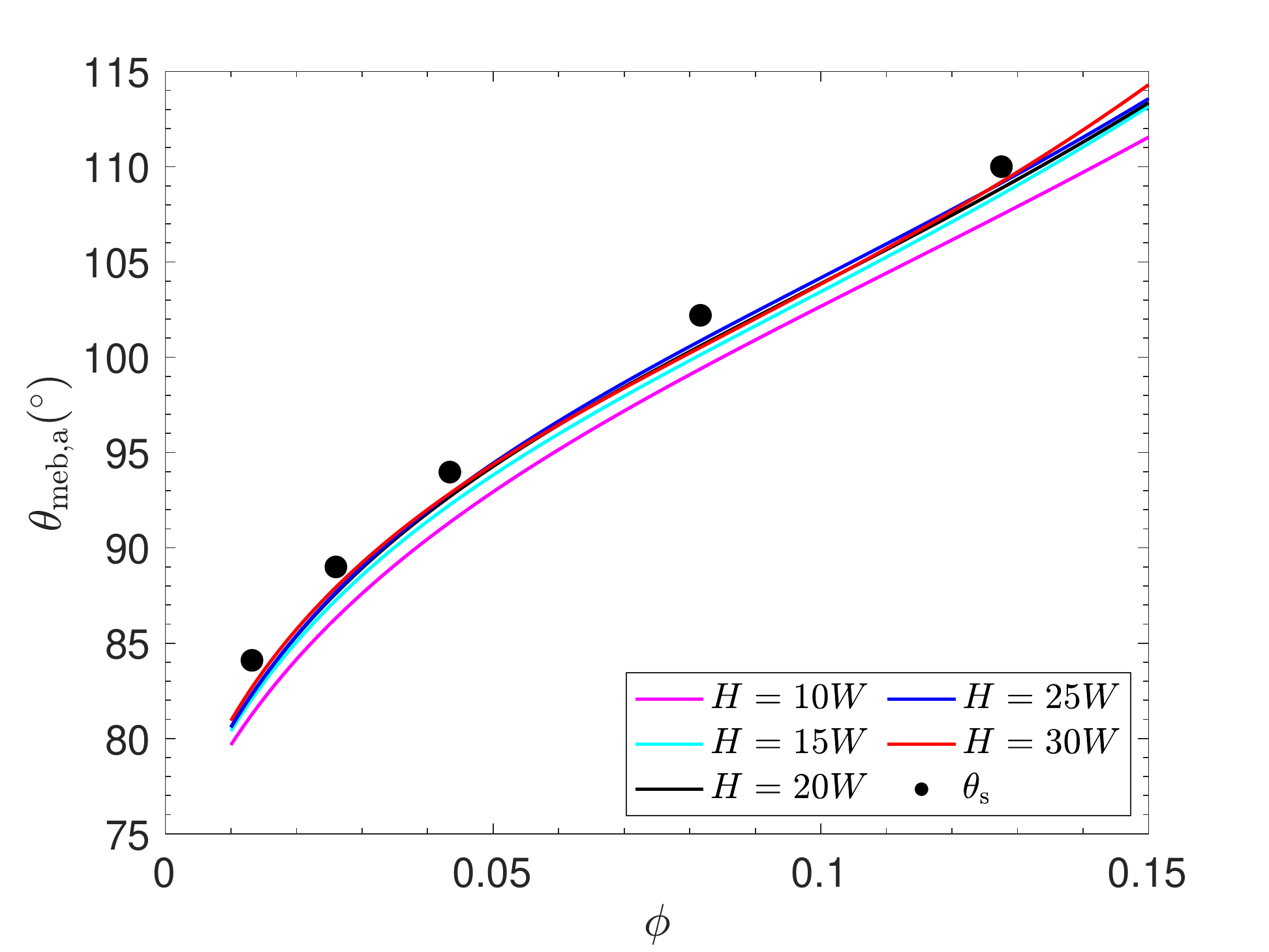}
		\caption{Variation in the advancing contact angle ($\theta_{\rm{meb,a}}$) calculated using equation (\ref{eqn:meb_equation3_ch6}) with total non-dimensional energy dissipation calculated from energy minimization simulations under a non-dilute surface assumption (equation (\ref{eqn:total_diss_non_dilute})) with pillar area fraction ($\phi$) for different domain heights on a surface with $\theta_{\rm{e}}=72$\tc and $h/a=0.5$. The (first) critical contact angles ($\theta_{\rm{s}}$) are shown as black circles.}
		\label{fig:th_meb_ch_ht_ch6}
\end{figure}

In figure \ref{fig:th_meb_ch_ht_ch6} we plot the variation in $\theta_{\rm{meb,a}}$ with the pillar area fraction ($\phi$) for different domain heights. For calculating the advancing angle, we have used the non-dilute form of total non-dimensional dissipation (equation (\ref{eqn:total_diss_non_dilute})) in equation (\ref{eqn:meb_equation3_ch6}) as it represents the numerical data most accurately. We observe that $\theta_{\rm{meb,a}}$ increases with the domain height and approaches $\theta_{\rm{s}}$ (shown as black circles) at higher domain heights.
The dependency of $\theta_{\rm{meb,a}}$ upon domain height originates from the dependency of the energy dissipation upon the domain height. We observe that the energy dissipation varies with the domain height and converges to a single value as the domain height is increased (refer to Appendix \ref{sec:channel_height:appendix} and figure \ref{fig:diss_ch_ht_ch6}). From figures \ref{fig:th_s_height_ch6} and \ref{fig:th_meb_ch_ht_ch6} we can therefore conclude that as $H \to \infty$, $\theta_{\rm{meb,a}} \to \theta_{\rm{s}}$ and also $\theta_{\rm{s}^{'}} \to \theta_{\rm{s}}$. Since the receding motion of a fluid is equivalent to the advancing motion of the surrounding fluid, therefore, the above discussion also holds for the receding mode. Hence, via observation, we have established a relationship between interface dynamics and energy conservation/dissipation for this particular periodic surface topology.

\subsection{Contact angle hysteresis}
\begin{figure}
    \centering
    \includegraphics[width=0.60\textwidth]{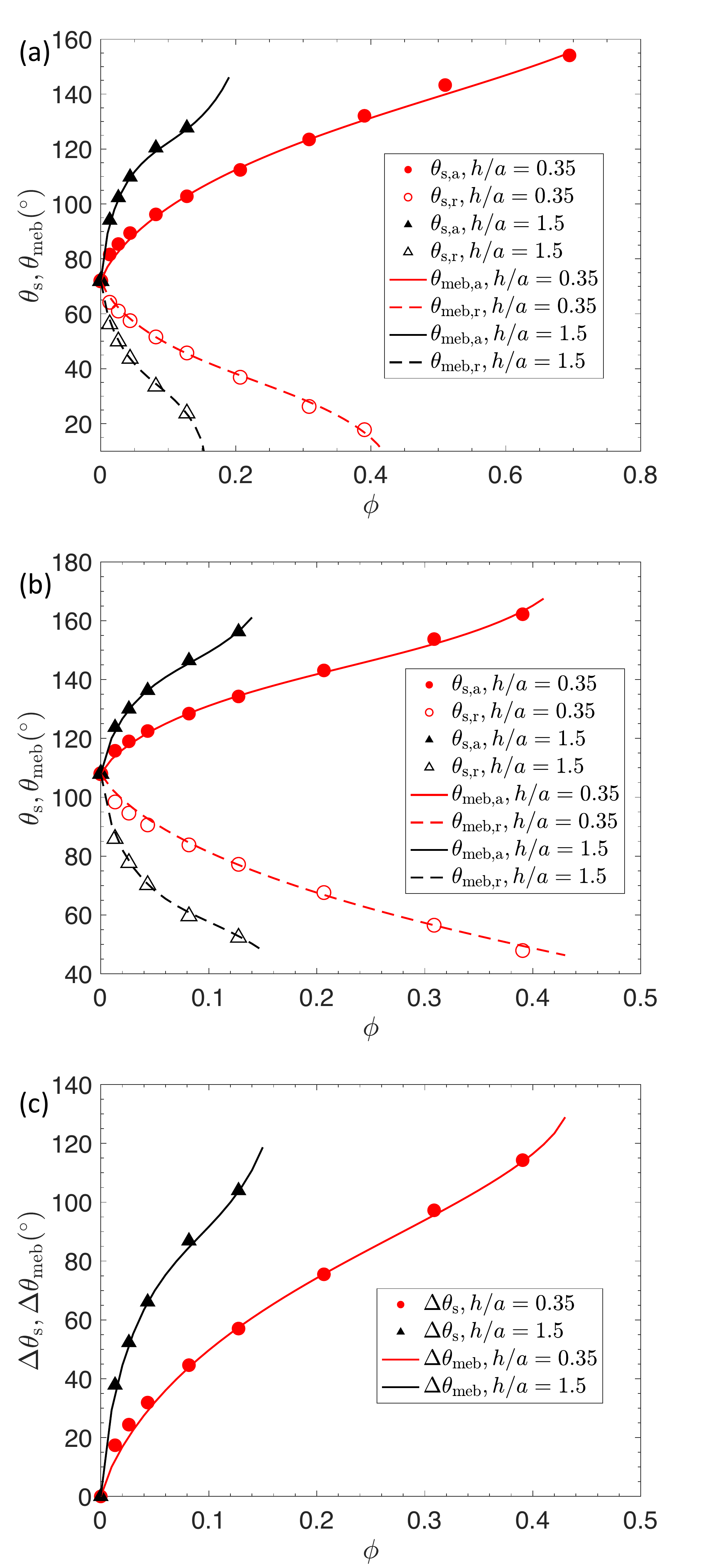}
    \caption{Variation in the advancing and receding contact angles obtained using energy minimization simulations ($\theta_{\rm{s,a}}$, $\theta_{\rm{s,r}}$) and mechanical energy balance equations (\ref{eqn:meb_equation3_ch6}) and (\ref{eqn:meb_rec_equation1_ch6})  ($\theta_{\rm{meb,a}}$, $\theta_{\rm{meb,r}}$) with the pillar area fraction ($\phi$) for pillar aspect ratios ($h/a$) 0.35 and 1.5 respectively and Young's angle, $\theta_{\rm{e}}=72$\tc (a) and 108\tc (b) respectively. (c) Shows the variation in contact angle hysteresis in degrees, obtained using energy minimization simulations (i.e. $\Delta \theta_{\rm{s}}=\theta_{\rm{s,a}}-\theta_{\rm{s,r}}$) and mechanical energy balance equations (\ref{eqn:meb_equation3_ch6}) and (\ref{eqn:meb_rec_equation1_ch6}) (i.e., $\Delta \theta_{\rm{meb}}=\theta_{\rm{meb,a}}-\theta_{\rm{meb,r}}$) with the pillar area fraction ($\phi$) for pillar aspect ratios $h/a=0.35$ and 1.5 respectively and $\theta_{\rm{e}}=72$\textdegree.}
    \label{fig:hysteresis_ch6}
\end{figure}

In figure \ref{fig:hysteresis_ch6}(a) we plot the (first) critical advancing ($\theta_{\rm{s,a}}$) and receding ($\theta_{\rm{s,r}}$) contact angles from the energy minimisation simulations with pillar area fraction ($\phi$) for $\theta_{\rm{e}}=72$\tc and pillar aspect ratios ($h/a$) of 0.35 and 1.5 respectively, along with the mechanical energy balance derived values ($\theta_{\rm{meb,a}}$, $\theta_{\rm{meb,r}}$) determined via equations (\ref{eqn:meb_equation3_ch6}) and (\ref{eqn:meb_rec_equation1_ch6}). The mechanical energy balance values use the non-dilute form of total non-dimensional energy dissipation for greater accuracy. Under these conditions, the critical and mechanical energy balance derived angles are very similar across all tests (refer to figure \ref{fig:th_meb_ch_ht_ch6}). We observe that the advancing contact angle increases with the pillar area fraction for both pillar aspect ratios, with higher advancing contact angles being achieved at higher aspect ratios for any value of $\phi$. Similarly, the receding contact angle decreases with increases in $\phi$, with lower value angles being found at higher pillar aspect ratios. We also observe that the increase in $\theta_{\rm{meb,a}}$ with $\phi$ is greater than the decrease of $\theta_{\rm{meb,r}}$ with $\phi$. This asymmetry in the contact angles is due to the asymmetry in the energy dissipation between the advancing and receding modes.
In figure \ref{fig:hysteresis_ch6}(b) we plot $\theta_{\rm{s,a}}$, $\theta_{\rm{s,r}}$, $\theta_{\rm{meb,a}}$ and $\theta_{\rm{meb,r}}$ with $\phi$, for $\theta_{\rm{e}}=108$\tc and pillar aspect ratios ($h/a$) of 0.35 and 1.5 respectively. Qualitatively, the contact angle variation with $\phi$ for $\theta_{\rm{e}}=72$\tc and $\theta_{\rm{e}}=108$\tc are similar, but for $\theta_{\rm{e}}=108$\tc we observe that the decrease in $\theta_{\rm{meb,r}}$ with $\phi$ is greater than the increase in $\theta_{\rm{meb,a}}$ with $\phi$. Since the receding motion of fluid-1 is equivalent to the advancing motion of fluid-2, therefore, the receding angle variations for $\theta_{\rm{e}}=72$\tc are equivalent to the variations in the advancing angle for $\theta_{\rm{e}}=108$\tc. Similarly, the advancing angle variation for $\theta_{e}=72$\tc is equivalent to the receding angle variations for $\theta_{\rm{e}}=108$\tc. 

In figure \ref{fig:hysteresis_ch6}(c) we plot the variation in contact angle hysteresis ($\Delta \theta_{\rm{s}}=\theta_{\rm{s,a}}-\theta_{\rm{s,r}}$ and $\Delta \theta_{\rm{meb}}=\theta_{\rm{meb,a}}-\theta_{\rm{meb,r}}$) with the pillar area fraction ($\phi$) for pillar aspect ratios ($h/a$) of 0.35 and 1.5 respectively and $\theta_{\rm{e}}=72$\textdegree. The contact angle hysteresis based on the mechanical energy balance is calculated using equations (\ref{eqn:meb_equation3_ch6}) and (\ref{eqn:meb_rec_equation1_ch6}) with the non-dilute form of total non-dimensional dissipation ($\overline{D}$). As expected, the contact angle hysteresis increases with both the pillar area fraction ($\phi$) and the pillar aspect ratio ($h/a$). However, as the area fraction is reduced, the hysteresis also decreases and eventually approaches zero as $\phi \to 0$. This is due to zero total non-dimensional energy dissipation, as the surface becomes very dilute (equation (\ref{eqn:diss_limit1})). In reality, there may be non-zero hysteresis even when $\phi \to 0$ due to the presence of roughness at smaller length scales, or other causes of inherent hysteresis.

\subsection{Comparison with experiments}
In this section we compare our numerical model with the experimental data of \cite{forsberg2010contact}. \Citeauthor{forsberg2010contact} measured advancing and receding contact angles on surfaces decorated with microscopic pillars of square cross-section with a fixed width of 20 $\mu \rm{m}$ and heights 7 $ \mu \rm{m}$ and 30 $ \mu \rm{m}$ respectively. The advancing and receding contact angles measured on a flat surface were $72$\tc and $59$\tc respectively. Before we proceed further, it is important to address the presence of inherent hysteresis on surfaces. In developing our model (part I) we assumed the surface to be perfectly smooth at length scales below $h_{\rm{rough}}$, chemically homogeneous and with no irreversibility in the creation or destruction of surfaces on a molecular scale. The presence of any one of these effects may result in a non-zero inherent hysteresis in the system. For example, all surfaces contain roughness at different length scales irrespective of the manufacturing methods used \citep{quere2008wetting}, which can result in a non-zero inherent hysteresis. 
%%%Figure: D vs phi for receding Forsberg case
 \begin{figure}
 	\centering
 	\includegraphics[width=0.65\textwidth]{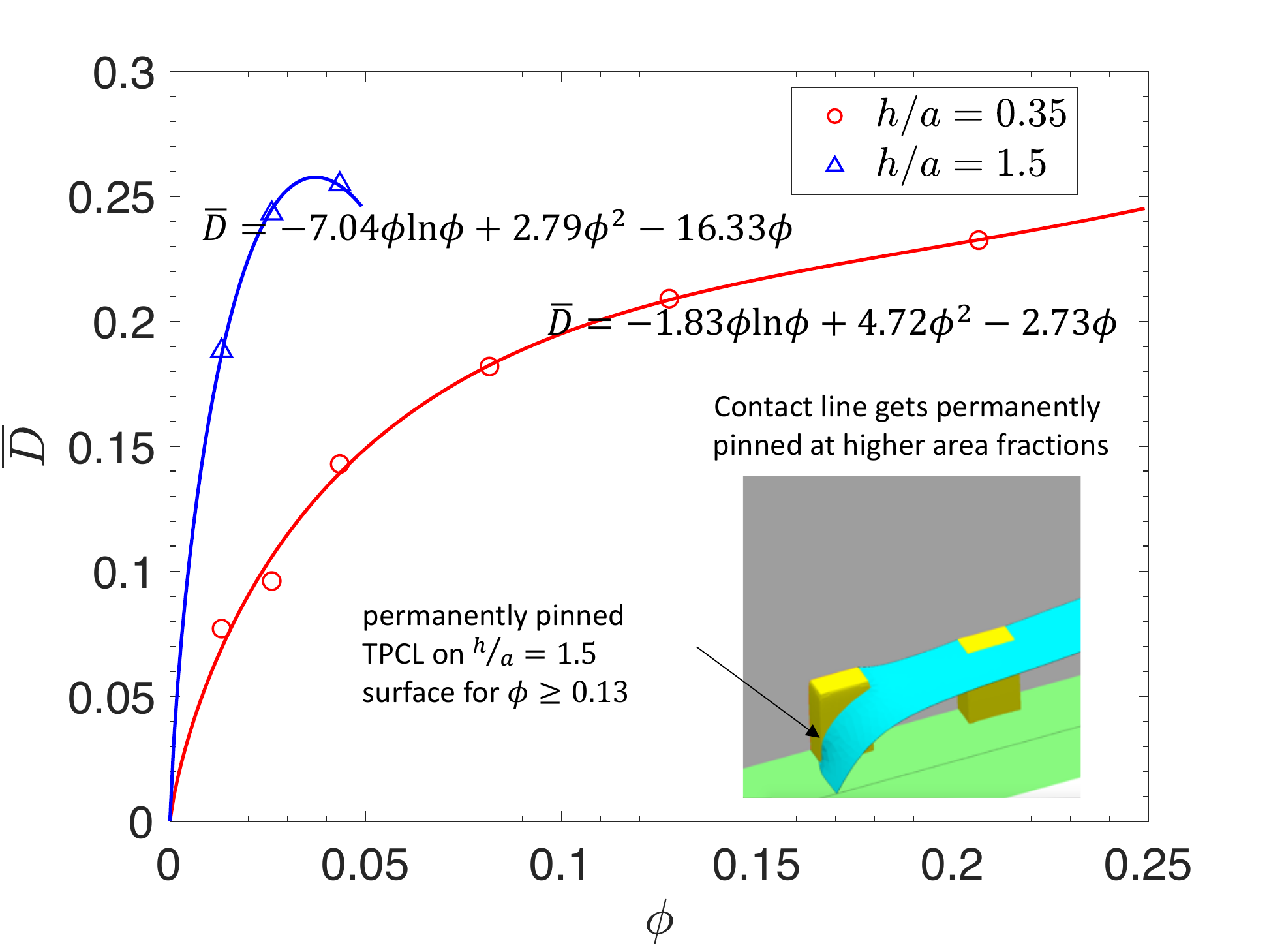}
 	\caption{Total non-dimensional dissipation ($\overline{D}$) for receding interface with $\theta_{\rm{R}}=59$\tc and pillar aspect ratios 0.35 and 1.5 respectively. Inset shows that at higher area fractions, we may observe a permanently pinned TPCL.}
 	\label{fig:rec59_diss_ch6}
 \end{figure}
In order to accommodate the inherent hysteresis on the surface, instead of $\theta_{\rm{e}}$ we use $\theta_{\rm{A}}$ and $\theta_{\rm{R}}$ for advancing and receding motion of the interface in our analysis \citep{he2004contact}, where $\theta_{\rm{A}}$ and $\theta_{\rm{R}}$ are the advancing and receding contact angles measured on the flat surface. Substituting $\theta_{\rm{e}}$ with $\theta_{\rm{A}}$ and $\theta_{\rm{R}}$ in equations (\ref{eqn:meb_equation3_ch6}) and (\ref{eqn:meb_rec_equation1_ch6}) respectively, gives
%%%%Equation: Modified advancing
\begin{equation}
\cos \theta_{\rm{meb,a}} = r \cos \theta_{\rm{A}} - \overline{D},
\label{eqn:advancing_eqn_general_ch6}
\end{equation}
%%%%Equation: Modified receding
\begin{equation}
\cos \theta_{\rm{meb,r}} = r \cos \theta_{\rm{R}} + \overline{D}.
\label{eqn:receding_eqn_general_ch6}
\end{equation}
The total non-dimensional energy dissipation for an advancing interface with $\theta_{\rm{A}}=72$\tc and pillar aspect ratios $0.35$ and $1.5$ is shown in figure \ref{fig:adv_diss_ch6} and the values of coefficients in the dissipation equation (\ref{eqn:total_diss_non_dilute}) are given in table \ref{tab:adv_diss_par_ch6}. In figure \ref{fig:rec59_diss_ch6} we show the variation of total non-dimensional dissipation ($\overline{D}$) with $\phi$ for the receding interface with $\theta_{\rm{R}}=59$\tc and pillar aspect ratios $0.35$ and $1.5$ respectively. We use equations (\ref{eqn:advancing_eqn_general_ch6}) and (\ref{eqn:receding_eqn_general_ch6}) and the dissipation parameters as tabulated in table \ref{tab:diss_par_forsberg_ch6} to calculate the advancing and receding contact angles. 
\begin{table}
  \begin{center}
\def~{\hphantom{0}}
  \begin{tabular}{lccc}
      \makecell{System}  & \quad $h/a=0.35$   & \quad \quad  $h/a=1.5$  \\[3pt]
      \makecell{Advancing interface\\$\theta_{\rm{A}}=72$\textdegree} & \makecell{$A=-1.80$\\$B=1.09$\\$C=0.77$} & \quad \quad \makecell{$A=-11.26$\\$B=53.16$\\$C=-20.95$}\\
      \makecell{Receding interface\\$\theta_{\rm{R}}=59$\textdegree} & \makecell{$A=-1.83$\\$B=4.72$\\$C=-2.73$} & \quad \quad \makecell{$A=-7.04$\\$B=2.79$\\$C=-16.33$}      
  \end{tabular}
  \caption{Fitting parameters for the total non-dimensional energy dissipation ($\overline{D}$) dependence on the pillar area fraction ($\phi$) for advancing and receding interface on a surface with $\theta_{\rm{A}}=72$\textdegree, $\theta_{\rm{R}}=59$\tc and pillar aspect ratio of 0.35 and 1.5 respectively. The form of the fitting equation is as suggested in equation (\ref{eqn:total_diss_non_dilute}).}
  \label{tab:diss_par_forsberg_ch6}
  \end{center}
\end{table}
 \begin{figure}
     \centering
     \includegraphics[width=\textwidth]{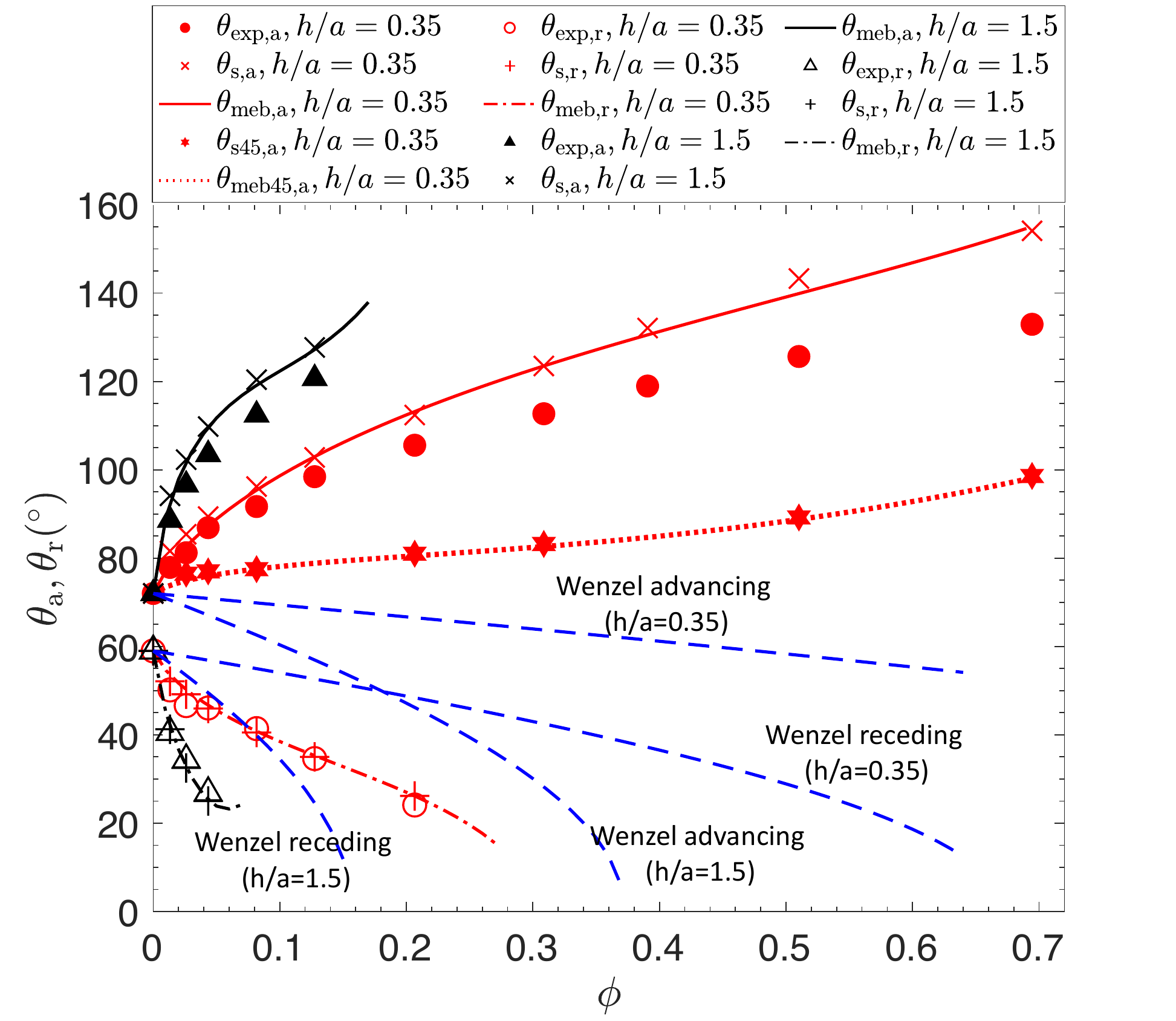}
     \caption{Variation in the advancing ($\theta_{\rm{a}}$) and receding ($\theta_{\rm{r}}$) contact angles (in degrees) with the pillar area fraction ($\phi$) for Wenzel wetting state. The experimental results of \cite{forsberg2010contact} ($\theta_{\rm{exp, a}}$, $\theta_{\rm{exp, r}}$), numerical simulations ($\theta_{\rm{s,a}}$, $\theta_{\rm{s,r}}$) and the mechanical energy balance equation (\ref{eqn:advancing_eqn_general_ch6}), (\ref{eqn:receding_eqn_general_ch6}) ($\theta_{\rm{meb,a}}$, $\theta_{\rm{meb,r}}$) are shown for a surface with $\theta_{\rm{A}}=72$\tc, $\theta_{\rm{R}}=59$\tc and pillar aspect ratio of 0.35 and 1.5 respectively. Wenzel's equation (\ref{eqn:wenzel_ch6}) for the advancing and receding interface is also plotted for comparison. The advancing contact angle based on the numerical simulations ($\theta_{\rm{s45,a}}$) and the balance of mechanical energy using equations (\ref{eqn:advancing_eqn_general_ch6}) ($\theta_{\rm{meb45,a}}$) for an interface moving at an angle of 45\tc to the surface periodicity direction ($\theta_{\rm{A}}=72$\textdegree, $h/a=0.35$) is also shown.}
     \label{fig:forsberg_ch6}
 \end{figure}

Figure \ref{fig:forsberg_ch6} shows experimental data from \cite{forsberg2010contact} along with our simulation results from equations (\ref{eqn:advancing_eqn_general_ch6}) and (\ref{eqn:receding_eqn_general_ch6}). The (first) critical contact angles ($\theta_{\rm{s}}$) obtained from the simulations for both the advancing and receding interface are also shown. Very good agreement between the simulation results and experimental data is observed at low pillar area fractions. At higher area fractions, the proposed equations tend to overpredict the advancing contact angles slightly. Receding contact angle values are in good agreement with our model for both the aspect ratios ($h/a=0.35$ and 1.5) for the entire range of experimental data, which is up to $\phi=0.21$ for $h/a=0.35$ and $\phi=0.04$ for $h/a=1.5$. It should be noted that at higher area fractions the receding TPCL can get permanently pinned at the start on the first pillar, causing the receding contact angle to approach $0$\tc and a film of liquid to be left within the pillars (see figure \ref{fig:rec59_diss_ch6} inset). The onset of an equilibrium morphology with a permanently pinned TPCL depends upon the pillar geometry and the receding contact angle on a flat surface ($\theta_{\rm{R}}$). For example on a surface with $h/a=1.5$ and $\theta_{\rm{R}}=59$\tc we observe that for $\phi \geq 0.13$ the TPCL gets permanently pinned. The morphological transitions in such cases have already been discussed in the context of an advancing interface (see figure \ref{fig:wetting_transition}), and even though the present numerical method is capable of dealing with these cases study of this transition is left to future work. Also, at a certain low Young's angle the TPCL can form a closed loop around the pillars \citep{semprebon2012advancing} such that a portion of the pillar is submerged under the liquid or the liquid may start exhibiting hemiwicking \citep{ishino2007wicking}. The study of such equilibrium interfacial morphologies is also left to future work.

\begin{figure}
 	\centering
 	\includegraphics[width=0.85\textwidth]{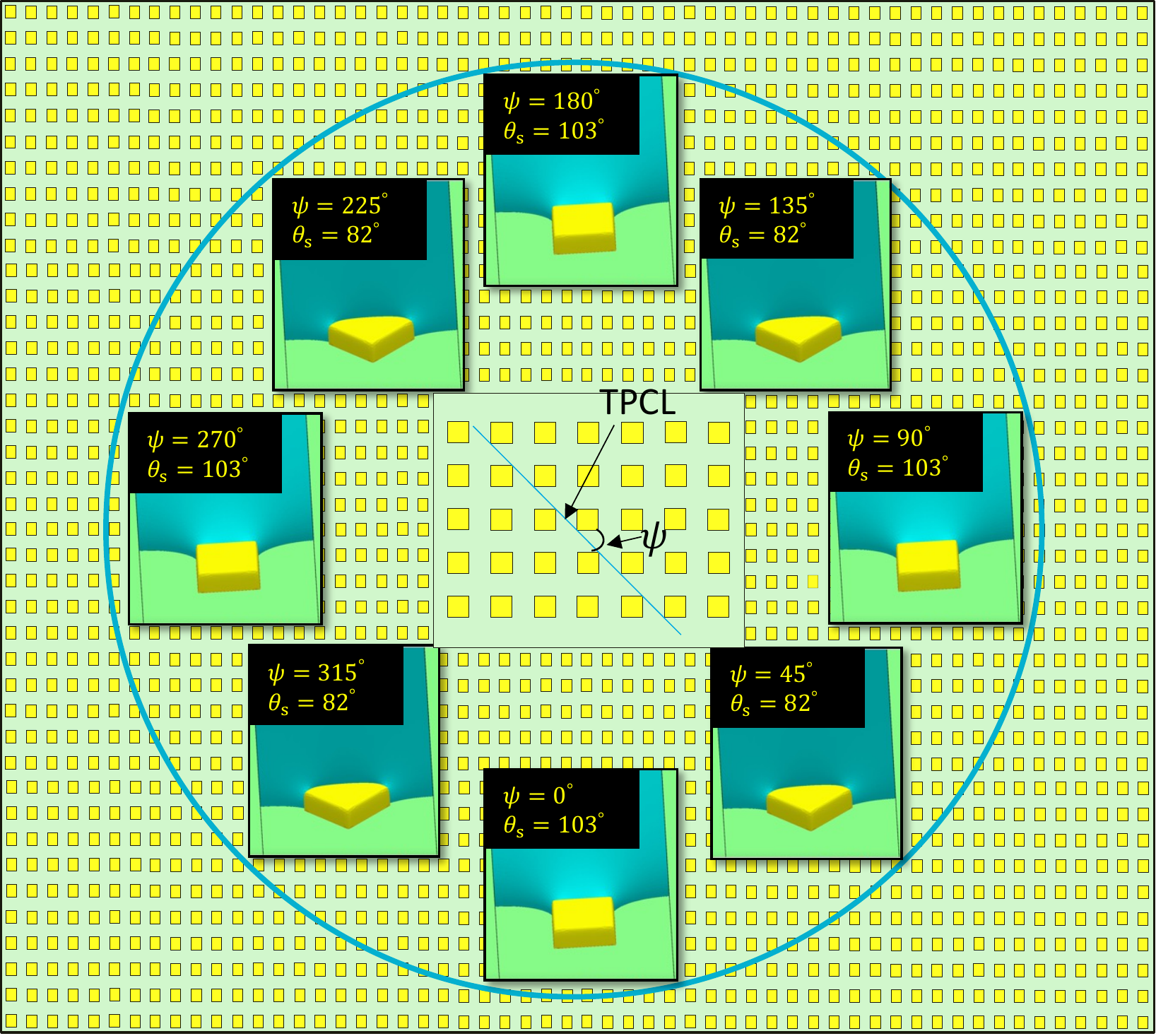}
 	\caption{Representation of the TPCL on a surface with a structured array of pillars. The TPCL can approach pillars from different directions (a typical scenario is shown in the inset where the TPCL is approaching pillars at an angle of $\psi$\textdegree), resulting in different equilibrium morphologies of the interface. Inset shows interface morphologies during the critical state for different approach ($\psi$) directions. Interface morphologies are simulated for a surface with $\theta_{\rm{e}}=72$\tc and $h/a=0.35$. The interface shapes in the inset only capture the effect of the pillar shape when the contact line approaches from a certain direction. In the actual scenario, the pillar shape as well as the inter-pillar distance would change depending on the direction of the approaching interface and we may end up getting very different interface morphologies. }
 	\label{fig:full_contact_line_ch6}
 \end{figure}

To understand the small differences between the simulated and experimental results shown in figure \ref{fig:forsberg_ch6}, particularly for the advancing state, it is important to note that during the spreading of a droplet, the interface can approach a pillar from different directions. Figure \ref{fig:full_contact_line_ch6} (inset) shows a TPCL approaching pillars at an angle $\psi$ relative to the periodicity direction of the structured surface. 
In order to show the impact of the interface movement direction with respect to the surface periodicity, in figure \ref{fig:forsberg_ch6} we also plot the critical contact angle ($\theta_{\rm{s45,a}}$) and the mechanical energy balance derived contact angle ($\theta_{\rm{meb45,a}}$) for an interface advancing at an angle ($\psi$) of 45\tc ($\theta_{\rm{e}}=72$\tc and $h/a=0.35$) that we have computed numerically using a slightly different computational domain (see Appendix \ref{app:dynamics_45_deg}). We observe that $\theta_{\rm{s45,a}}$ values for $\psi=45$\tc are less than the equivalent $\theta_{\rm{s,a}}$ values calculated for an interface advancing in the direction of the surface periodicity. Given that the actual advancing or receding contact angle of a droplet spreading on a structured surface would be an average value for the interface advancing in different directions relative to the surface periodicity (\cite{dorrer2008drops} made a similar suggestion) it is encouraging to see that the experimentally measured values for $\theta_{\rm{exp,a}}$ are bounded by the equivalent $\theta_{\rm{meb,a}}$ and $\theta_{\rm{meb45,a}}$, results. In fact, analysis of the \cite{forsberg2010contact} study suggests that the advancing and receding contact angles were measured in such a way that they capture the TPCL movement in the direction of surface periodicity. Therefore it is encouraging that the $\theta_{\rm{exp,a}}$ scatter leans towards the $\theta_{\rm{s,a}}$ values rather than the $\theta_{\rm{s45,a}}$ values. We have shown the results for $\psi=45$\tc since the surface is periodic along that direction and we can use periodic boundary conditions in the simulations (as detailed in Appendix \ref{app:dynamics_45_deg}). However, for other directions of interface motion we cannot use periodic boundary conditions and a much wider computational domain is required to simulate the interface dynamics. These simulations, while feasible, are left to future study.

\section{Conclusions}
We have presented a novel numerical method to simulate the microscale interface dynamics on a structured surface. The present model can predict contact angle hysteresis during homogeneous wetting (Wenzel state) of micro-structured surfaces from just the knowledge of surface topology. By using superquadrics to model the pillar geometry, we are able to simulate the interface dynamics over pillars with rounded edges, closely. 
%This unique approach allows us to control the pillar's edge roundness for mimicking the actual pillar geometry more closely. 
Using a surface minimization approach to approximate fluid-driven interface movement, we obtain the energy dissipation during the advancing and receding motion of the interface. 

With our simulations that utilize a constant Young's angle, at low pillars area fractions we observe a logarithmic dependence of total non-dimensional energy dissipation ($\overline{D}$) and non-dimensional energy dissipation per pillar ($\overline{D}_1$) on the pillar area fraction ($\phi$), broadly consistent with the \cite{joanny1984model} analysis of pinning on sparsely spaced (dilute) strong defects. At higher area fractions a correction term ($B\phi$) was added to capture the predicted dissipation. Both $\overline{D}_1$ and $\overline{D}$ depend upon the pillar aspect ratio ($h/a$) as well, with higher dissipation observed at higher aspect ratios.  We simulated the receding motion of an interface with Young's angle ($\theta_{\rm{e}}$) by simulating the advancing motion of surrounding fluid with Young's angle (180\textdegree - $\theta_{\rm{e}}$). We observed a logarithmic relationship between $\overline{D}_1$ {\&} $\phi$ and $\overline{D}$ {\&} $\phi$ for the receding interface as well. The dissipation for the receding interface also increases with the pillar aspect ratio. At certain higher aspect ratios and/or higher area fractions we observed permanent pinning of the TPCL indicating the maximum range of validity of our Wenzel-based results. 
%While beyond the scope of the present study, the receding contact angles for such cases would approach 0 degrees, causing the liquid film to remain within the forest of pillars. 
We used the roughness scale mechanical energy balance from part I to obtain the equation for advancing and receding contact angles on surfaces with no inherent hysteresis (equations (\ref{eqn:meb_equation3_ch6}), (\ref{eqn:meb_rec_equation1_ch6})) and on surfaces with a finite inherent hysteresis (equations (\ref{eqn:advancing_eqn_general_ch6}), (\ref{eqn:receding_eqn_general_ch6})), based on the results of energy dissipation calculations.

Finally, we compared our model with the experimental data of \cite{forsberg2010contact} and obtained very good agreement between simulation and experimental results across all relevant area fractions, but particularly at low area fractions. At higher area fractions we find the model slightly overpredicts the advancing contact angle. However, we  also observed that the maximum stable contact angle ($\theta_{\rm{s}}$) for the interface depends upon the direction from which the interface approaches a pillar. Isolated simulations using an advance direction of ($\psi=45$\textdegree) to the surface periodicity direction show that neither the $\psi=0$\tc or 45\tc is the best representation of the interface advancement, but that the experimental results are in fact `bracketed' by the simulation results at these two angles. We conclude that, orientation effects should be more fully considered in future work. In this work, we have considered only a square pillar geometry using a very limited number of Young's angles and pillar aspect ratios. It would be interesting to see the effect of different pillar geometries and Young's angle on contact angle hysteresis. Again, this is left to future work. Lastly, the micro-scale interface dynamics and the CAH calculated on pillared surfaces are observed to depend upon the pillar aspect ratio rather than their actual size. Therefore, the present model for simulating interface advancement and the dissipation equations presented in \S\ref{sec:dissipation_advancing} are valid for any pillar sizes for which the separation of length scales can be observed in the system. Even though we developed equations (\ref{eqn:diss_dilute}) to (\ref{eqn:total_diss_non_dilute}) for the roughness in the form of micron sized pillars, the same equations can be used on surfaces with nanometric roughness as well, provided that the defects are strong enough to pin the interface, and that the various velocity and length scale constraints detailed in part I are satisfied.

\section*{Acknowledgements}
One of the authors (P.K.) acknowledges the financial support from the University of Melbourne in form of the Melbourne research Scholarships program. P.K. also acknowledges the support from Melbourne India Postgraduate Program (MIPP).

\section*{Declaration of interests}
The authors report no conflict of interest.

\section*{Author ORCIDs}
Pawan Kumar, https://orcid.org/0000-0002-8654-8514 \\
D.J.E. Harvie, https://orcid.org/0000-0002-8501-1344

\appendix
\section{Effect of step size on the interface dynamics}
\label{sec:step-size:appendix}
\begin{figure}
    \centering
    \includegraphics[width=0.85\textwidth]{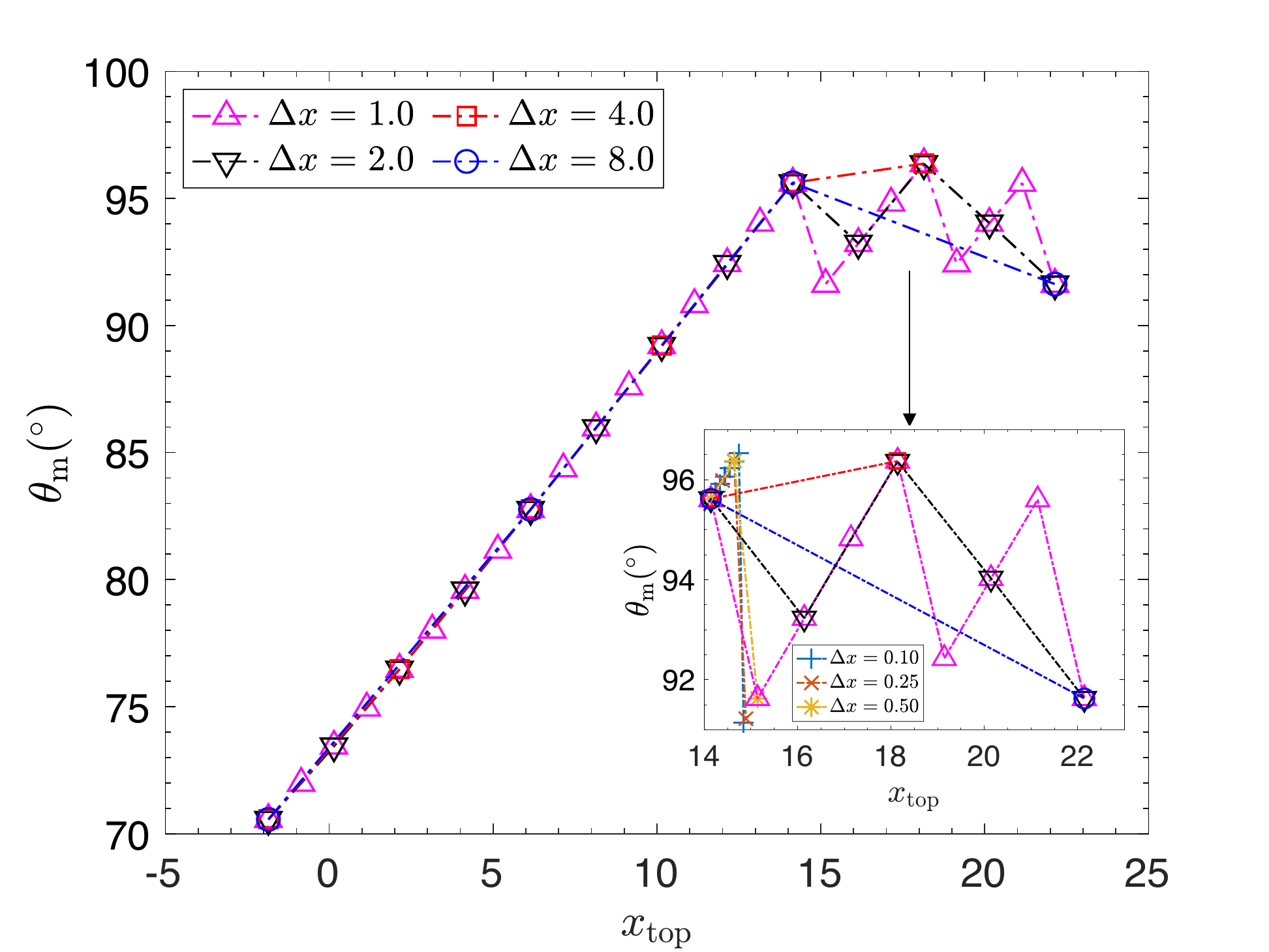}
    \caption{Variation in the macroscopic contact angle ($\theta_{\rm{m}}$) with the position of the interface ($x_{\rm{top}}$) for an advancing interface simulated by incremental advance method with a step size of $\Delta x=1.0,2.0,4.0$ and 8.0 respectively. The inset shows a zoomed view of the variation in $\theta_{\rm{m}}$ with $x_{\rm{top}}$ around a first critical state which also shows the variation with three additional step sizes of 0.50, 0.25 and 0.10 respectively, capturing only the first and second critical states. The simulations are performed with $\theta_{\rm{e}}=72$\textdegree, $h/a=0.35$, $\phi=0.08$, $W=3.5$ and $H=10W$.}
    \label{fig:delx_pillar_skip}
\end{figure}

In \S\ref{sec:int_dyn_sim} we discussed the method of incremental advance for simulating the interface dynamics on a rough surface. There we commented that a small $\Delta x$ is desired for accurately capturing the interface movement. In figure \ref{fig:delx_pillar_skip} we show the variation in macroscopic contact angle ($\theta_{\rm{m}}$) with the interface top position ($x_{\rm{top}}$) as a function of $\Delta x$. Specifically we have used seven different step sizes ($\Delta x=0.10,0.25,0.50,1.0,2.0,4.0$ and 8.0 respectively) for simulating the interface advancement using the method of incremental advance on a surface with $\phi=0.08$, $h/a=0.5$ and $\theta_{\rm{e}}=72$\tc within a domain which is $3.5$ units wide and $35$ units high. Even though we observe the same critical angle ($\theta_{\rm{s}}$) for $\Delta x=1.0,2.0,4.0$ and 8.0, this, however, may not be the most accurate prediction of $\theta_{\rm{s}}$. Since the accuracy with which $\theta_{\rm{s}}$ is calculated depends upon the magnitude of the step size, we observe a different $\theta_{\rm{s}}$ when a smaller step size is used. For example, we show the $\theta_{\rm{m}}$ vs. $x_{\rm{top}}$ variation capturing only the first and second critical states of the interface when step sizes of $\Delta x=0.50,0.25$ and 0.10 are used (shown as inset). We can see that the $\theta_{\rm{s}}$ value depicting the critical state of the interface depends upon the magnitude of $\Delta x$ with a better prediction of the critical state for a smaller $\Delta x$. Another observation from the figure is that for $\Delta x=4.0$, $\theta_{\rm{s}^{'}}$ is greater than $\theta_{\rm{s}}$, which is due to an inaccurate prediction of $\theta_{\rm{s}}$ when a bigger step size ($\Delta x=4.0$) is used.

\begin{figure}
    \centering
    \includegraphics[width=\textwidth]{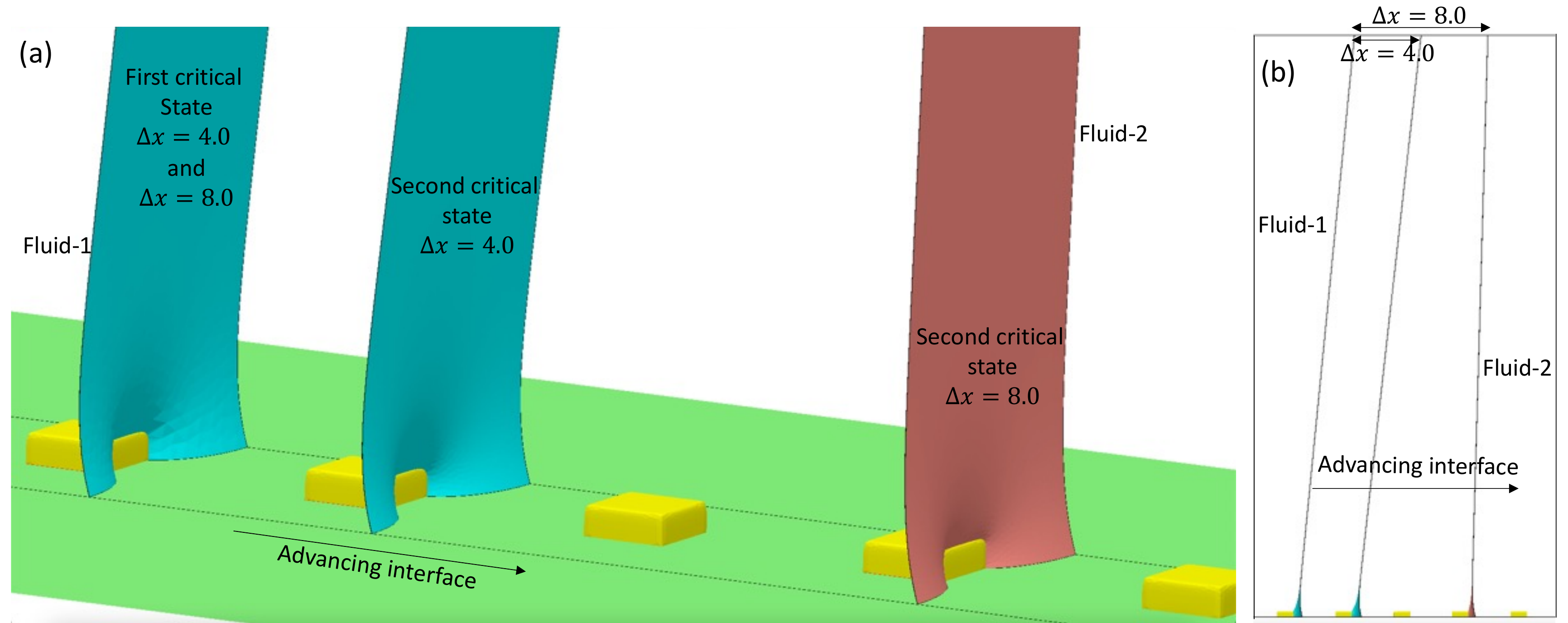}
    \caption{(a) Equilibrium interface morphologies capturing the TPCL jump during the interface advancement for two different step sizes, $\Delta x=4.0$ and $\Delta x=8.0$. The equilibrium interface morphology representing the second critical state for $\Delta x=8.0$ is shown red in colour. For $\Delta x=8.0$, the TPCL skips two pillars in the advancement direction before getting pinned on the third pillar. (b) Equilibrium interface morphologies depicting TPCL jump for $\Delta x=4.0$ and 8.0 respectively, showing the full domain height. The simulations are performed with $\theta_{\rm{e}}=72$\textdegree, $h/a=0.35$, $\phi=0.08$, $W=3.5$ and $H=10W$.}
    \label{fig:morphology_pillar_skip}
\end{figure}

Another interesting result that we may observe when a bigger step size is used for interface advancement is that the interface may not be able to find an equilibrium position on the next pillar in the direction of interface advancement and may skip a few pillars before the TPCL can finally pin again at a suitable location. We observed such behaviour for a step size of $\Delta x=8.0$, where the TPCL after depinning from the first pillar skipped two pillars in the direction of advancement before pinning again at the third pillar. Equilibrium interface morphologies depicting this behaviour are shown in figure \ref{fig:morphology_pillar_skip}.

\section{Equilibrium interface morphologies during a typical advancement}
\label{sec:morphologies_appendix}

\begin{figure}
    \centering
    \includegraphics[width=\textwidth]{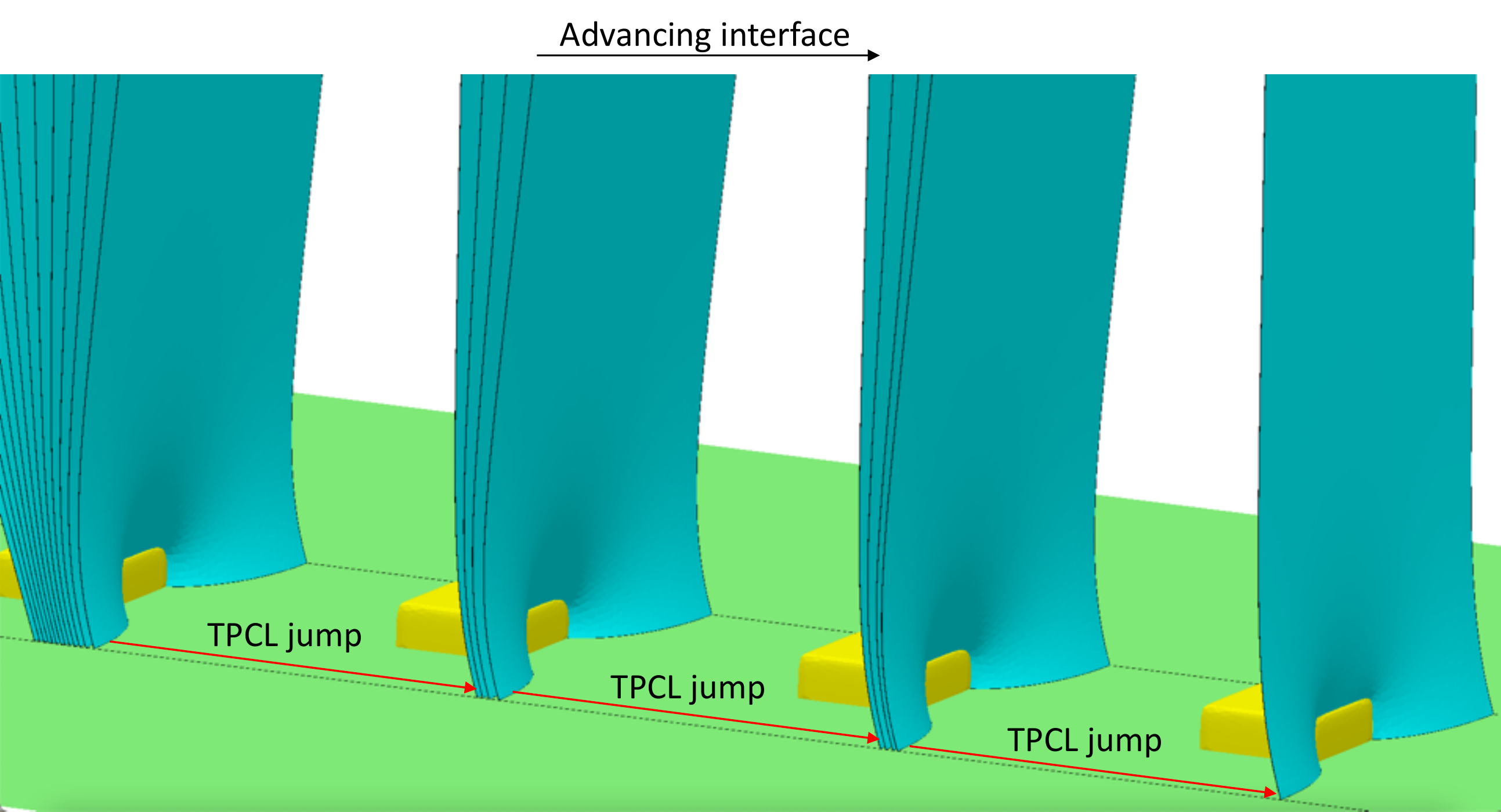}
    \caption{Equilibrium interface morphologies depicting a typical advancing motion. The TPCL moves in  a \textit{stick-slip} manner executing jumps, three such jumping events are shown in the figure. The simulations are performed with $\theta_{\rm{e}}=72$\textdegree, $h/a=0.35$, $\phi=0.08$, $W=3.5$ and $H=10W$.}
    \label{fig:morphology_advancing_interface}
\end{figure}

Equilibrium interface morphologies depicting a typical interface advancement are shown in figure \ref{fig:morphology_advancing_interface}. The simulation run starts with an equilibrium morphology with $x_{\rm{top}}$ chosen such that $\theta_{\rm{m}}=\theta_{\rm{e}}$. As the interface is advanced, the TPCL remains pinned on the pillar until the first critical state is reached ($\theta_{\rm{m}}=\theta_{\rm{s}}$). Upon further advancement of the interface, the TPCL depins from the first pillar and the interface only finds an equilibrium morphology (second critical state) when the TPCL gets pinned on the next pillar in the direction of advancement ($\theta_{\rm{m}}=\theta_{\rm{s}^{'}}$). The TPCL remains pinned on this pillar until the first critical state is reached again upon further interface advancement (i.e. $\theta_{\rm{m}}$ increases from $\theta_{\rm{s}^{'}}$ to $\theta_{\rm{s}}$) at which it executes a jumping event. In figure \ref{fig:morphology_advancing_interface} we have shown three TPCL jumping events depicting its typical \textit{stick-slip} motion.

\section{The critical-state method for simulating interfacial dynamics}
\label{sec:second-critical:appendix}

\begin{figure}
    \centering
        \begin{subfigure}[b]{\textwidth}
            \centering
            \includegraphics[width=\textwidth]{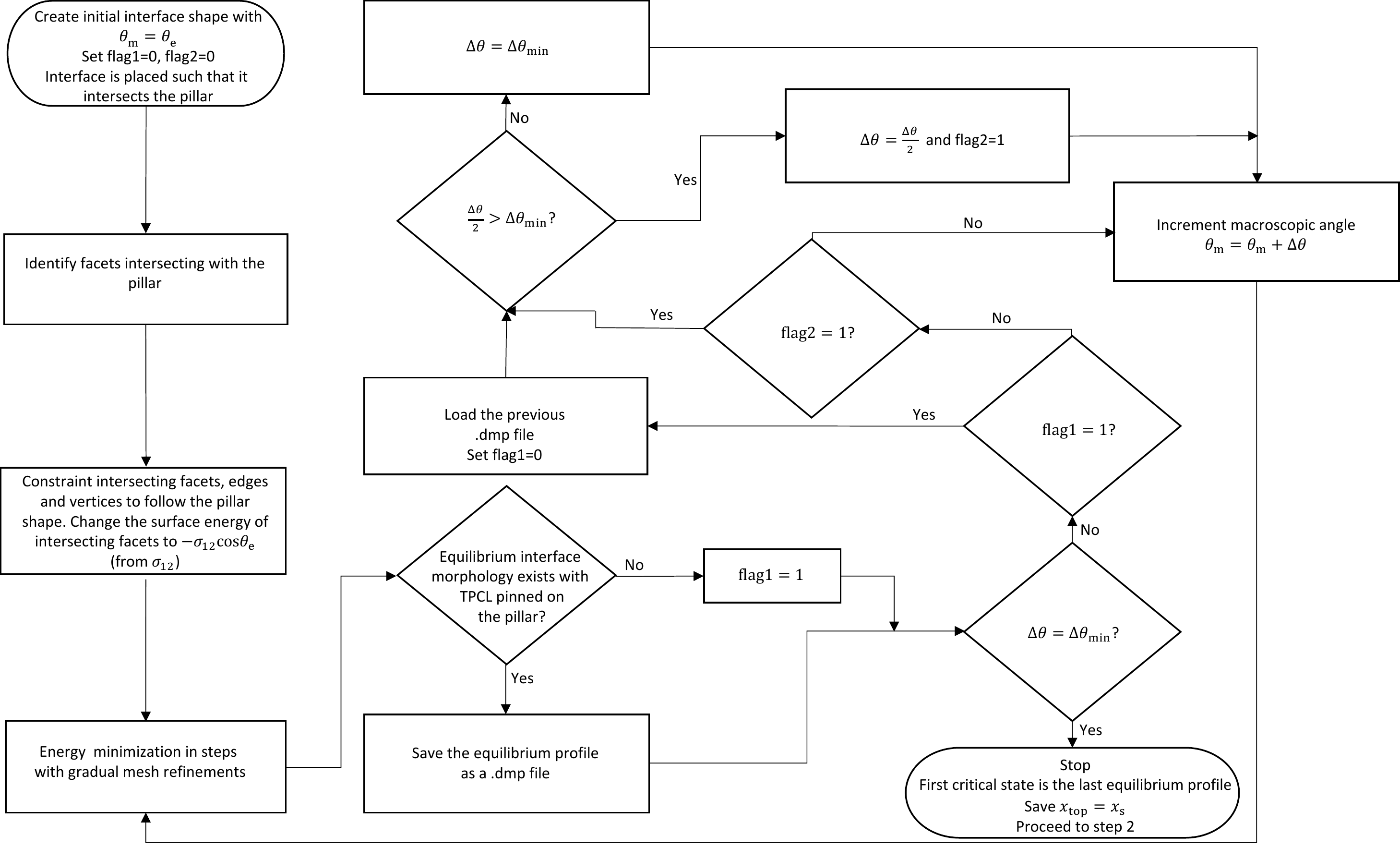}
            \caption{Step1 - Obtaining the first critical state of the interface.}
            \label{fig:critical_algo1}
        \end{subfigure}
        \hfill
        \hfill\hfill
    \begin{subfigure}[b]{\textwidth}
        \centering
            \includegraphics[width=0.76\textwidth]{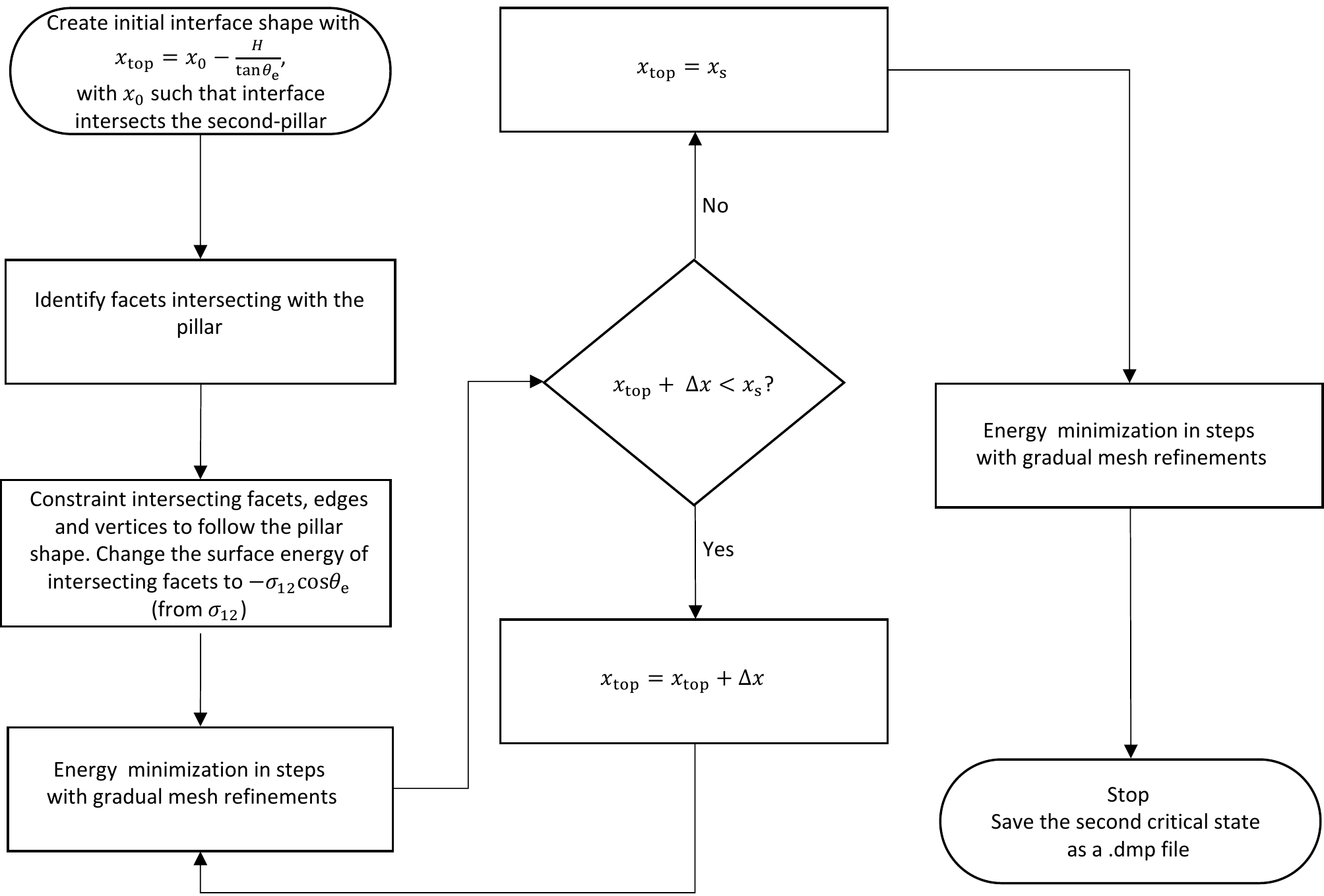}
            \caption{Step2 - Obtaining the second critical state of the interface.}
            \label{fig:critical_algo2}
        \end{subfigure}
        \caption{Flowchart of the algorithm used for simulating interfacial dynamics using the critical-state method.}
    \label{fig:critical_state_algo}
\end{figure}

In this section, we explain the critical-state method for simulating the interface dynamics on a rough surface. This method involves two different steps for obtaining the first and second critical states respectively. For the first critical state, we start with an initial equilibrium morphology obtained by minimizing the total energy within the domain consisting of a flat surface inclined at Young's angle relative to the domain base. To carry out energy minimization, the boundary conditions used are $\theta_{\rm{m}}=\theta_{\rm{e}}$ at the domain top, equation (\ref{eqn:surface_normals_ch6}) along the TPCL and the orthogonality between the interface and the domain walls (equation (\ref{eqn:bc_wall})). If $d$ is the inter-pillar distance and the center of the pillar is situated at $y=0$, the domain walls are situated at $y=-d/2$ and $y=d/2$ respectively, on planes parallel to the $x-z$ plane (and at $y=0$ and $y=d/2$ if a half domain is used, see figure \ref{fig:physical_model_ch6} for domain geometry and coordinate system). Also, initially, the interface is located near the pillar such that it is intersecting it. The intersecting facets of the pillar and the interface are identified and the intersecting facets, edges and vertices (of the interface) are constrained to follow the shape of the pillar, i.e. they are constrained to follow equation (\ref{eqn:superquad_ch6}). Since the portion of the interface constrained to follow the pillar's profile represents the fluid-1/solid interface instead of the fluid-1/fluid-2 interface, its surface energy is changed from $\sigma_{12}$ to $-\sigma_{12} \cos \theta_{\rm{e}}$. Now the energy minimization is carried out in a number of steps with gradual mesh refinements and an equilibrium interface morphology is obtained. This is the initial equilibrium shape with $\theta_{\rm{m}}=\theta_{\rm{e}}$. To simulate the advancing motion of the interface, $\theta_{\rm{m}}$ is increased by a small amount ($\Delta \theta$), which is the step size of interface advancement and the energy minimization with gradual refinements is carried out again to obtain the equilibrium interface morphology with $\theta_{\rm{m}}=\theta_{\rm{m}}+\Delta \theta$, with the rest of the boundary conditions unchanged. This process is repeated to simulate the interface advancement. The interface advancement is stopped when the $\theta_{\rm{m}}$ reaches a value such that any further advancement results in the depinning of the TPCL, i.e. the interface cannot exist in equilibrium with the TPCL pinned on the pillar. This equilibrium interface morphology is the first critical morphology and the macroscopic angle is the (first) critical angle ($\theta_{\rm{s}}$) (see \cite{forsberg2010contact}, \cite{semprebon2012advancing} for a discussion on the advancing contact angle being the same as the maximum macroscopic contact angle for which the interface can be in mechanical equilibrium while pinned on a pillar). The accurate prediction of the first critical interface morphology depends upon the step size and a smaller step size gives a better approximation. Therefore, once a first critical state is obtained by using a certain $\Delta \theta$, the step size is reduced to $\Delta \theta/2$ and the first critical interface morphology is obtained again. To this end, the first critical interface morphology obtained using the step size of $\Delta \theta$ serves as the starting point and the process of successive energy minimization and interface advancement is repeated to obtain a better estimate of the first critical interface morphology. This process is repeated again until we reach the minimum step size, $\Delta \theta_{\rm{min}}$ (for example, 0.1\tc which is used in this study). The discussion here covers the first step of the method, which is obtaining the first critical interface morphology. Figure \ref{fig:critical_algo1} shows the algorithm for this step of the method.

Once the first critical interfacial morphology is obtained to the desired level of accuracy, we proceed to obtain the second critical state of the interface. Based on the assumption of a slow-moving interface, the second critical state is the equilibrium interface morphology calculated such that the interface top remains at the same location, i.e. $x_{\rm{top}}=x_{\rm{s}}$ ($x_{\rm{s}}$ is obtained from the first step) while the TPCL pins on the next pillar along the advancing direction. To obtain the second critical state we start with an initial equilibrium interface morphology which is obtained by minimizing the total energy within the domain containing a flat interface inclined at Young's angle relative to the domain bottom and positioned close to the second pillar in the advancement direction such that the interface is touching the pillar. The boundary conditions used here are $x_{\rm{top}}=x_{\rm{initial}}$ at the top, equation (\ref{eqn:surface_normals_ch6}) along the TPCL and equation (\ref{eqn:bc_wall}) along the domain walls. Here, $x_{\rm{initial}}$ is the initial value for positioning the interface such that $x_{\rm{initial}}=x_0 - H/ \tan \theta_{\rm{e}}$, this ensures that we start with macroscopic contact angle being same as Young's angle. The intersecting facets of the second pillar and the interface are identified and the facets, edges and vertices (of the interface) are constrained to follow the profile of the intersecting pillar. Also, the surface energy of the intersecting facets of the interface is changed to $-\sigma_{\rm{12}} \cos \theta_{\rm{e}}$ (from $\sigma_{12}$) followed by energy minimization with gradual mesh refinements. The equilibrium interface morphology thus obtained is the initial morphology which is used to arrive at the second critical morphology by gradual advancements and energy minimization (with gradual mesh refinements). The interface is advanced by moving the top, i.e. $x_{\rm{top}}=x_{\rm{top}} + \Delta x$, where $\Delta x$ is the step size. The interface advancement is carried out until $x_{\rm{top}}+\Delta x < x_{\rm{s}}$, and when this happens, $x_{\rm{top}}$ is set as equal to $x_{\rm{s}}$ followed by the energy minimization with gradual mesh refinements. The equilibrium interface morphology thus obtained (at $x_{\rm{top}}=x_{\rm{s}}$) is the second critical morphology. This concludes the second and final step of the process. A flow chart depicting the algorithm used for obtaining the second critical state is shown in figure \ref{fig:critical_algo2}. 

\begin{figure}
    \centering
    \includegraphics[width=0.65\textwidth]{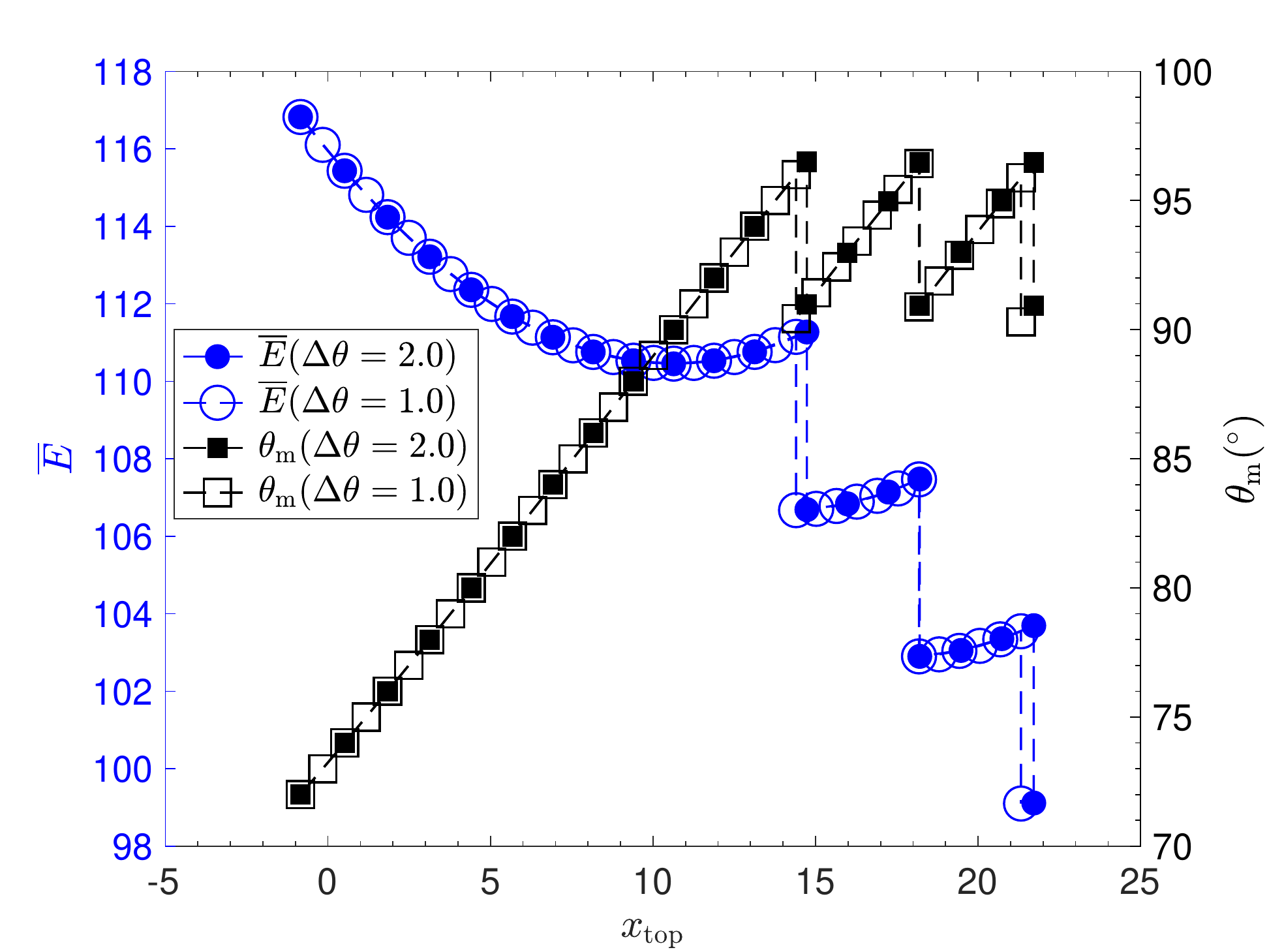}
    \caption{Variation in the total non-dimensional energy within the simulation domain ($\overline{E}$, shown as blue circles) and macroscopic contact angle ($\theta_{\rm{m}}$, shown as black squares) with the position of the interface ($x_{\rm{top}}$) for an advancing interface simulated by critical-state method with step sizes of $\Delta \theta=2.0$\tc (shown by filled symbols) and $\Delta \theta=1.0$\tc (shown by empty symbols). For $\Delta \theta=2.0$\tc we use $\Delta \theta_{\rm{min}}=0.1$\tc. The simulations are performed with $\theta_{\rm{e}}=72$\textdegree, $h/a=0.35$, $\phi=0.08$, $W=3.5$ and $H=10W$.}
    \label{fig:effect_of_del_th}
\end{figure}

In figure \ref{fig:effect_of_del_th} we show a typical variation in the total non-dimensional energy within the simulation domain ($\overline{E}$)\footnote{When the interface top is constrained by an angle ($\theta_{\rm{m}}$), additional energy equal to $x_{\rm{top}} W \cos\theta_{\rm{m}}$ is added to the total energy by SE, which needs to be subtracted from the total SE energy to get the true interfacial energy within the simulation domain.} and the macroscopic contact angle ($\theta_{\rm{m}}$) with the interface position ($x_{\rm{top}}$) during the interface advancement as captured by the critical-state method. Simulation results obtained with two different step sizes ($\Delta \theta$) i.e., $\Delta \theta=2.0$\tc (filled symbols) $\Delta \theta=1.0$\tc (empty symbols) are shown in the figure. We find that irrespective of the step size used, a very similar variation in the total non-dimensional energy and the macroscopic angle is observed. However, the critical states identified by the method can be different for different step sizes used (i.e., $\Delta \theta$, or more specifically $\Delta \theta_{\rm{min}}$ used in the simulation). For example, in figure \ref{fig:effect_of_del_th} we have used two different $\Delta \theta_{\rm{min}}$, 0.1\tc for the simulation run with $\Delta \theta=2.0$\tc and 1.0\tc for the run with $\Delta \theta=1.0$\tc respectively. We can observe two different first critical states (two different $\theta_{\rm{s}}$) identified by the method for two different $\Delta \theta_{\rm{min}}$ values, with the better approximation of the critical-state being achieved at the smaller $\Delta \theta_{\rm{min}}$. Since the second critical state depends upon the interface position during the first critical state ($x_{\rm{s}}$), a smaller $\Delta \theta_{\rm{min}}$ ensures a better approximation of the second critical state as well.

\section{Comparison between the incremental advance and critical-state method}
\label{sec:algo_comparison:appendix}

\begin{figure}
    \centering
    \includegraphics[width=0.65\textwidth]{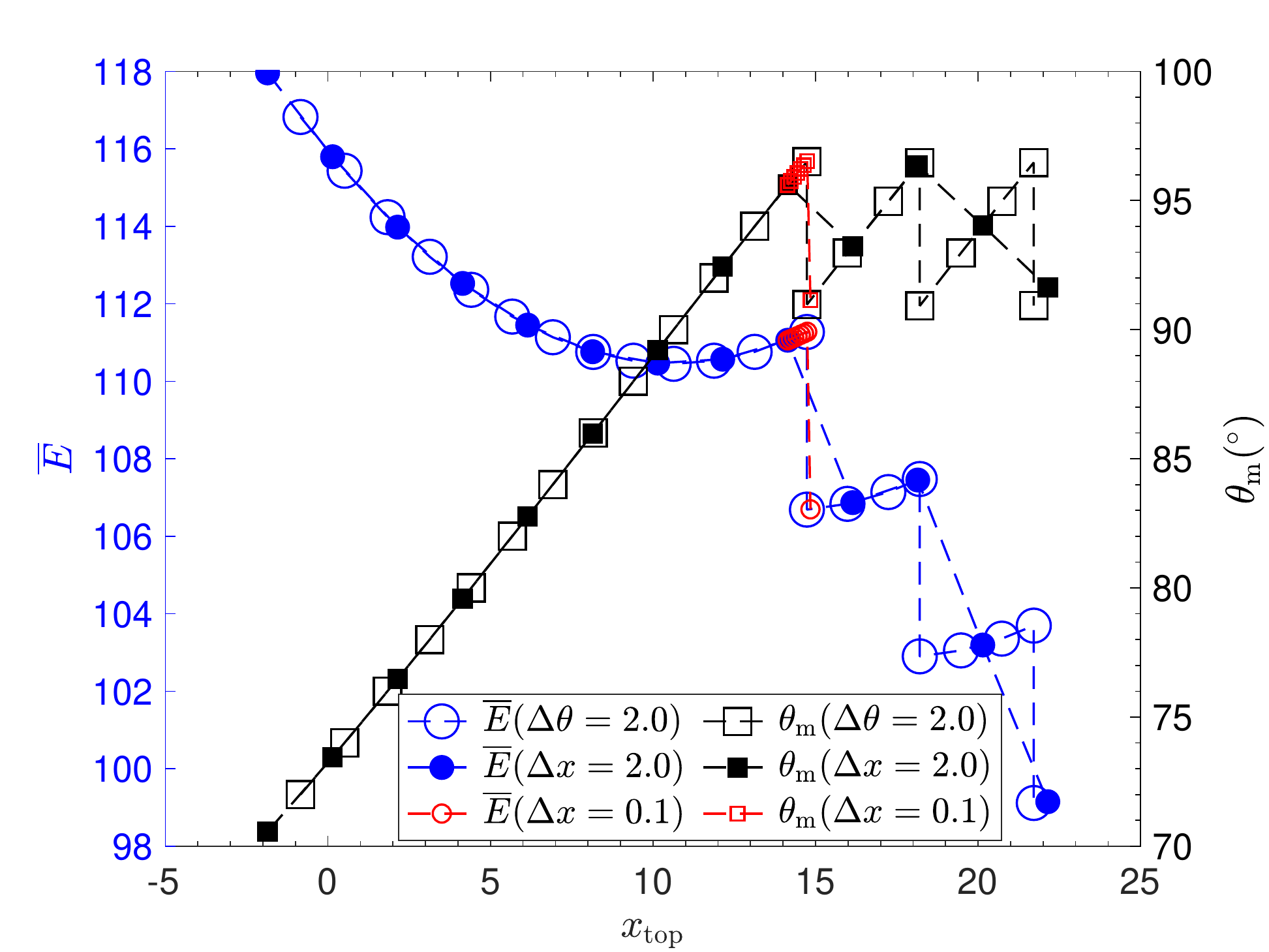}
    \caption{Variation in the total non-dimensional energy within the simulation domain ($\overline{E}$, shown as blue circles) and macroscopic contact angle ($\theta_{\rm{m}}$, shown as black squares) with the interface position ($x_{\rm{top}}$) for an advancing interface simulated by incremental advance method with a step size ($\Delta x$) of 2.0 (shown by filled symbols) and by critical-state method with a step size ($\Delta \theta$) 2.0 (shown by empty symbols). $\overline{E}$ and $\theta_{\rm{m}}$ values capturing the first and second critical states for $\Delta x=0.1$ are also shown in red colour for the incremental advance method. The simulations are performed with $\theta_{\rm{e}}=72$\textdegree, $h/a=0.35$, $\phi=0.08$, $W=3.5$ and $H=10W$.}
    \label{fig:approach_equivalence}
\end{figure}

In this work, we have demonstrated two methods, namely the incremental advance method and the critical-state method for simulating the advancing motion of an interface. Here we demonstrate that both methods capture the same interfacial dynamics. In figure \ref{fig:approach_equivalence} we show the variation in total non-dimensional energy within the simulation domain ($\overline{E}$) and the macroscopic contact angle ($\theta_{\rm{m}}$) with the interface position ($x_{\rm{top}}$) for an advancing interface simulated by incremental advance (filled symbols) and critical-state (empty symbols) methods respectively. The step sizes used for simulating the interface advancement are $\Delta x=2.0$ for the incremental advance method and $\Delta \theta=2.0$\tc, $\Delta \theta_{\rm{min}}=0.1$\tc for the critical-state method. We observe a very similar variation in $\overline{E}$ and $\theta_{\rm{m}}$ with $x_{\rm{top}}$ predicted via both the methods. The only difference is in the precise identification of the critical states, which is due to the finite step size used in the simulations. For example, when a smaller $\Delta x$ is used in the incremental advance method, the first critical state identified by the incremental advance method approaches that identified by the critical-state method. In figure \ref{fig:approach_equivalence} we show the variation in $\overline{E}$ and $\theta_{\rm{m}}$ with $x_{\rm{top}}$ capturing the first and second critical states when a step size of $\Delta x=0.10$ is used in the incremental advance method as red circles and squares respectively. We observe that the first and second critical states as captured by the incremental advance method approach to the corresponding states captured by the critical-state method when a smaller $\Delta x$ (0.1) is used. We therefore expect that this small difference between the two methods should vanish when a very small step size, i.e. $\Delta x \approx 0$ and $\Delta \theta_{\rm{min}} \approx 0$ is used. The critical-state method is preferred over the incremental advance method since we are only interested in the first and second critical states for calculating the dissipation in energy during interface advancement and hence this method is more efficient for a given accuracy level. However, there is a limitation to this method in that that it can only be used on structured surfaces when the interface is advancing in the direction of surface periodicity and all the pillars are identical.

\begin{figure}
    \centering
    \includegraphics[width=0.65\textwidth]{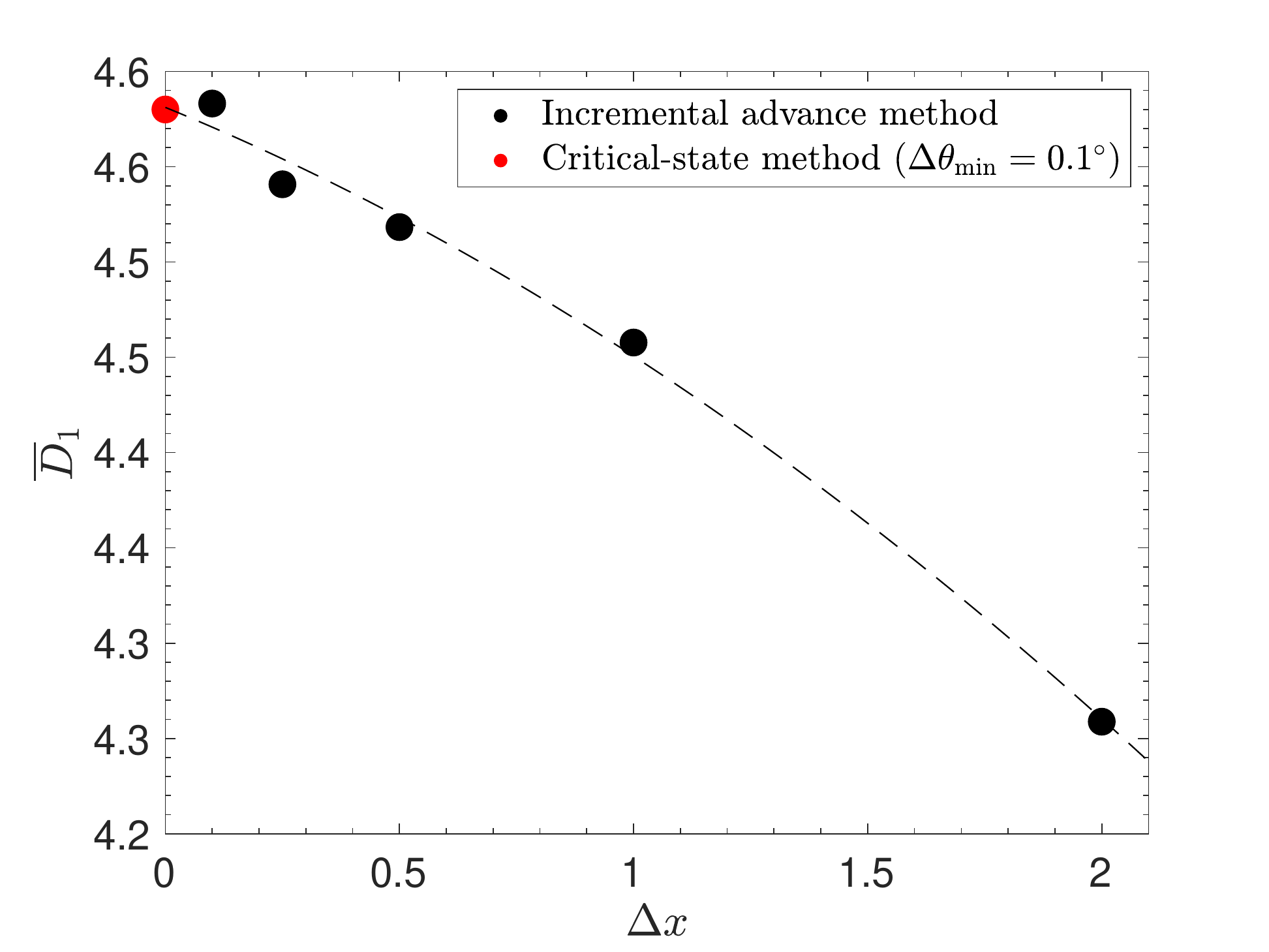}
    \caption{Variation in the non-dimensional energy dissipation per pillar ($\overline{D}_1$) with the step size ($\Delta x$) calculated using the incremental advance method for simulating interface dynamics. $\overline{D}_1$ calculated using the critical-state method for a $\Delta \theta_{\rm{min}}=0.1$\tc is a also shown as a red circle. The simulations are performed with $\theta_{\rm{e}}=72$\textdegree, $h/a=0.35$, $\phi=0.08$, $W=3.5$ and $H=10W$.}
    \label{fig:del_x_vs_diss}
\end{figure}

In the preceding discussion, we demonstrated that the incremental advance and critical-state methods generate the same critical interface topologies as the step sizes used in both methods approach zero. Therefore, we expect that the energy dissipation calculated by the two methods should approach the same value for small step sizes. In figure \ref{fig:del_x_vs_diss} we plot the variation in the non-dimensional energy dissipation per pillar ($\overline{D}_1$) calculated by the method of incremental advance with the step size ($\Delta x$). The $\overline{D}_1$ value calculated by the critical-state method for a $\Delta \theta_{\rm{min}}$ of 0.1\tc is also plotted (in red colour). We observe that as the step size ($\Delta x$) is reduced, the non-dimensional dissipation per pillar ($\overline{D}_1$) increases and approaches the $\overline{D}_1$ calculated by the method of critical-state. Hence we conclude that for an infinitesimal step size ($\Delta \theta \to 0$ and $\Delta x \to 0$), the dissipation values calculated by either of the two methods will be the same for an interface advancing on a structured surface in the direction of surface periodicity.

\section{Effect of mesh resolution}\label{sec:mesh_resolution:appendix}
Surface Evolver uses the finite element discretization method with each surface divided into a large number of triangular facets. The larger the number of facets, the better the representation of the curved surface, but at the same time the time required for convergence increases. We say the solution has converged when there is minimal change in the  total energy ($E$) of the system with further iterations. Specifically, if $E^n_g$ represents the total energy during the $n{\rm{th}}$ iteration then our convergence criteria can be represented by
\begin{equation}
|{E^{{n}}_g - E^{n+1}_g}|< \epsilon, 
\label{eqn:convergence_criterion_ch6}
\end{equation}
where $\epsilon \ll 1$. We have used  a value of $\epsilon = 10^{-7}$. 
Since the morphology of the interface pinned on a pillar is distorted in a small region local to the pillar, we choose a non-uniform mesh for representing the interface with the mesh being finer in regions closer to the pillar. Specifically, the mesh resolution is varied according to the following relation
\begin{equation}
    A_{\rm{f}}(r)= 
\begin{cases}
    A_{\rm{f}}(0)r^{n},& \text{if } r> 1\\
    A_{\rm{f}},              & \text{otherwise}.
\end{cases}
\label{eqn:mesh_refinement_scheme_ch6}
\end{equation}
Here, $A_{\rm{f}}(r)$ is the target area of each mesh facet at a distance $r$ from the center of the pillar's rear face (measured from the centroid of the facet), $A_{\rm{f}}(0)$ is the minimum area of the mesh facets and $n$ is a constant ($n>1$). $A_{\rm{f}}(0)$  and $n$ are the user-set constants that control mesh size. Equation (\ref{eqn:mesh_refinement_scheme_ch6}) is implemented via a user-written subroutine in Surface Evolver. 
%%%FIgure: Gradient Mesh
\begin{figure}
\centering
\includegraphics[width=0.85\textwidth]{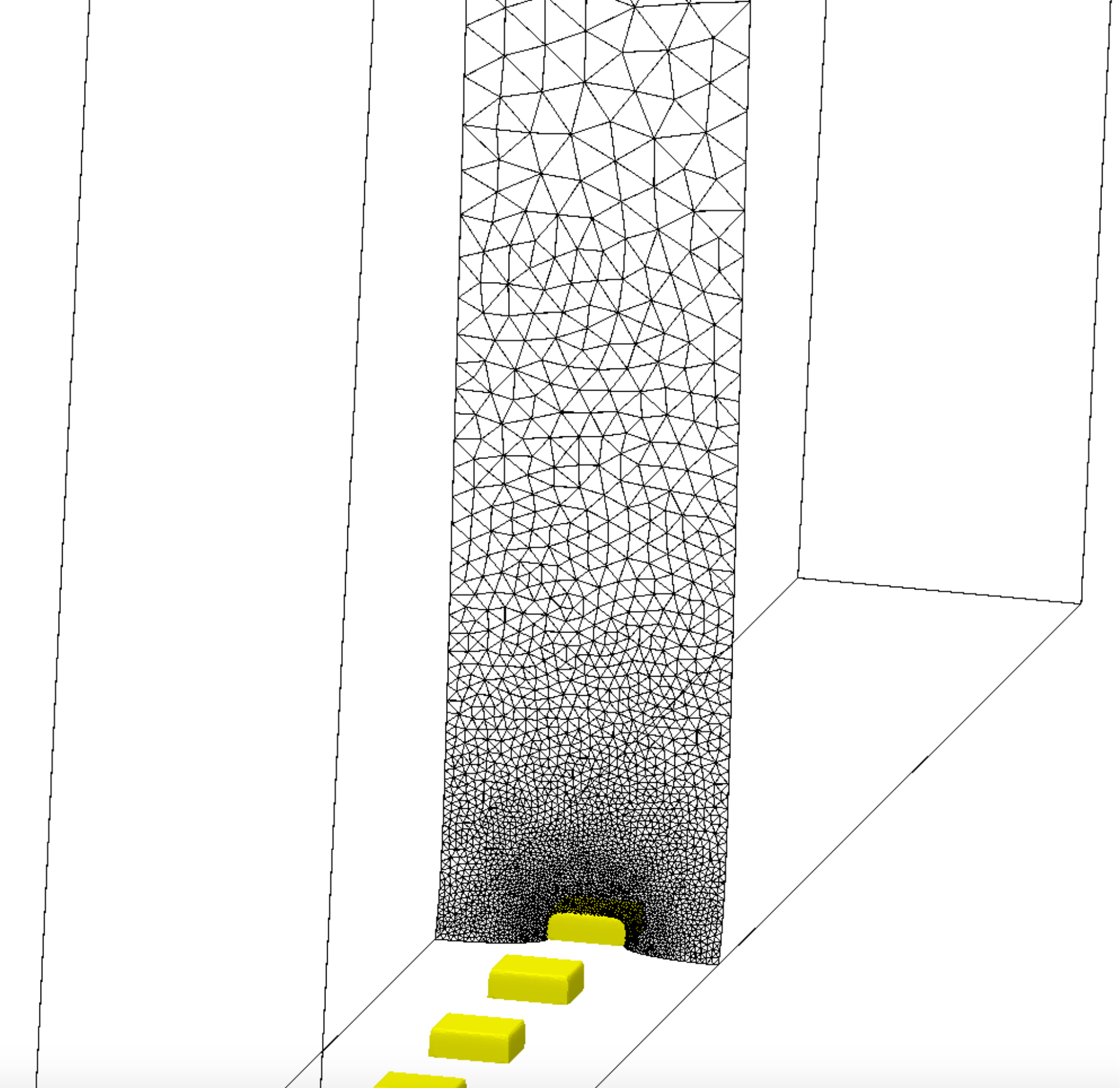}
\caption{Typical mesh used for representing the interface morphology. Mesh is very fine near the pillar to capture the fine details of the interface curvature. At distances away from the pillar, the interface is approximately planar and hence a relatively coarser is used. Mesh resolution is based on the relation given in equation (\ref{eqn:mesh_refinement_scheme_ch6}).}
\label{fig:interface_mesh_ch6}
\end{figure}
Figure \ref{fig:interface_mesh_ch6} shows a typical interface mesh having a high resolution near the pillar and a much coarser resolution at distances away from the pillar. Low values of $A_{\rm{f}}(0)$ and $n>1$ give the best representations of curved surfaces near the pillar but at the same time increase computational expense.
\begin{figure}
    \centering
    \includegraphics[width=0.65\textwidth]{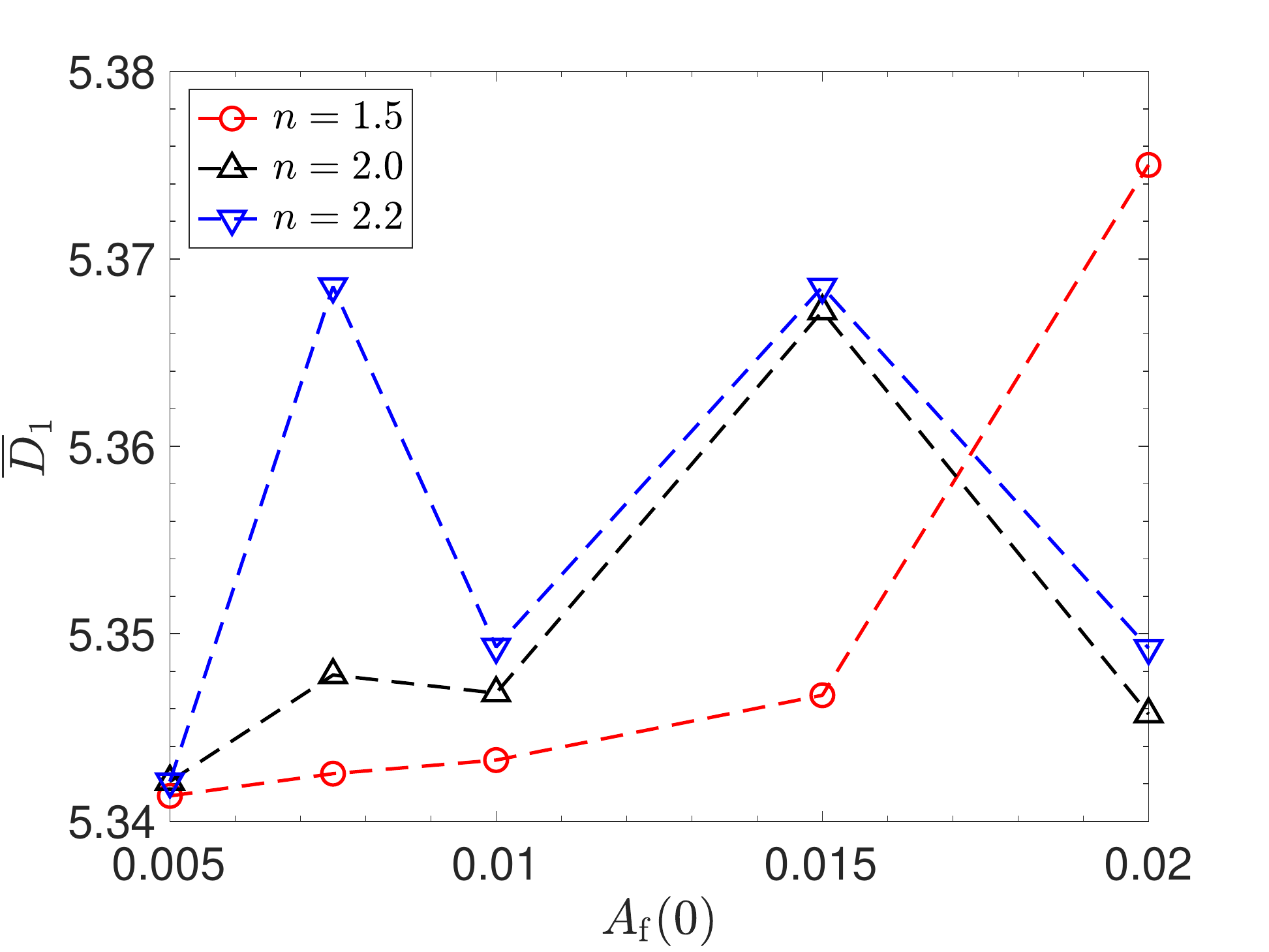}
    \caption{Variation in the non-dimensional energy dissipation per pillar $\overline{D}_1$ with the minimum non-dimensional mesh facet area ($A_{\rm{f}}(0)$) for different values of the coefficient $n$ in equation (\ref{eqn:mesh_refinement_scheme_ch6}). The simulations are done for a surface with $\theta_{\rm{e}}=72$\textdegree, $h/a=0.35$ and $\phi=0.08$. The percentage change in $\overline{D}_1$ with the change in $A_{\rm{f}}(0)$ reduces as $A_{\rm{f}}(0)$ is reduced. Percentage change in $\overline{D}_1$ as $A_{\rm{f}}(0)$ is reduced from 0.0075 to 0.005 is 0.11\% for $n=2.0$. In this work we have used $A_{\rm{f}}(0)=0.005$ and $n=2.0$.}
    \label{fig:mesh_sensitivity_ch6}
\end{figure}

In figure \ref{fig:mesh_sensitivity_ch6} we show the variation in non-dimensional dissipation per pillar ($\overline{D}_1$) with variations to the minimum mesh facet area ($A_{\rm{f}}(0)$) and exponential constant $n$. The dissipation ($\overline{D}_1$) changes with both $A_{\rm{f}}(0)$ and $n$, however this change diminishes as both $A_{\rm{f}}(0)$ and $n$ are reduced. For $A_{\rm{f}}(0)$=0.005 and $n=2.0$, the percentage change in $\overline{D}_1$ as $A_{\rm{f}}(0)$ is reduced from 0.0075 to 0.005 is 0.11\%. For the current work we have used $A_{\rm{f}}(0)=0.005$ and $n=2.0$.

\section{Effect of simulation domain height}\label{sec:channel_height:appendix}
\begin{figure}
	\centering
		\includegraphics[width=1.0\textwidth]{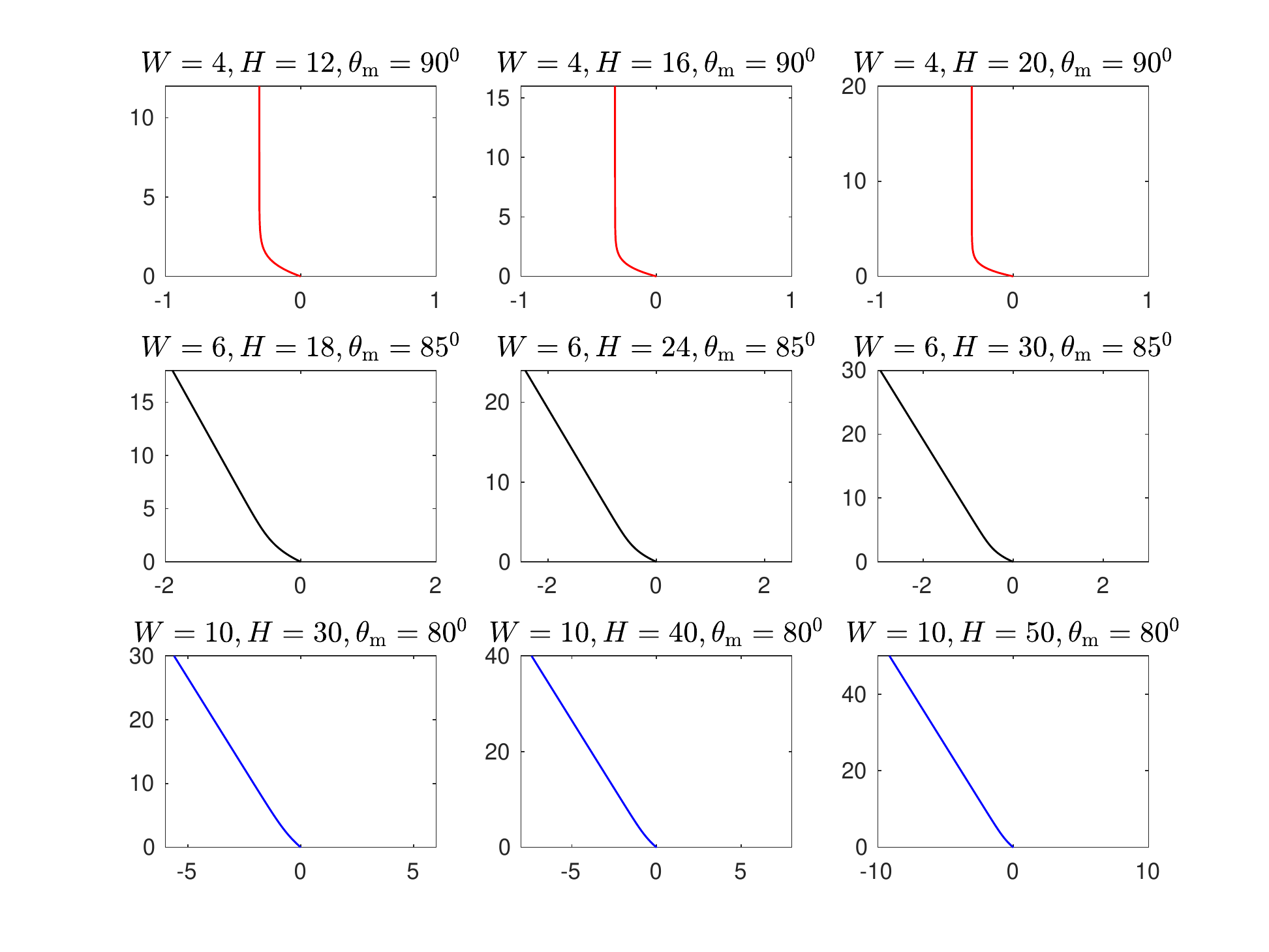}
		\caption{Projection of equilibrium interface profiles on domain wall for $\theta_{\rm{e}}=72$\tc and different values of domain width ($W$), height ($H$) and macroscopic angle ($\theta_{\rm{m}}$). The interface morphology is distorted only in a small region near the contact line and is approximately planar as we move away from it. For $H/W \geq 3$ we observe the interface profile to be approximately planar near the top for any value of $\theta_{\rm{m}}$.}
		\label{fig:channel_height_ch6}
\end{figure}

%%%theta channel height
Simulation domain width ($W$) depends upon the distance between pillars in the direction parallel to the TPCL. For a structured surface with pillars arranged in a square array, the domain width is related to the area fraction of pillars via $W = 1/ \sqrt{\phi}=d$. However, there is no strict limitation on domain height ($H$), but it should be high enough to ensure the interface is approximately planar near the top consistent with the separation of length scales required for the mechanical energy balance application as discussed in part I. Figure \ref{fig:channel_height_ch6} shows the interface profile projected on the domain wall for different $W$, $H$ and $\theta_{\rm{m}}$. We observe that for $H/W \geq 3$ (at least) we can get an approximately planar interface profile near the top. 
%%%Figure: interface projection
%%%%Figure: diss with phi, h/a 0.5 and th_Y 72deg
\begin{figure}
	\centering
		\includegraphics[width=0.65\textwidth]{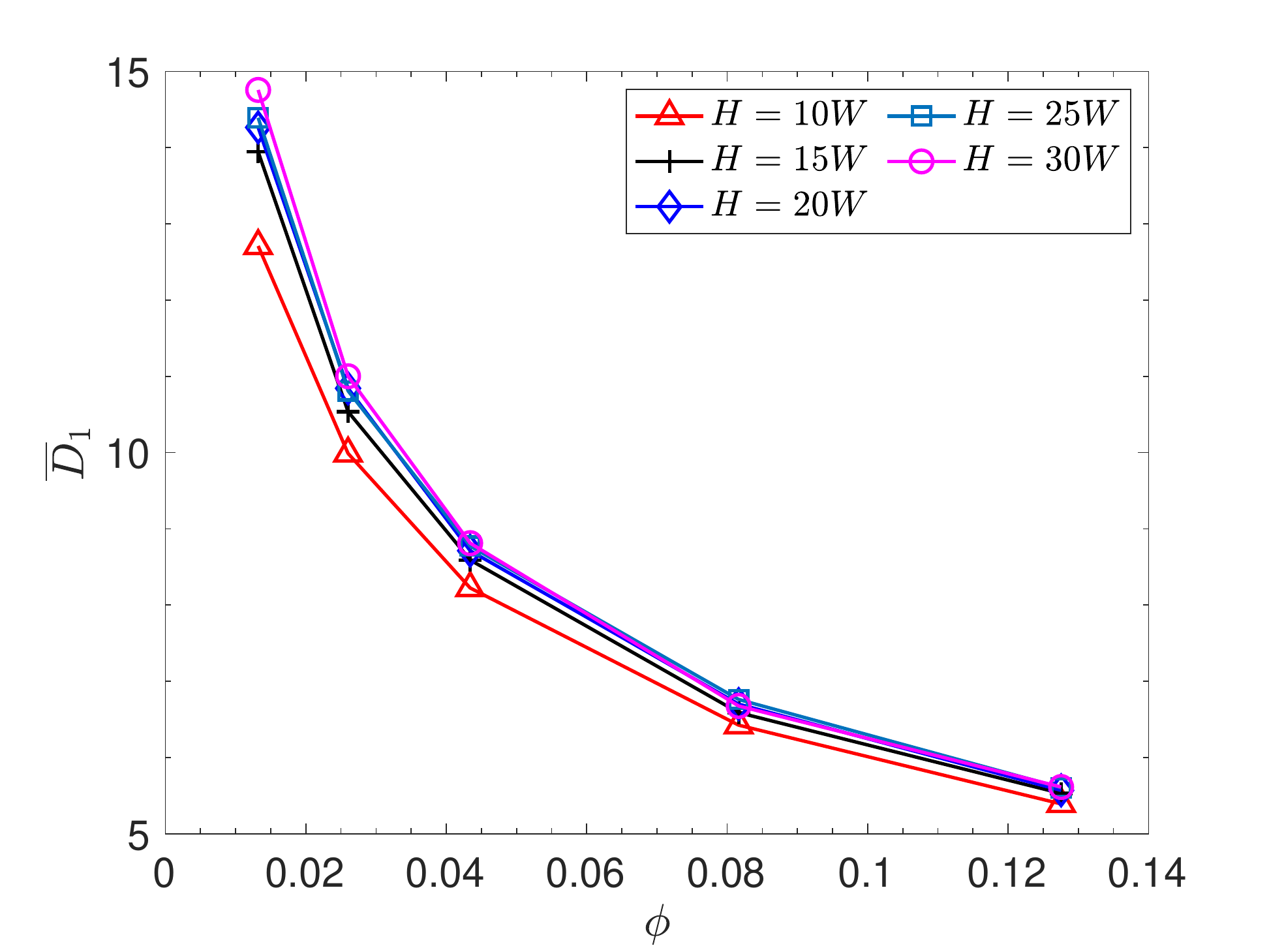}
		\caption{Variation of non-dimensional energy dissipation per pillar ($\overline{D}_1$) with pillar area fraction ($\phi$) for different domain heights on a surface with $\theta_{\rm{e}}=72$\tc and $h/a=0.5$.}
		\label{fig:diss_ch_ht_ch6}
\end{figure}
The height of the simulation domain also affects the energy dissipation during the interface motion. In figure \ref{fig:diss_ch_ht_ch6} we plot the variation in non-dimensional energy dissipation per pillar ($\overline{D}_1$) with the pillar area fraction ($\phi$) for different domain heights (expressed as a ratio of the domain width). We observe that $\overline{D}_1$ for the entire range of $\phi$, increases with the domain height and approaches a constant value as the domain height becomes large. 

\section{Simulating interface dynamics at 45\tc to surface periodicity}\label{app:dynamics_45_deg}

\begin{figure}
	\centering
		\includegraphics[width=0.65\textwidth]{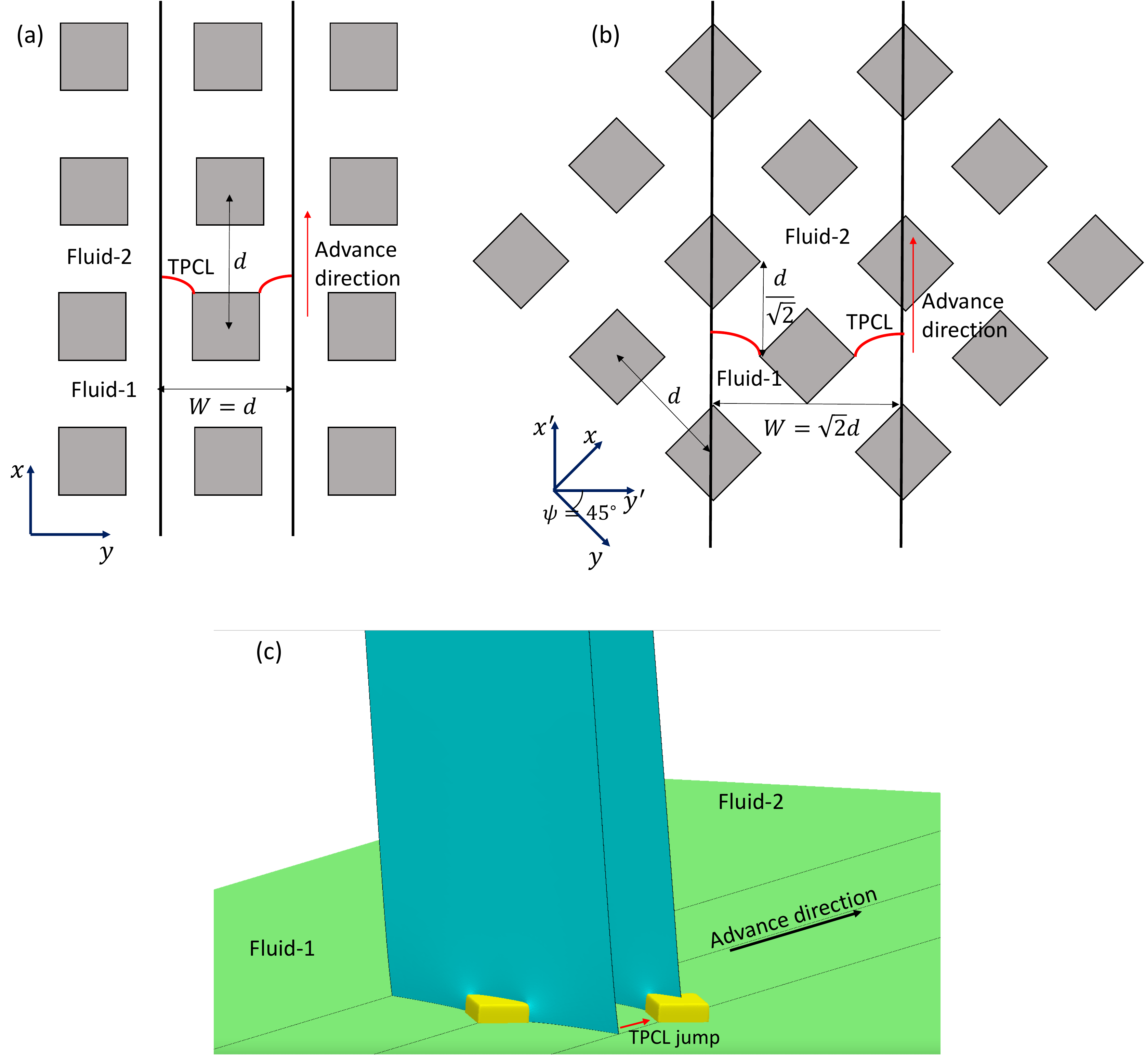}
		\caption{Schematic of a typical interface advance in the direction of surface periodicity (a) and at 45\tc to it (b). A full domain width is shown in (a) and (b). (c) Shows the equilibrium interface morphologies capturing the first and second critical states during a typical TPCL jumping event when the interface is advancing at 45\tc to the surface periodicity direction.}
		\label{fig:rotated_pillar}
\end{figure}

Here, we show the methodology for simulating the interface dynamics on a structured surface with pillars arranged in a square array and the interface advancing at 45\tc (i.e. $\psi=45$\textdegree, refer to figure \ref{fig:full_contact_line_ch6}) to the surface periodicity. If $d$ is the inter-pillar distance, then in direction of $\psi=45$\tc the inter-pillar distance along the domain width is $\sqrt{2}d$ and $d/\sqrt{2}$ in the direction of the interface advancement (shown in figure \ref{fig:rotated_pillar}(b)). Since the surface is periodic along 45\tc as well, therefore we can use periodic boundary conditions at the sides. The simulation methodology is the same as that discussed in Appendix \ref{sec:second-critical:appendix} with the only difference being the domain width now is $\sqrt{2}d$ instead of $d$. Also, the pillar is rotated by 45\tc (shown in figure \ref{fig:rotated_pillar}) and the superquadrics to achieve the particular pillar geometry can be represented by the following equation
\begin{equation}
\begin{split}
&\left(\frac{(x-x_1)\cos\psi + (z-z_1)\sin\psi}{s_1}\right)^{\epsilon_1}+\left(\frac{y-y_1}{s_2}\right)^{\epsilon_2}\\
    &+\left(\frac{-(x-x_1)\sin\psi + (z-z_1)\cos\psi}{s_3}\right)^{\epsilon_3}=1.
\label{eqn:superquad_rotated}
\end{split}
\end{equation}

\newpage 
\bibliographystyle{jfm}
\bibliography{bibliography}
\end{document}